\documentclass[a4paper,11pt]{article}
\pdfoutput=1

\usepackage{jheppub}

\usepackage[T1]{fontenc}

\title{\boldmath Chiral conformal field theory for topological states and the anyon eigenbasis on the torus}

\author[1,2,3]{Hua-Chen Zhang,}
\author[4]{Ying-Hai Wu,}
\author[1,2,5]{Tao Xiang}
\author[3]{and Hong-Hao Tu}

\affiliation[1]{Beijing National Laboratory for Condensed Matter Physics \& Institute of Physics, Chinese Academy of Sciences, Beijing 100190, China}
\affiliation[2]{School of Physical Sciences, University of Chinese Academy of Sciences, Beijing 100049, China}
\affiliation[3]{Institut f\"ur Theoretische Physik, Technische Universit\"at Dresden, 01062 Dresden, Germany}
\affiliation[4]{School of Physics and Wuhan National High Magnetic Field Center, Huazhong University of Science and Technology, Wuhan 430074, China}
\affiliation[5]{Beijing Academy of Quantum Information Sciences, Beijing 100193, China}

\emailAdd{hzhang@iphy.ac.cn}
\emailAdd{yinghaiwu88@hust.edu.cn}
\emailAdd{txiang@iphy.ac.cn}
\emailAdd{hong-hao.tu@tu-dresden.de}

\abstract{Model wave functions constructed from (1+1)D conformal field theory (CFT) have played a vital role in studying chiral topologically ordered systems. There usually exist multiple degenerate ground states when such states are placed on the torus. The common practice for dealing with this degeneracy within the CFT framework is to take a full correlator on the torus, which includes both holomorphic and antiholomorphic sectors, and decompose it into several conformal blocks. In this paper, we propose a pure chiral approach for the torus wave function construction. By utilizing the operator formalism, the wave functions are written as chiral correlators of holormorphic fields restricted to each individual topological sector. This method is not only conceptually much simpler, but also automatically provides us the anyon eigenbasis of the degenerate ground states (also known as the ``minimally entangled states''). As concrete examples, we construct the full set of degenerate ground states for SO($n$)$_1$ and SU($n$)$_1$ chiral spin liquids on the torus, the former of which provide a complete wave function realization of Kitaev's sixteenfold way of anyon theories. We further characterize their topological orders by analytically computing the associated modular $S$ and $T$ matrices.}


\makeatletter
\gdef\@fpheader{}
\makeatother
\begin{document} 
\maketitle
\flushbottom

\section{Introduction}
\label{sec:intro}

The study of strongly correlated quantum many-body systems has been a central theme of physics since the inception of quantum mechanics, where a large variety of interesting phenomena occur due to strong interactions between spins, bosons, and fermions. The fact that Hilbert-space dimensions grow exponentially with the number of constituents makes it notoriously hard to compute physical quantities exactly in general, and strong correlation precludes the success of perturbative and/or mean-field methods. Despite these difficulties in analytical and numerical studies, significant progresses have been made in understanding such systems. One crucial approach is the construction of many-body trial wave functions, which usually are not the~\emph{exact} ground states of physically realistic Hamiltonians but capture essential aspects of the interesting physics. Laughlin made a great contribution along this line~\cite{laughlin1983} by writing down a trial wave function for the fractional quantum Hall (FQH) state at filling factor $1/3$~\cite{tsui1982}. This wave function is not the exact ground state of Coulomb potential but very close to it in finite-size systems, and many subsequent works have further demonstrated that it correctly describes the fundamental properties of the $1/3$ state. Haldane pointed out that the Laughlin wave function is the zero-energy ground state of a certain pseudopotential Hamiltonian~\cite{Haldane1983}. The trial wave function approach has been indispensible for studying FQH physics since Laughlin's success, and many wave functions have been shown to provide information about the ground states and the low-energy excitations.

In a large number of quantum many-body systems, for which the FQH states are prominent examples, the characterizations of different phases and the transitions between them are beyond the Landau paradigm based on spontaneous symmetry breaking and emergence of local order parameters. In fact, various FQH states have exactly the same symmetry, and cannot be distinguished from each other by any local order parameter. To this end, new quantities that can probe the physics have to be considered. For gapped phases, one key observation was made by Wen~\cite{wen1989b,wen1990a}: when the system is placed on a closed surface with non-zero genus (e.g., the torus), there are multiple degenerate ground states with the degeneracy depending only on the genus. This topological degeneracy does not originate from symmetries of the system and is exact only in the thermodynamic limit. Systems with such a property are said to exhibit topological order, whose low-energy effective theories are topological quantum field theories (TQFTs)~\cite{atiyah1988,witten1989}. For two-dimensional (2D) systems with topological order, the degenerate ground states on the torus and the Berry phases of them under modular transformations of the torus are believed to provide a complete characterization of the bulk anyon theory (the chiral central charge is determined up to a modulo of 8)~\cite{keski-vakkuri1993,kitaev2006b}. One could define different bases for the ground-state subspace by taking linear combinations of the topologically degenerate ground states. However, there exists a specific basis called the anyon eigenbasis (also known as the basis of ``minimally entangled states''~\cite{zhang2012}), which is of special importance. As the name implies, the elements in this basis are in one-to-one correspondence with the types of anyons, namely gapped excitations above the ground states. The term ``anyon'' originates from the fact that the braiding statistics obeyed by these excitations can be neither bosonic nor fermionic. To study the properties of these anyonic excitations and characterize the topological order, it is desirable to construct trial wave functions for the anyon eigenbasis on the torus.

Apart from possessing non-trivial topological orders, many FQH states are also chiral. This means that time-reversal and reflection symmetries are broken by the external magnetic fields, and if the surface on which the system resides has edges (e.g., the two edges of a cylinder), there will be gapless modes localized in vicinity of the edges that propagate only in one direction. The edge modes of chiral FQH states are described by $(1+1)$D conformal field theories (CFTs). A deep insight due to Moore and Read~\cite{moore1991} has proven to be very useful in constructing trial wave functions for such chiral topological orders: they noticed that a number of FQH states can be expressed as conformal blocks, i.e., chiral correlators in certain CFTs~\cite{hansson2017}. Their observation is a powerful manifestation of the intriguing bulk-edge correspondence of topological states, which connects the wave function for the $2$D bulk topological order and the $(1+1)$D edge theory described by CFT. It is valid for many chiral topological orders and allows for a systematic construction of trial wave functions for such systems. This method has been generalized from the continuum to the lattice, both on the plane~\cite{nielsen2012,tu2013a,tu2014a,tu2014b,bondesan2014,nielsen2015,glasser2015,hackenbroich2017,herwerth2017,manna2018,manna2019,quella2020,jaworowski2020,jaworowski2021} and on the torus~\cite{nielsen2014b,herwerth2015,deshpande2016,zhang2021}. One instructive progress is to express the resulting wave functions as infinite-dimensional matrix product states~\cite{cirac2010}. In some cases, even the exact parent Hamiltonians for these states can be derived using the CFT null field technique~\cite{nielsen2011}, which can shed light on the realization of such states in experiments.

Based on aesthetic and practical considerations, one aspect of the previous constructions on the torus is not satisfactory. Even for chiral topological states, one has to compute the multipoint non-chiral correlator first, break it up into a sum of the products of chiral (holomorphic) and antichiral (antiholomorphic) components, and retain only the chiral components as model wave functions (i.e., the topologically degenerate ground states). The ``breaking up'' procedure appears to be somewhat artificial as no clear physical meaning can be ascribed to it. Moreover, the wave functions obtained in this approach are in general~\emph{not} the anyon eigenbasis, which are linear combinations of the CFT correlators. To overcome this drawback, we propose another approach that directly generates trial wave functions for the anyon eigenbasis of chiral topological states in lattice systems from various CFTs. In contrast to the approach mentioned before, we use~\emph{chiral} CFTs from the beginning and compute certain multipoint correlators on the torus. We assert that the wave functions constructed from these correlators are~\emph{automatically} the anyon eigenbasis. This claim is verified for several CFTs that admit free-field representations: compactified boson with special radii, the $\mathrm{SU}(n)_1$ Wess-Zumino-Witten (WZW) models, and the $\mathrm{SO}(n)_1$ WZW models. The wave functions on the torus are constructed explicitly and their topological orders are probed by studying how they transform under modular transformations of the torus, which demonstrates that they are indeed the anyon eigenbasis. We note that the wave functions associated with the compactified boson CFT reproduce the results reported in ref.~\cite{deshpande2016}. The wave functions constructed from the $\mathrm{SU}(n)_1$ and $\mathrm{SO}(n)_1$ WZW models are new results of this paper. Another highlight is that the wave functions constructed from the $\mathrm{SO}(n)_1$ WZW models provide a specific realization of Kitaev's sixteenfold way~\cite{kitaev2006b}, which is a classification of 2D topological orders described by free or weakly interacting fermions coupled to $\mathbb{Z}_{2}$ gauge fields~\cite{kells2011,wang2019,zhang2020,fuchs2020,chulliparambil2020,chulliparambil2021}. In a previous work of the same authors~\cite{zhang2021}, it was shown that the wave functions constructed from the $\mathrm{SO}(3)_1$ WZW model are equivalent to certain resonating valence bond (RVB) states in terms of fermionic partons. We will prove that similar equivalences hold for the $\mathrm{SO}(n)_1$ WZW model with $n \geq 3$. This generalizes the construction in ref.~\cite{tu2013a} from the plane to the torus.

The rest of this paper is structured as follows. In section~\ref{sec:general}, we briefly review the commonly adopted approach for constructing chiral topological states on the torus, which relies on conformal blocks of non-chiral CFTs. The chiral approach to be used in this paper is introduced as a simplified alternative. In section~\ref{sec:son}, our method is illustrated using the $\mathrm{SO}(n)_1$ WZW models, which leads to a set of wave functions that realize Kitaev's sixteenfold way. In section~\ref{sec:boson}, we use the compactified boson CFT to recover some known results for the Laughlin states, and then discuss the general $\mathrm{SU}(n)_1$ WZW model. For both $\mathrm{SO}(n)_1$ and $\mathrm{SU}(n)_1$ wave functions, we characterize their topological orders by analytically calculating the modular matrices. Section~\ref{sec:summary} summarizes this work and gives some outlook. The technical details about some calculations are relegated to three appendices. In appendix~\ref{appdx:derivation-correlator}, an operator formalism is employed to derive multipoint chiral correlators of free fermion and compactified boson CFTs on the torus. In appendix~\ref{appdx:RVB}, we provide an RVB formulation for the $\mathrm{SO}(n)_1$ wave functions. In appendix~\ref{appdx:derivation-modular}, some details about computing the modular matrices are explained.

\section{The chiral approach}
\label{sec:general}

This section presents the main result of this work, i.e., a method that produces model wave functions of chiral topological states on the torus directly from chiral correlators. We also provide arguments that these states correspond to the anyon eigenbasis.

\begin{figure}
\centering
\includegraphics[width=0.60\textwidth]{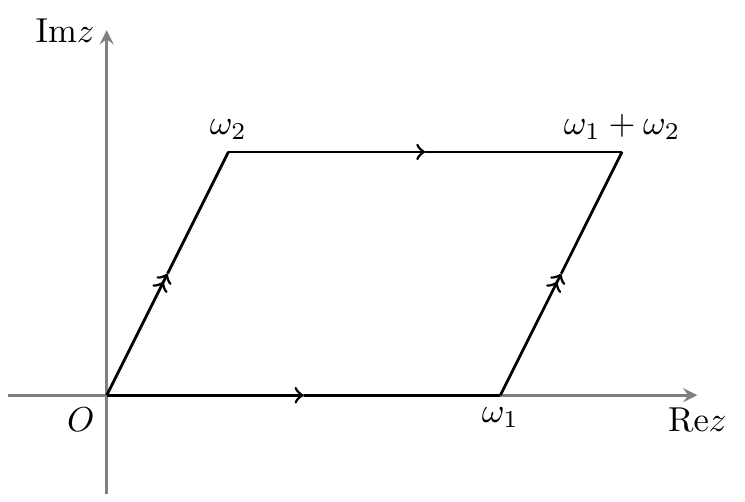}
\caption{Parametrization of the torus. Two noncollinear complex numbers, $\omega_{1}$ and $\omega_{2}$, define a parallelogram in the complex plane; the torus is then obtained by identifying the opposite sides of the parallelogram. The modular parameter $\tau$ is defined as the aspect ratio of the parallelogram,~$\tau = \omega_{2}/\omega_{1}$. Without loss of generality, we choose $\omega_{1}$ to be real and positive, and $\mathrm{Im}\tau>0$.}
\label{Figure1}
\end{figure}

To start with, we first review some basic concepts of $2$D CFTs that would be useful in our construction. All information about a CFT on the torus is encoded in the partition function
\begin{equation}
\label{eq:partition-function-definition}
Z(\tau) = \mathrm{Tr}\left( q^{L_{0}-c/24}\bar{q}^{\bar{L}_{0}-c/24} \right),
\end{equation}
where $c$ is the central charge of the CFT, $\tau$ is the modular parameter of the torus (see figure~\ref{Figure1}), $q = \mathrm{e}^{2 \pi \mathrm{i} \tau}$ ($\bar{q} = \mathrm{e}^{-2 \pi \mathrm{i} \bar{\tau}}$), and the zeroth Virasoro generators $L_{0}$ and $\bar{L}_{0}$ are defined by the Laurent expansions
\begin{equation}
\label{eq:virasoro-generators-definition}
    T(z)=\underset{n\in\mathbb{Z}}{\sum}~L_{n}z^{-n-2}, \qquad \bar{T}(\bar{z})=\underset{n\in\mathbb{Z}}{\sum}~\bar{L}_{n}\bar{z}^{-n-2}
\end{equation}
of the holomorphic and antiholomorphic energy-momentum tensors in terms of the complex coordinates $z$ or $\bar{z}$.

For model wave functions of 2D \emph{gapped} chiral topological states, it is sufficient to restrict ourselves to \emph{rational unitary} CFTs, which describe gapless edge excitations of these systems~\cite{wen1990b} and reflect the bulk topological order due to the remarkable bulk-edge correspondence~\cite{moore1991}. Further simplication can be made by assuming that the CFTs have \emph{diagonal} partition functions
\begin{equation}
    Z(\tau) = \sum_{a}\mathrm{ch}_{a}(\tau) \cdot \overline{\mathrm{ch}}_{a}(\bar{\tau})
\end{equation}
with chiral character
\begin{equation}
\label{eq:character-definition}
    \mathrm{ch}_{a}(\tau)=\mathrm{Tr}_{\mathcal{H}_{a}}\left(q^{L_{0}-c/24}\right)
\end{equation}
and its antichiral counterpart $\overline{\mathrm{ch}}_{a}(\bar{\tau})$. The number of characters in a rational CFT is \emph{finite}, i.e., the index $a$ belongs to a finite set. While the trace in the partition function~\eqref{eq:partition-function-definition} should be computed over the entire Hilbert space $\bigoplus_{a} \mathcal{H}_{a} \otimes \mathcal{\bar{H}}_{a}$, that in the chiral character~\eqref{eq:character-definition} is defined in the subspace $\mathcal{H}_{a}$, which is spanned by the so-called (chiral) primary state (denoted by $\vert a \rangle$) together with its descendants. Mathematically, $\mathcal{H}_{a}$ is the representation space with the highest weight state $\vert a \rangle$ of the chiral algebra (e.g., Virasoro algebra for Virasoro minimal models and Kac-Moody algebra for WZW models). The state-operator correspondence relates primary and descendant states to local operators, e.g., $\vert a \rangle = \lim_{z \rightarrow 0}\phi_{a}(z) \vert 0 \rangle$, where $\vert 0 \rangle$ is the vacuum and $\phi_{a}(z)$ is the (chiral) primary field associated with $\vert a \rangle$. By definition, a primary state is an eigenstate of $L_{0}$,~$L_{0} \vert a \rangle = h_{a} \vert a \rangle$, where the associated eigenvalue $h_{a}$ is the conformal weight of $\phi_{a}$.

The basic building block leading to the wave functions is the multipoint correlation functions of the primary fields~\cite{moore1991}. Specifically, we shall consider ground-state wave functions, for which the primary fields used in the construction are simple currents (i.e., these primary fields have a unique fusion outcome). For manifolds with zero genus such as the plane or the sphere, it suffices to consider only the chiral sector. In contrast, if one considers the torus geometry, the commonly adopted approach has to use the full correlators with both chiral and antichiral sectors (see ref.~\cite{hansson2017} for a review). These correlators of simple currents can be obtained by inserting $N$ fields into the partition function~\eqref{eq:partition-function-definition},
\begin{equation}
\label{eq:unnormalized-correlator-definition}
    \langle \phi_{s_{1}}(z_{1},\bar{z}_{1}) \cdots \phi_{s_{N}}(z_{N},\bar{z}_{N}) \rangle^{\prime} = \mathrm{Tr}\left( \phi_{s_{1}}(z_{1},\bar{z}_{1}) \cdots \phi_{s_{N}}(z_{N},\bar{z}_{N}) q^{L_{0}-c/24}\bar{q}^{\bar{L}_{0}-c/24} \right),
\end{equation}
where the index $s_{i}$ ($i = 1, \ldots, N$) specifies the field inserted at the position $z_{i}$. Here and throughout this article, the prime sign attached to a correlator indicates that it is~\emph{unnormalized} (i.e., without dividing by the partition function). For FQH systems in the continuum, an extra field representing the background charge should be included to fulfil the charge neutral condition such that the correlator~\eqref{eq:unnormalized-correlator-definition} would not vanish. To avoid this kind of system-specific issues, we shall focus on \emph{lattice} systems that have the same setup as in refs.~\cite{nielsen2014b,deshpande2016}. The coordinates $z_i$ in eq.~\eqref{eq:unnormalized-correlator-definition} represent the complex coordinates of the lattice sites, and $s_i$ are identified with the states $\vert s_{i} \rangle$ in the local Hilbert space at site $i$.

Loosely speaking, the primary fields $\phi(z,\bar{z})$ are related to the chiral ones via $\phi(z,\bar{z}) = \phi(z) \bar{\phi}(\bar{z})$, so the correlator~\eqref{eq:unnormalized-correlator-definition} takes the form
\begin{equation}
\label{eq:breaking-up}
    \langle \phi_{s_{1}}(z_{1},\bar{z}_{1}) \cdots \phi_{s_{N}}(z_{N},\bar{z}_{N}) \rangle^{\prime} = \sum_{\nu} \vert \mathcal{F}_{\nu}(\{s_{i}\};\{z_{i}\}) \vert^{2}.
\end{equation}
The chiral components $\mathcal{F}_{\nu}(\{s_{i}\};\{z_{i}\})$ are the so-called conformal blocks, which form a set of linearly independent functions. It was demonstrated in refs.~\cite{nielsen2014b,deshpande2016} that model wave functions for chiral topological states are given by conformal blocks as follows:
\begin{equation}
\label{eq:ansatz-previous}
    \vert \psi_{\nu} \rangle = \sum_{s_{1},\ldots,s_{N}} \mathcal{F}_{\nu}(\{s_{i}\};\{z_{i}\}) \vert s_{1},\ldots,s_{N} \rangle.
\end{equation}
For the lattice Laughlin states on the torus, this approach allows us to obtain analytical expressions of the model wave functions and explicitly compute their modular $S$ matrix~\cite{nielsen2014b,deshpande2016}. It is worth noting that there is a freedom in choosing the conformal blocks in eq.~\eqref{eq:breaking-up}, where a unitary basis rotation $\mathcal{F}_{\nu} \rightarrow \tilde{\mathcal{F}}_{\nu} = \sum_{\nu'} U_{\nu \nu'} \mathcal{F}_{\nu'}$ would give rise to a new set of linearly independent functions $\tilde{\mathcal{F}}_{\nu}$. This would result in a unitary rotation in the wave functions defined in eq.~\eqref{eq:ansatz-previous}, but the new ones span the same ground-state subspace and no physical effects are induced.

In recent years, it has become increasingly clear that the anyon eigenbasis on the torus or the cylinder (also known as the basis of ``minimally entangled states''~\cite{zhang2012}) is very useful for characterizing topological orders. Each state in this set has a definite anyon flux threading an incontractible loop of the system. If the anyon for one such state has topological spin $\theta_{a}$ (with $a$ labelling the anyon types) and the underlying TQFT has chiral central charge $c_{-}$, the state would transform to itself with an additional phase $\theta_{a}\mathrm{e}^{-2\pi\mathrm{i}c_{-}/24}$ under the Dehn twist (modular $T$ transformation). The quantum dimensions associated with all possible anyons in a system can be extracted from the entanglement entropy of the anyon eigenbasis~\cite{zhang2012}. For chiral topological states, the entanglement spectra of the anyon eigenbasis directly reflect the conformal towers of the corresponding edge CFT~\cite{li2008,qi2012}, which allows for numerical computations of conformal data. These properties have been widely exploited in numerical studies to probe topological states~\cite{cincio2013,zaletel2013,tu2013b,zhu2014,he2014,poilblanc2016,wu2020,szasz2020,chen2020,chenbb2021,chen2021}.

Despite the importance of anyon eigenbasis, a sufficiently general systematic method for constructing these states is still lacking. It would be very much desirable if the CFT approach can provide some insights into this problem. In the method mentioned above, both chiral and antichiral sectors are used, so the anyon eigenbasis amounts to particular decompositions of the full correlator into the conformal blocks. However, even for the CFTs which have free-field representations, the analytical expressions of the conformal blocks are quite complicated, let alone a large number of other CFTs whose conformal blocks do not have a closed-form expression. To fill this gap, we develop a method that generates the anyon eigenbasis directly. As such, the wave function ansatz
\begin{equation}
\label{eq:ansatz}
    \vert \psi_{a} \rangle = \sum_{s_{1},\ldots,s_{N}} \psi_{a}(s_{1},\ldots,s_{N}) \vert s_{1},\ldots,s_{N} \rangle
\end{equation}
is proposed, where the superposition coefficients
\begin{align}
\label{eq:wave-function-coefficients}
    \psi_{a}(s_{1},\ldots,s_{N}) & = \langle \phi_{s_{1}}(z_{1}) \cdots \phi_{s_{N}}(z_{N}) \rangle^{\prime}_{a} \nonumber \\
    & \equiv \mathrm{Tr}_{\mathcal{H}_{a}}\left(\phi_{s_{1}}(z_{1}) \cdots \phi_{s_{N}}(z_{N}) q^{L_{0}-c/24}\right)
\end{align}
are (unnormalized) \emph{chiral correlators evaluated in a specific sector} (denoted by $\mathcal{H}_{a}$). These chiral correlators can be viewed as chiral characters~\eqref{eq:character-definition} with primary fields inserted at each lattice site.

\begin{figure}
\centering
\includegraphics[width=0.70\textwidth]{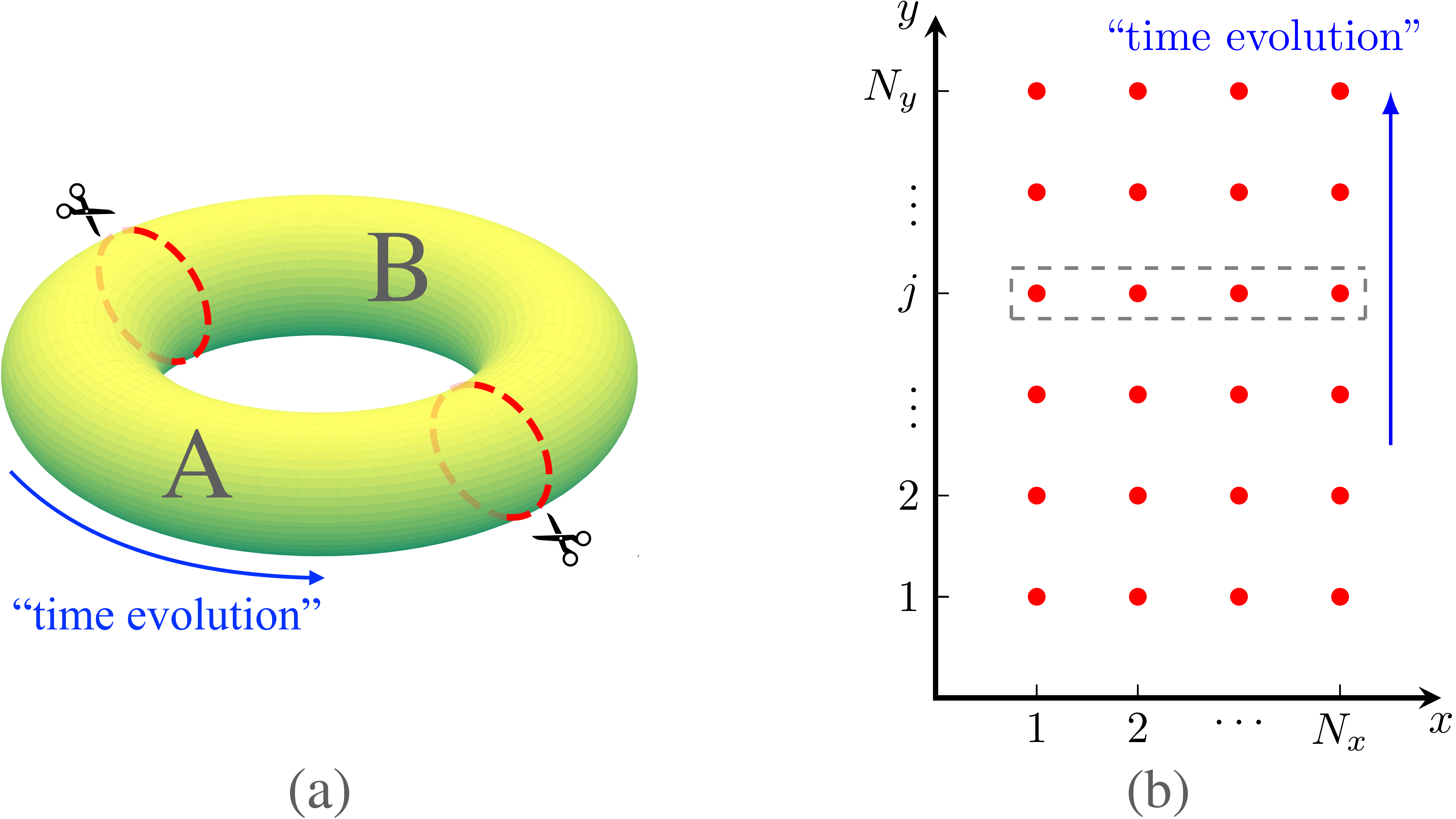}
\caption{Entanglement structure of the wave function defined in~\eqref{eq:wave-function-coefficients}. An entanglement cut (shown as the dashed lines in (a)) divides the torus into two halves (A and B). Then, this state is expected to admit the Schmidt decomposition $\vert\psi_a\rangle = \sum_{k} \lambda^a_k \vert\phi^a_{\mathrm{A},k}\rangle \vert\phi^a_{\mathrm{B},k}\rangle$, where $\vert\phi^a_{\mathrm{A}(\mathrm{B}),k}\rangle$ reside within $\mathcal{H}_{a}$. Intuitively, this can be understood by considering the ``space-time'' picture illustrated in (b). In the path integral formalism, the chiral correlator~\eqref{eq:wave-function-coefficients} is formulated as the chiral character with primary fields inserted on the ``space-time'' lattice. Nevertheless, by suitably choosing $N_{x}$ such that the primary fields on each ``time slice'' $j$ ($= 1, 2, \ldots, N_{y}$) fuse into the identity, the whole evolution is restricted to the same sector of the chiral CFT Hilbert space, namely $\mathcal{H}_{a}$, which agrees with the entanglement cut  being supported by states in this sector only.}
\label{Figure2}
\end{figure}

The intuition that guides us to this proposal is the connection between the CFT primary fields and anyons as well as the entanglement structure brought by eq.~\eqref{eq:wave-function-coefficients}. The model wave functions constructed here belong to the family of ``infinite-dimensional matrix product states'' (IDMPSs)~\cite{cirac2010,nielsen2012,zaletel2012,estienne2013,tu2015}. In the usual matrix product states (MPSs), the wave-function coefficients (in certain basis) are products of matrices defined on each lattice site, which act on the auxiliary Hilbert spaces that mediate entanglement between different bipartitions. The finite-dimensional matrices are replaced by ``infinite-dimensional matrices'' (primary fields) in IDMPSs, and the auxiliary space is the infinite-dimensional Hilbert space of CFTs. Nevertheless, IDMPSs inherit certain entanglement properties of MPSs. Without loss of generality, let us consider an $N_{x}{\times}N_{y}$ square lattice on the torus.~$N_{x}$ is suitably chosen such that the primary fields on each row of the lattice fuse into the identity sector. As shown in figure~\ref{Figure2}(a), an entanglement cut is introduced to divide the torus into two halves denoted by A and B. The state $\vert\psi_a\rangle$ is Schmidt decomposed as $\vert\psi_a\rangle = \sum_{k} \lambda^a_k \vert\phi^a_{\mathrm{A},k}\rangle \vert\phi^a_{\mathrm{B},k}\rangle$. One makes the crucial observation that the Schmidt vectors $\vert\phi^a_{\mathrm{A}(\mathrm{B}),k}\rangle$ must reside in the $a$-sector of the chiral CFT Hilbert space, regardless of their explicit expressions. The reason is that the primary fields in subsystems A and B fuse into the identity sector due to the form of eq.~\eqref{eq:wave-function-coefficients}. If $\vert\psi_a\rangle$ indeed belongs to the anyon eigenbasis, its entanglement spectrum $-\ln(\lambda^a_{k})$ would directly correspond to the spectrum of the chiral CFT in the $a$-sector~\cite{li2008,qi2012}, provided that $N_x$ and $N_y$ are much larger than the correlation length. This is unfortunately very difficult to prove in general, although some arguments in ref.~\cite{dubail2012} may be useful.

The same conclusion can also be achieved if we take the chiral correlator~\eqref{eq:wave-function-coefficients} as a ``space-time'' path integral, where $x$ ($y$) is the ``space'' (``Euclidean time'') direction [see figure~\ref{Figure2}(b)] and $q^{L_{0}-c/24}$ generates the ``time'' evolution of a 1D chiral CFT Hamiltonian. In eq.~\eqref{eq:wave-function-coefficients}, the initial states of the path integral reside in the sector $\mathcal{H}_{a}$ (since the trace is restricted to this sector), and the evolution itself cannot change the sector. At certain ``time'' intervals, the evolution hits (a row of) primary fields. The fields at each ``time'' slice fuse into the identity sector (this is made possible by appropriate choices of $N_{x}$ as mentioned before), so they cannot change the sector during the evolution. Therefore, the whole evolution is restricted to the $a$-sector, which agrees with our previous analysis that entanglement is only present between states in the $a$-sector.

While the entanglement structure can serve as good indications, a stronger justification of eq.~\eqref{eq:wave-function-coefficients} is still needed. In section~\ref{sec:son} and section~\ref{sec:boson}, we will check this proposal using several concrete examples. Specifically, model wave functions for $\mathrm{SO}(n)_1$ and $\mathrm{SU}(n)_1$ chiral spin liquids are constructed from the $\mathrm{SO}(n)_1$ and $\mathrm{SU}(n)_1$ WZW models, respectively. Furthermore, their modular $S$ and $T$ matrices are computed analytically. As we shall see, the modular matrices coincide with those of the CFTs from which the wave functions are constructed, and the modular $T$ matrices are diagonal. This confirms that our wave functions are indeed the anyon eigenbasis and provides a strong support for the validity of our proposal.

\section{States from the~\texorpdfstring{$\mathrm{SO}(n)_1$}{SO(n)1} WZW model and Kitaev's sixteenfold way}
\label{sec:son}

In this section, we explicitly calculate the wave functions defined by eq.~\eqref{eq:wave-function-coefficients} for the $\mathrm{SO}(n)_1$ WZW model. For odd (even) $n$, there are three (four) degenerate ground states on the torus. By analytically computing the modular $S$ and $T$ matrices, we show that these wave functions realize the $\mathrm{SO}(n)_1$ topological order, which is also known as Kitaev's sixteenfold way~\cite{kitaev2006b}. In particular, the diagonal form of the modular $T$ matrix confirms that our chiral approach indeed generates the anyon eigenbasis.

With central charge $c=n/2$, the $\mathrm{SO}(n)_1$ WZW model has a free field representation in terms of an $n$-component massless fermion~\cite{witten1984}. For our purpose, it is helpful to review the CFT of a single free massless Majorana fermion, which is just the Ising CFT. As already mentioned, the chiral part is sufficient for our construction.

\subsection{Ising CFT and free massless Majorana fermion}

On the complex plane parametrized by coordinate $z$, the mode expansion of the free fermion field $\chi(z)$ (with conformal weight $h = 1/2$) is
\begin{equation}
    \chi(z)=\underset{n}{\sum}~\chi_{n}z^{-n-\frac{1}{2}},
\end{equation}
and the modes satisfy $\{\chi_{n},\chi_{m}\}=\delta_{n+m,0}$.
For $n\in\mathbb{Z}+1/2$, the fermion has periodic (P) boundary condition: $\chi(\mathrm{e}^{2\pi\mathrm{i}}z)=\chi(z)$; for $n\in\mathbb{Z}$, it has antiperiodic (A) boundary condition: $\chi(\mathrm{e}^{2\pi\mathrm{i}}z)=-\chi(z)$. These two cases are known as the Neveu-Schwarz (NS) sector and the Ramond (R) sector, respectively.

The energy-momentum tensor of this free fermion is
\begin{equation}
\label{eq:EMtensor-fermion}
    T(z)=-\frac{1}{2}:\chi(z)\partial_{z}\chi(z): \, ,
\end{equation}
where $:\ldots:$ stands for normal-ordering. The normal-ordering prescription is chosen such that $\langle T(z) \rangle = 0$ for the NS sector and, accordingly, $\langle T(z) \rangle = 1/(16z^{2})$ for the R sector. The Laurent expansion of $T(z)$ yields the Virasoro generators
\begin{equation}
\label{eq:virasoro-generator-fermion}
    L_{n}=\frac{1}{2}~\underset{r}{\sum}~\left(r+\frac{1}{2}\right):\chi_{n-r}\chi_{r}:
\end{equation}
with
\begin{equation}
\label{eq:virasoro-generator-NS}
    L_{0}=\underset{r\in\mathbb{Z}+1/2,r>0}{\sum}r\chi_{-r}\chi_{r}
\end{equation}
for the NS sector and
\begin{equation}
\label{eq:virasoro-generator-R}
    L_{0}=\underset{r\in\mathbb{Z},r>0}{\sum}r\chi_{-r}\chi_{r}+\frac{1}{16}
\end{equation}
for the R sector.

The Ising CFT has three chiral primary fields: the identity field $\boldsymbol{1}$, the fermion field $\chi$, and the twist field $\sigma$. The corresponding chiral characters are given by
\begin{equation}
\label{eq:character-ising}
    \mathrm{ch}_{a}(\tau)= q^{-\frac{1}{48}}~\mathrm{Tr}_{\mathcal{H}_{a}}\left(q^{L_{0}}\right)
\end{equation}
with $a=\boldsymbol{1}, \, \chi, \, \sigma$. Let us consider first the chiral character of the identity sector $a=\boldsymbol{1}$, whose representation space is built by acting the Virasoro generators $L_{-n}$ with $n > 0$ on the ground state $\vert 0 \rangle$ of the NS sector. As the Virasoro generators~\eqref{eq:virasoro-generator-fermion} are bilinear in the fermion modes, the number of fermions must be even in this sector. This means that the chiral character of the identity sector can be written as
\begin{equation}
    \mathrm{ch}_{\boldsymbol{1}}(q) = q^{-\frac{1}{48}} ~\mathrm{Tr}_{\textrm{NS}} \left(\frac{1+(-1)^F}{2}q^{L_0}\right),
\end{equation}
where $(-1)^F$ is the fermion parity operator. Similarly, the states in the fermion sector $a = \chi$ are built by acting $L_{-n}$ on the one-fermion state $\chi_{-\frac{1}{2}} \vert 0 \rangle$ in the NS sector, which dictates that the number of fermions must be odd in this sector. Then, the chiral character of the fermion sector is
\begin{equation}
    \mathrm{ch}_{\chi}(q) = q^{-\frac{1}{48}} ~\mathrm{Tr}_{\textrm{NS}} \left(\frac{1-(-1)^F}{2}q^{L_0}\right).
\end{equation}
Finally, we consider the $\sigma$ sector. The operator product expansion (OPE) of $\chi$ with $\sigma$ reads~\cite{ginsparg1988}
\begin{equation}
\label{eq:OPE-twist}
    \chi(z) \sigma(w) \sim (z - w)^{-\frac{1}{2}} \mu(w) + \cdots,
\end{equation}
where $\mu$ is another field with the same conformal weight as $\sigma$. The square-root in this expression means that, when the fermion field is transported around $\sigma$, its sign gets flipped. In other words, inserting this twist field would change the boundary condition of the fermion field from P to A, or vice versa. Consequently, in the chiral character of the representation $\sigma$, the trace is taken over the R sector to produce
\begin{equation}
\label{eq:character-sigma}
    \mathrm{ch}_{\sigma}(q) = q^{-\frac{1}{48}} ~\mathrm{Tr}_{\textrm{R}} \left(\frac{1+(-1)^F}{2}q^{L_0}\right).
\end{equation}
One may wonder what happens if we project into the sector with odd fermion parity, instead of that with even fermion parity as in~\eqref{eq:character-sigma}. This would amount to the insertion of the field $\mu$ appearing in~\eqref{eq:OPE-twist}. However, since $\mathrm{Tr}_{\textrm{R}} ((-1)^{F} q^{L_0}) = 0$ as we will see below, this result is actually the same as $\mathrm{ch}_{\sigma}(q)$. We note that the projection into sectors with definite fermion parity using $[1 \pm (-1)^{F}]/2$ is known as the Gliozzi-Scherk-Olive (GSO) projection in string theory~\cite{polchinski2007}.

With this information in possession, we are ready to calculate the chiral correlators on the torus. As we have mentioned in section~\ref{sec:general}, the torus is obtained by identifying the two ends of a cylinder. The (infinite) cylinder, in turn, is related to the complex plane by the radial map
\begin{equation}
\label{eq:radial-map}
    z=\mathrm{e}^{-2\pi\mathrm{i}w},
\end{equation}
where the complex coordinate of the plane (cylinder) is $z$ ($w$). Under this map, the chiral primary field $\chi(z)$ transforms to
\begin{equation}
\label{eq:fermion-field-cylinder}
    \chi_{\textrm{cyl.}}(w)=\left(\frac{\partial z}{\partial w}\right)^{h}\chi(z)=(-2\pi\mathrm{i})^{\frac{1}{2}}~\underset{n}{\sum}~\chi_{n}~\mathrm{e}^{2\pi\mathrm{i}nw}.
\end{equation}
The circumference of the cylinder is $1$, and we have $\chi_{\textrm{cyl.}}(w+1)=-\chi_{\textrm{cyl.}}(w)$ when $n\in\mathbb{Z}+1/2$ and $\chi_{\textrm{cyl.}}(w+1)=\chi_{\textrm{cyl.}}(w)$ when $n\in\mathbb{Z}$. Note that the case is opposite to that on the plane: the NS sector has A boundary condition and the R sector has P boundary condition. For notational convenience, we drop the subscript ``cyl.'' of $\chi_{\textrm{cyl.}}(w)$ from now on and simply use $\chi(z)$ to represent fermion fields on the cylinder/torus. Recall that in the ``space-time'' picture introduced in section~\ref{sec:general}, a theory on the cylinder is naturally obtained by ``time-evolving'' a Hamiltonian defined on a ``spatial'' circle, and taking the trace amounts to compactifying also the ``time'' direction and converting the cylinder into the torus. Hence, the correlators we are going to calculate can then be expressed as
\begin{equation}
\label{eq:fermion-correlator}
    \langle \chi(z_1) \chi(z_2) \cdots \rangle^{\prime}_{a}
    = q^{-\frac{1}{48}}~\mathrm{Tr}_{\mathcal{H}_{a}} \left( \chi(z_1) \chi(z_2) \cdots q^{L_{0}} \right).
\end{equation}

As a warm-up, we begin with the ``zero-point'' case, which is, of course, nothing but the chiral characters~\eqref{eq:character-ising}. It is instructive to revisit these characters from a slightly different point of view. Since the fermion parity operator $(-1)^{F}$ anticommutes with the fermions, inserting it in $\mathrm{Tr}$ has the effect of changing the fermion boundary condition along the direction of~\emph{axis} of the cylinder from A to P. Recalling that the NS (R) sector corresponds to A (P) boundary condition along the direction of~\emph{circumference} of the cylinder, there are in total four fermionic spin structures according to the boundary conditions of the fermion along the two global loops of the torus: AA, AP, PA and PP, where the first (second) letter stands for the boundary condition along the circumference (axis) direction of the cylinder. The chiral characters can thus be represented as the following combinations of the holomorphic partition functions with definite spin structure:
\begin{align}
\label{eqs:chiral-character-ising-begin}
    \mathrm{ch}_{\boldsymbol{1}}(\tau) &= \frac{1}{2} \left( Z_{\textrm{AA}}(\tau) + Z_{\textrm{AP}}(\tau) \right), \\
    \mathrm{ch}_{\chi}(\tau) &= \frac{1}{2} \left( Z_{\textrm{AA}}(\tau) - Z_{\textrm{AP}}(\tau) \right), \\
    \mathrm{ch}_{\sigma}(\tau) &= \frac{1}{\sqrt{2}} \left( Z_{\textrm{PA}}(\tau) + Z_{\textrm{PP}}(\tau) \right),
\label{eqs:chiral-character-ising-end}
\end{align}
in which
\begin{align}
\label{eqs:chiral-partition-function-begin}
    Z_{\textrm{AA}}(\tau) &\equiv q^{-\frac{1}{48}} ~\mathrm{Tr}_{\textrm{NS}} \left(q^{L_0}\right) = \sqrt{\frac{\vartheta_{3}(\tau)}{\eta(\tau)}}, \\
    Z_{\textrm{AP}}(\tau) &\equiv q^{-\frac{1}{48}} ~\mathrm{Tr}_{\textrm{NS}} \left((-1)^{F}q^{L_0}\right) = \sqrt{\frac{\vartheta_{4}(\tau)}{\eta(\tau)}}, \\
    Z_{\textrm{PA}}(\tau) &\equiv \frac{1}{\sqrt{2}} q^{-\frac{1}{48}}~\mathrm{Tr}_{\textrm{R}} \left(q^{L_0}\right) = \sqrt{\frac{\vartheta_{2}(\tau)}{\eta(\tau)}}, \\
    Z_{\textrm{PP}}(\tau) &\equiv \frac{1}{\sqrt{2}} q^{-\frac{1}{48}}~\mathrm{Tr}_{\textrm{R}} \left((-1)^{F}q^{L_0}\right) = 0,
\label{eqs:chiral-partition-function-end}
\end{align}
where $\vartheta_{2}(\tau), \vartheta_{3}(\tau), \vartheta_{4}(\tau)$ are (standard) Jacobi's theta functions and $\eta(\tau)$ is Dedekind's eta function. For the definition of these special functions and the derivation of~\eqref{eqs:chiral-partition-function-begin} to~\eqref{eqs:chiral-partition-function-end}, see appendix~\ref{appdx:derivation-correlator}. Note that the holomorphic partition function $Z_{\textrm{PP}}(\tau)$ vanishes due to the existence of the fermionic zero mode in the R sector. This zero mode is shared by the holomorphic and the anti-holomorphic components. Consequently, there is an overcounting in carrying out the trace in the R sector, and a factor of $1/\sqrt{2}$ is introduced to remedy this. However, for the PP boundary condition, the holomorphic partition function with the insertion of one fermion field (the position of which is not important), namely the (unnormalized) single-point chiral correlator, does not vanish. This is because the trace over the zero mode is precisely compensated by the inserted field, giving the result
\begin{equation}
\label{eq:chiral-partition-function-pp}
    Z^{\prime}_{\textrm{PP}}(\tau) \equiv q^{-\frac{1}{48}}~\mathrm{Tr}^{\prime}_{\textrm{R}} \left((-1)^{F}q^{L_0}\right) = \eta(\tau),
\end{equation}
where $\mathrm{Tr}^{\prime}$ indicates that the zero mode is excluded in the trace. More generally, for the AA, AP and PA boundary conditions, only even-point correlators of fermion fields are non-vanishing, while for the PP boundary condition, only odd-point ones are non-vanishing. This fact, as we will see in the next subsection, plays a crucial role in the construction of the series of wave functions from the $\mathrm{SO}(n)_1$ WZW models.

For the general even-point case, the approach of GSO projection can also be used to identify the correlators with the insertion of fermion fields; for $a = \boldsymbol{1},~\chi,~\sigma$, those are
\begin{align}
    \langle \chi(z_1) \cdots \chi(z_{2N}) \rangle^{\prime}_{\boldsymbol{1}} &= \frac{1}{2} \left( \langle \chi(z_1) \cdots \chi(z_{2N}) \rangle^{\prime}_{\textrm{AA}} + \langle \chi(z_1) \cdots \chi(z_{2N}) \rangle^{\prime}_{\textrm{AP}} \right), \\
    \langle \chi(z_1) \cdots \chi(z_{2N}) \rangle^{\prime}_{\chi} &= \frac{1}{2} \left( \langle \chi(z_1) \cdots \chi(z_{2N}) \rangle^{\prime}_{\textrm{AA}} - \langle \chi(z_1) \cdots \chi(z_{2N}) \rangle^{\prime}_{\textrm{AP}} \right), \\
    \langle \chi(z_1) \cdots \chi(z_{2N}) \rangle^{\prime}_{\sigma} &= \frac{1}{\sqrt{2}} \langle \chi(z_1) \cdots \chi(z_{2N}) \rangle^{\prime}_{\textrm{PA}}.
\end{align}
Here we have
\begin{equation}
\label{eq:multi-point-1}
    \langle \chi(z_1) \cdots \chi(z_{2N}) \rangle^{\prime}_{\nu} = Z_{\nu}(\tau)~\mathrm{Pf} \left( g_{\nu}(z_{i}-z_{j} \vert \tau) \right),
\end{equation}
where the index $\nu = \textrm{AA, AP, PA}$ labels the spin structures and $g_{\nu}(z_{i}-z_{j} \vert \tau)$ is the corresponding two-point correlator. For $i,j = 1,2, \ldots,2N$, these two-point correlators are arranged into a $(2N) \times (2N)$ antisymmetric matrix whose $(i,j)$ entry is $g_{\nu}(z_{i}-z_{j} \vert \tau)$ for $i \neq j$ and $0$ for $i = j$; $\mathrm{Pf} \left( g_{\nu}(z_{i}-z_{j} \vert \tau) \right)$ is the Pfaffian of this matrix. Note that $g_{\nu}(z_{i}-z_{j} \vert \tau)$ can be represented in terms of Jacobi's theta functions. In appendix~\ref{appdx:derivation-correlator}, we give the explicit expressions of $g_{\nu}(z_{i}-z_{j} \vert \tau)$ and the derivation of~\eqref{eq:multi-point-1}.

\subsection{Anyon eigenbasis from chiral correlators of the~\texorpdfstring{$\mathrm{SO}(n)_1$}{SO(n)1} WZW models}

Having studied the $n=1$ case (Ising CFT) in detail, we are well in a position to obtain the chiral correlators on the torus for the $\mathrm{SO}(n)_1$ WZW models with general $n$. The next step is to insert primary fields in the chiral correlators and interpret the results as the many-body wave functions in the IDMPS form. It is straightforward to compute these wave functions because the primary fields of the $\mathrm{SO}(n)_1$ WZW model are directly related to those of the Ising CFT.

Let us first review the description of primary fields of the $\mathrm{SO}(n)_1$ WZW models~\cite{francesco1997}. For $n = 2r + 1$, as in the case of the Ising CFT (i.e., the $r = 0$ case), the theory has three primary fields, $\boldsymbol{1},~\boldsymbol{v},~\boldsymbol{s}$. Note that here we use the same symbols for all $r$: $\boldsymbol{1}$ is the identity field, $h_{\boldsymbol{1}} = 0$; $\boldsymbol{v}$ belongs to the vector representation and is the fermion field itself, $h_{\boldsymbol{v}} = 1/2$; $\boldsymbol{s}$ belongs to the spinor representation and can be constructed as a product of $(2r+1)$ $\sigma$- and $\mu$-fields of the Ising CFT, $h_{\boldsymbol{s}} = (2r + 1)/16$. Since the $\mathrm{SO}(n)_1$ WZW model can be formulated as a free field theory of an $n$-component massless fermion, the characters of these representations can be expressed in terms of the holomorphic partition functions of the Ising CFT on the torus:
\begin{align}
    \mathrm{ch}_{\boldsymbol{1}}(\tau) &= \frac{1}{2} \left( \left( Z_{\textrm{AA}}(\tau) \right)^{2r+1} + \left( Z_{\textrm{AP}}(\tau) \right)^{2r+1} \right), \\
    \mathrm{ch}_{\boldsymbol{v}}(\tau) &= \frac{1}{2} \left( \left( Z_{\textrm{AA}}(\tau) \right)^{2r+1} - \left( Z_{\textrm{AP}}(\tau) \right)^{2r+1} \right), \\
    \mathrm{ch}_{\boldsymbol{s}}(\tau) &= \frac{1}{\sqrt{2}} \left( Z_{\textrm{PA}}(\tau) \right)^{2r+1}.
\end{align}
For $n = 2r$, on the other hand, the theory has four primary fields: $\boldsymbol{1},~\boldsymbol{v},~\boldsymbol{s}_{+},~\boldsymbol{s}_{-}$. Again, $\boldsymbol{1}$ and $\boldsymbol{v}$ are the identity field and fermion field, respectively. Both of the spinor fields, $\boldsymbol{s}_{+}$ and $\boldsymbol{s}_{-}$, have conformal weight $h_{\boldsymbol{s}_{+}} = h_{\boldsymbol{s}_{-}} = r/8$. The characters corresponding to these primary fields read
\begin{align}
    &\mathrm{ch}_{\boldsymbol{1}}(\tau) = \frac{1}{2} \left( \left( Z_{\textrm{AA}}(\tau) \right)^{2r} + \left( Z_{\textrm{AP}}(\tau) \right)^{2r} \right), \\
    &\mathrm{ch}_{\boldsymbol{v}}(\tau) = \frac{1}{2} \left( \left( Z_{\textrm{AA}}(\tau) \right)^{2r} - \left( Z_{\textrm{AP}}(\tau) \right)^{2r} \right), \\
    &\mathrm{ch}_{\boldsymbol{s}_{+}}(\tau) = \mathrm{ch}_{\boldsymbol{s}_{-}}(\tau) = \frac{1}{2} \left( Z_{\textrm{PA}}(\tau) \right)^{2r}.
\end{align}
In the following, we introduce a ``color'' index $A$ and write $\chi^{A}$ ($A = 1, 2,\ldots, n$) for the components of the fermion field. Due to that the $\mathrm{SO}(n)_1$ WZW model has three/four primaries for odd/even $n$, we anticipate three/four-fold degenerate ground states for the corresponding topological order on the torus. Below we shall see how to obtain them.

We consider a lattice with $2N$ ($N$: integer) sites labelled by $i = 1,2,\ldots,2N$. On each site lives an $\mathfrak{so}(n)$-spin, of which the Hilbert space is spanned by $n$ basis states $\vert A_{i} \rangle$ with $A_{i} = 1, 2,\ldots, n$. The many-body wave functions are expanded in the basis
\begin{equation}
\label{eq:local-spin-basis}
    \{\vert A_{1},A_{2},\ldots,A_{2N}\rangle\}.
\end{equation}
According to our general formalism~\eqref{eq:ansatz} and~\eqref{eq:wave-function-coefficients}, the ansatz for the anyon eigenbasis in sector $a$ is
\begin{equation}
\label{eq:MESs-son}
    \vert\psi_{a}\rangle=~\sum_{A_{1},\ldots,A_{2N}=1}^{n}\langle\chi^{A_{1}}(z_{1})\cdots\chi^{A_{2N}}(z_{2N})\rangle_{a}^{\prime}~\vert A_{1},\ldots,A_{2N}\rangle.
\end{equation}
The correlators for $n = 2r + 1$ in the three sectors $a = \boldsymbol{1},~\boldsymbol{v},~\boldsymbol{s}$ are
\begin{align}
    \langle \chi^{A_{1}}(z_{1})\cdots\chi^{A_{2N}}(z_{2N}) \rangle^{\prime}_{\boldsymbol{1}} &= \frac{1}{2} \left( \langle \chi^{A_{1}}(z_{1})\cdots\chi^{A_{2N}}(z_{2N}) \rangle^{\prime}_{\textrm{AA}} + \langle \chi^{A_{1}}(z_{1})\cdots\chi^{A_{2N}}(z_{2N}) \rangle^{\prime}_{\textrm{AP}} \right), \\
    \langle \chi^{A_{1}}(z_{1})\cdots\chi^{A_{2N}}(z_{2N}) \rangle^{\prime}_{\boldsymbol{v}} &= \frac{1}{2} \left( \langle \chi^{A_{1}}(z_{1})\cdots\chi^{A_{2N}}(z_{2N}) \rangle^{\prime}_{\textrm{AA}} - \langle \chi^{A_{1}}(z_{1})\cdots\chi^{A_{2N}}(z_{2N}) \rangle^{\prime}_{\textrm{AP}} \right), \\
    \langle \chi^{A_{1}}(z_{1})\cdots\chi^{A_{2N}}(z_{2N}) \rangle^{\prime}_{\boldsymbol{s}} &= \frac{1}{\sqrt{2}} \langle \chi^{A_{1}}(z_{1})\cdots\chi^{A_{2N}}(z_{2N}) \rangle^{\prime}_{\textrm{PA}},
\end{align}
where
\begin{align}
\label{eq:MESs-odd-n}
    \langle \chi^{A_{1}}(z_{1})\cdots\chi^{A_{2N}}(z_{2N}) \rangle^{\prime}_{\nu} &= \xi \prod_{A=1}^{2r+1}\langle\chi^{A}(z_{i_{1}^{(A)}})\cdots\chi^{A}(z_{i_{2N_{A}}^{(A)}})\rangle_{\nu}^{\prime} \nonumber \\
    &= \xi \left( Z_{\nu} \right)^{2r+1} \prod_{A=1}^{2r+1} \mathrm{Pf}_{A} \left( g_{\nu}(z_{i}-z_{j} \vert \tau) \right)
\end{align}
for $\nu = $ AA, AP, PA. The lattice sites populated by $\vert A \rangle$ are denoted as $i_1^{(A)}<\cdots<i_{2N_{A}}^{(A)}$, with $\sum_{A=1}^{2r+1} (2N_{A}) = 2N$. Note that we have grouped the fermion fields with the same color together, and $\xi = \mathrm{sgn}(i_{1}^{(1)},\ldots,i_{2N_{1}}^{(1)},\ldots,i_{1}^{(2r+1)},\ldots,i_{2N_{2r+1}}^{(2r+1)})$ is the sign arising due to the permutation of this grouping. $\mathrm{Pf}_{A} \left( g_{\nu}(z_{i}-z_{j} \vert \tau) \right)$
is the Pfaffian of a $(2N_{A}) \times (2N_{A})$ antisymmetric matrix, whose diagonal entries are zero and off-diagonal entries are $g_{\nu}(z_{i}-z_{j} \vert \tau)$, in which the complex coordinates are restricted to the sites populated by the spin state $\vert A \rangle$.\\
The correlators for $n = 2r$ in the four sectors $a = \boldsymbol{1},~\boldsymbol{v},~\boldsymbol{s}_{+},~\boldsymbol{s}_{-}$ are
\begin{align}
    & \langle \chi^{A_{1}}(z_{1})\cdots\chi^{A_{2N}}(z_{2N}) \rangle^{\prime}_{\boldsymbol{1}} = \frac{1}{2} \left( \langle \chi^{A_{1}}(z_{1})\cdots\chi^{A_{2N}}(z_{2N}) \rangle^{\prime}_{\textrm{AA}} + \langle \chi^{A_{1}}(z_{1})\cdots\chi^{A_{2N}}(z_{2N}) \rangle^{\prime}_{\textrm{AP}} \right), \\
    & \langle \chi^{A_{1}}(z_{1})\cdots\chi^{A_{2N}}(z_{2N}) \rangle^{\prime}_{\boldsymbol{v}} = \frac{1}{2} \left( \langle \chi^{A_{1}}(z_{1})\cdots\chi^{A_{2N}}(z_{2N}) \rangle^{\prime}_{\textrm{AA}} - \langle \chi^{A_{1}}(z_{1})\cdots\chi^{A_{2N}}(z_{2N}) \rangle^{\prime}_{\textrm{AP}} \right), \\
    & \langle \chi^{A_{1}}(z_{1})\cdots\chi^{A_{2N}}(z_{2N}) \rangle^{\prime}_{\boldsymbol{s}_{+}} = \frac{1}{2} \left( \langle \chi^{A_{1}}(z_{1})\cdots\chi^{A_{2N}}(z_{2N}) \rangle^{\prime}_{\textrm{PA}} + \langle \chi^{A_{1}}(z_{1})\cdots\chi^{A_{2N}}(z_{2N}) \rangle^{\prime}_{\textrm{PP}} \right), \\
    & \langle \chi^{A_{1}}(z_{1})\cdots\chi^{A_{2N}}(z_{2N}) \rangle^{\prime}_{\boldsymbol{s}_{-}} = \frac{1}{2} \left( \langle \chi^{A_{1}}(z_{1})\cdots\chi^{A_{2N}}(z_{2N}) \rangle^{\prime}_{\textrm{PA}} - \langle \chi^{A_{1}}(z_{1})\cdots\chi^{A_{2N}}(z_{2N}) \rangle^{\prime}_{\textrm{PP}} \right),
\end{align}
where
\begin{align}
    &\langle \chi^{A_{1}}(z_{1})\cdots\chi^{A_{2N}}(z_{2N}) \rangle^{\prime}_{\textrm{PP}}~=~\xi^{\prime} \prod_{A=1}^{2r}\langle\chi^{A}(z_{i_{1}^{(A)}})\cdots\chi^{A}(z_{i_{2N_{A}+1}^{(A)}})\rangle_{\textrm{PP}}^{\prime} \nonumber \\
    &=~\xi^{\prime} \left( Z^{\prime}_{\textrm{PP}} \right)^{2r} \prod_{A=1}^{2r} \left( \sum_{M_{A}=1}^{2N_{A}+1} (-1)^{M_{A}-1} \mathrm{Pf}_{A}\left(g^{\prime}_{\textrm{PP}}(z_{i}-z_{j}\vert\tau)\right) \right).
\label{eq:wave-function-PP-sector}
\end{align}
For the PP sector, the lattice sites with spin state $\vert A \rangle$ are ordered as $i_1^{(A)}<\cdots<i_{2N_{A}+1}^{(A)}$ with $\sum_{A=1}^{2r} (2N_{A}+1) = 2N$. Here, one of the fermion fields with each color $A$ must be picked out to provide the zero mode and compensate the trace over it~[c.f.~\eqref{eq:odd-point-correlator}]. The position of this fermion field is labelled by its site index $i^{(A)}_{M_{A}}$, where $M_{A} = 1, \ldots, 2N_{A}+1$. The sign factor is $\xi^{\prime} = \mathrm{sgn}(i_{1}^{(1)},\ldots,i_{2N_{1}+1}^{(1)},\ldots,i_{1}^{(2r)},\ldots,i_{2N_{2r}+1}^{(2r)})$, and the Pfaffian factor $\mathrm{Pf}_{A}\left(g^{\prime}_{\textrm{PP}}(z_{i}-z_{j}\vert\tau)\right)$ is restricted to $i,j=i_{1}^{(A)},\cdots,\hat{i}_{M_{A}}^{(A)},\cdots,i_{2N_{A}+1}^{(A)}$ (the hat `` $\hat{}$ '' on an object indicates that the object is to be omitted). In appendix~\ref{appdx:RVB}, we prove that for all $n\geq 3$, these wave functions can be represented as RVB states with fermionic partons, which generalize our earlier results for $n=3$~\cite{zhang2021}.

As we have seen, the $\mathrm{SO}(n)_1$ WZW model with odd (even) $n$ naturally gives three (four) wave functions on the torus, which are conjectured to be the topologically degenerate ground states. The difference between the cases of even and odd $n$ can be traced back to the fact that an even number ($2N$) can always be expressed as the sum of an even number ($n = 2r$) of odd numbers ($2N_{A} + 1$). This allows one to use the non-vanishing odd-point correlators in the PP sector to construct two different wave functions labelled by $\boldsymbol{s}_{+}$ and $\boldsymbol{s}_{-}$ when $n$ is even. This result is of course in accordance with Kitaev's sixteenfold way, but direct revelation of the topological order possessed by these states is still very much in need. To this end, we compute the modular $S$ and $T$ matrices of these wave functions. Up to this stage, we have not made any assumptions about the geometry of the lattice. For simplicity in the calculation of the modular matrices, we take the system to be a square lattice embedded in a square defined by two complex vectors $\omega_1=1$ and $\omega_2=\mathrm{i}$ with modular parameter $\tau=\mathrm{i}$ (see figure~\ref{Figure1}). Thanks to the fact that $Z_{\nu}(\tau)$ and $g_{\nu}(z_{i}-z_{j} \vert \tau)$ are expressed in terms of special functions, whose modular-transformation properties are known, the modular matrices can be derived analytically. These properties are listed in appendix~\ref{appdx:derivation-modular}; here we present only the result.

For the $n=2r+1$ case, the modular $S$ transformation changes the states as
\begin{align}
    \vert \psi_{\boldsymbol{1}} \rangle &\rightarrow \frac{1}{2} \vert \psi_{\boldsymbol{1}} \rangle + \frac{1}{2} \vert \psi_{\boldsymbol{v}} \rangle + \frac{1}{\sqrt{2}} \vert \psi_{\boldsymbol{s}} \rangle, \\
    \vert \psi_{\boldsymbol{v}} \rangle &\rightarrow \frac{1}{2} \vert \psi_{\boldsymbol{1}} \rangle + \frac{1}{2} \vert \psi_{\boldsymbol{v}} \rangle - \frac{1}{\sqrt{2}} \vert \psi_{\boldsymbol{s}} \rangle, \\
    \vert \psi_{\boldsymbol{s}} \rangle &\rightarrow \frac{1}{\sqrt{2}} \vert \psi_{\boldsymbol{1}} \rangle - \frac{1}{\sqrt{2}} \vert \psi_{\boldsymbol{v}} \rangle;
\end{align}
and the modular $T$ transformation changes them as
\begin{align}
    \vert \psi_{\boldsymbol{1}} \rangle &\rightarrow \mathrm{e}^{-\frac{(2r+1)\pi\mathrm{i}}{24}} \vert \psi_{\boldsymbol{1}} \rangle, \\
    \vert \psi_{\boldsymbol{v}} \rangle &\rightarrow -\mathrm{e}^{-\frac{(2r+1)\pi\mathrm{i}}{24}} \vert \psi_{\boldsymbol{v}} \rangle, \\
    \vert \psi_{\boldsymbol{s}} \rangle &\rightarrow \mathrm{e}^{\frac{(2r+1)\pi\mathrm{i}}{12}} \vert \psi_{\boldsymbol{s}} \rangle.
\end{align}
Hence the modular matrices read
\begin{equation}
\mathcal{S} = \frac{1}{2}
\begin{pmatrix}
1 & 1 & \sqrt{2}\\
1 & 1 & -\sqrt{2}\\
\sqrt{2} & -\sqrt{2} & 0
\end{pmatrix}, \qquad
\mathcal{T} = \mathrm{e}^{-\frac{(2r+1)\pi\mathrm{i}}{24}}
\begin{pmatrix}
1 & 0 & 0\\
0 & -1 & 0\\
0 & 0 & \mathrm{e}^{\frac{(2r+1)\pi\mathrm{i}}{8}}
\end{pmatrix}.
\end{equation}
For the $n=2r$ case, the modular $S$ transformation changes the state as
\begin{align}
    &\vert \psi_{\boldsymbol{1}} \rangle \rightarrow \frac{1}{2} \vert \psi_{\boldsymbol{1}} \rangle + \frac{1}{2} \vert \psi_{\boldsymbol{v}} \rangle + \frac{1}{2} \vert \psi_{\boldsymbol{s}_{+}} \rangle + \frac{1}{2} \vert \psi_{\boldsymbol{s}_{-}} \rangle, \\
    &\vert \psi_{\boldsymbol{v}} \rangle \rightarrow \frac{1}{2} \vert \psi_{\boldsymbol{1}} \rangle + \frac{1}{2} \vert \psi_{\boldsymbol{v}} \rangle - \frac{1}{2} \vert \psi_{\boldsymbol{s}_{+}} \rangle - \frac{1}{2} \vert \psi_{\boldsymbol{s}_{-}} \rangle, \\
    &\vert \psi_{\boldsymbol{s}_{+}} \rangle \rightarrow \frac{1}{2} \vert \psi_{\boldsymbol{1}} \rangle - \frac{1}{2} \vert \psi_{\boldsymbol{v}} \rangle + \frac{1}{2}\mathrm{i}^{-r} \vert \psi_{\boldsymbol{s}_{+}} \rangle - \frac{1}{2}\mathrm{i}^{-r} \vert \psi_{\boldsymbol{s}_{-}} \rangle,
    \\
    &\vert \psi_{\boldsymbol{s}_{-}} \rangle \rightarrow \frac{1}{2} \vert \psi_{\boldsymbol{1}} \rangle - \frac{1}{2} \vert \psi_{\boldsymbol{v}} \rangle - \frac{1}{2}\mathrm{i}^{-r} \vert \psi_{\boldsymbol{s}_{+}} \rangle + \frac{1}{2}\mathrm{i}^{-r} \vert \psi_{\boldsymbol{s}_{-}} \rangle;
\end{align}
the modular $T$ transformation changes them as
\begin{align}
    &\vert \psi_{\boldsymbol{1}} \rangle \rightarrow \mathrm{e}^{-\frac{r\pi\mathrm{i}}{12}} \vert \psi_{\boldsymbol{1}} \rangle, \\
    &\vert \psi_{\boldsymbol{v}} \rangle \rightarrow -\mathrm{e}^{-\frac{r\pi\mathrm{i}}{12}} \vert \psi_{\boldsymbol{v}} \rangle, \\
    &\vert \psi_{\boldsymbol{s}_{+}} \rangle \rightarrow \mathrm{e}^{\frac{r\pi\mathrm{i}}{6}} \vert \psi_{\boldsymbol{s}_{+}} \rangle, \\
    &\vert \psi_{\boldsymbol{s}_{-}} \rangle \rightarrow \mathrm{e}^{\frac{r\pi\mathrm{i}}{6}} \vert \psi_{\boldsymbol{s}_{-}} \rangle.
\end{align}
Hence the modular matrices read
\begin{equation}
\mathcal{S} = \frac{1}{2}
\begin{pmatrix}
1 & 1 & 1 & 1\\
1 & 1 & -1 & -1\\
1 & -1 & \mathrm{i}^{-r} & -\mathrm{i}^{-r}\\
1 & -1 & -\mathrm{i}^{-r} & \mathrm{i}^{-r}
\end{pmatrix}, \qquad
\mathcal{T} = \mathrm{e}^{-\frac{r\pi\mathrm{i}}{12}}
\begin{pmatrix}
1 & 0 & 0 & 0\\
0 & -1 & 0 & 0\\
0 & 0 & \mathrm{e}^{\frac{r\pi\mathrm{i}}{4}} & 0 \\
0 & 0 & 0 & \mathrm{e}^{\frac{r\pi\mathrm{i}}{4}} \\
\end{pmatrix}.
\end{equation}

As the modular $T$ matrices are diagonal in both cases, the wave functions defined by~\eqref{eq:MESs-son} indeed form anyon eigenbasis. The diagonal entries of the modular $T$ matrix are known to encode the topological spins $\theta_{a}$ of the anyonic quasiparticles (labelled by $a$) as well as the chiral central charge $c_{-}$ of the edge theory via $\mathcal{T}_{a a} = \mathrm{e}^{-2\pi\mathrm{i}c_{-}/24} \theta_{a}$. From the above results, we obtain for $n=2r+1$: $\theta_{\boldsymbol{1}} = 1$, $\theta_{\boldsymbol{v}} = -1$, $\theta_{\boldsymbol{s}} = \mathrm{e}^{\frac{(2r+1)\pi\mathrm{i}}{8}}$ and $c_{-} = (2r+1)/2$; for $n=2r$: $\theta_{\boldsymbol{1}} = 1$, $\theta_{\boldsymbol{v}} = -1$, $\theta_{\boldsymbol{s}_{+}} = \theta_{\boldsymbol{s}_{-}} = \mathrm{e}^{\frac{r\pi\mathrm{i}}{4}}$ and $c_{-} = r$. We note, however, that with the modular $T$ transformation, the chiral central charge is determined up to a modulo of 8. These are in agreement with the $\mathrm{SO}(n)_1$ anyon theories, i.e., Kitaev's sixteenfold way~\cite{kitaev2006b}.

For completeness, the fusion rules of the anyons are obtained from the modular $S$ matrix using Verlinde's formula~\cite{verlinde1988}. For $n=2r+1$, the non-trivial fusion rules are
\begin{equation}
\label{eq:non-abelian-fusion-rules}
    \boldsymbol{v} \times \boldsymbol{v} = \boldsymbol{1}, \quad
    \boldsymbol{v} \times \boldsymbol{s} = \boldsymbol{s}, \quad
    \boldsymbol{s} \times \boldsymbol{s} = \boldsymbol{1} + \boldsymbol{v}.
\end{equation}
For $n=2r$, the cases with odd $r$ and even $r$ differ from each other. For the former case, the non-trivial fusion rules are
\label{eq:abelian-fusion-rules}
\begin{equation}
\begin{split}
    \boldsymbol{v} \times \boldsymbol{v} = \boldsymbol{1}, \quad
    \boldsymbol{v} \times \boldsymbol{s}_{+} = \boldsymbol{s}_{-}, \quad
    \boldsymbol{v} \times \boldsymbol{s}_{-} = \boldsymbol{s}_{+}, \\
    \boldsymbol{s}_{+} \times \boldsymbol{s}_{+} = \boldsymbol{v}, \quad
    \boldsymbol{s}_{-} \times \boldsymbol{s}_{-} = \boldsymbol{v}, \quad
    \boldsymbol{s}_{+} \times \boldsymbol{s}_{-} = \boldsymbol{1},
\end{split}
\end{equation}
while for the latter case, they are
\begin{equation}
\begin{split}
    \boldsymbol{v} \times \boldsymbol{v} = \boldsymbol{1},  \quad
    \boldsymbol{v} \times \boldsymbol{s}_{+} = \boldsymbol{s}_{-}, \quad
    \boldsymbol{v} \times \boldsymbol{s}_{-} = \boldsymbol{s}_{+}, \\
    \boldsymbol{s}_{+} \times \boldsymbol{s}_{+} = \boldsymbol{1}, \quad
    \boldsymbol{s}_{-} \times \boldsymbol{s}_{-} = \boldsymbol{1}, \quad
    \boldsymbol{s}_{+} \times \boldsymbol{s}_{-} = \boldsymbol{v}.
\end{split}
\end{equation}
These fusion rules also coincide with the non-Abelian ($n=2r+1$) and Abelian ($n=2r$) series of Kitaev's sixteenfold way.

\section{States from free bosons}
\label{sec:boson}

In this section, we apply the formalism developed in section~\ref{sec:general} to some CFTs that admit free boson representations: a single compactified boson with radius $R = \sqrt{p}$ ($p>1$ is an integer) and the~$\mathrm{SU}(n)_1$ WZW model. The former theory leads to lattice Laughlin states at filling factor $1/p$, which has already been constructed using the ``breaking-up'' approach in ref.~\cite{deshpande2016}. In contrast, the wave functions on the torus constructed from the~$\mathrm{SU}(n)_1$ WZW model are new results, which generalize previous results on the plane~\cite{tu2014b,bondesan2014}. The modular matrices for these states are derived analytically, which confirm that they have topological order described by the~$\mathrm{SU}(n)_1$ TQFT.

\subsection{Lattice Laughlin states at filling factor~\texorpdfstring{$1/p$}{1/p}}
\label{subsec:laughlin}

In this subsection, we construct the lattice Laughlin states at filling factor $1/p$ on the torus, in which the $p=2$ case reproduces the famous Kalmeyer-Laughlin state~\cite{kalmeyer1987} and is one of the first examples of chiral spin liquids. Each lattice site is either empty or occupied by one fermion (hard-core boson) for odd (even) $p$. These wave functions can be constructed from the CFT of a compactified boson with radius $R = \sqrt{p}$~\cite{moore1991,tu2014a,deshpande2016}, so we briefly review some basics of the operator formalism of this CFT.

Consider a boson field $\varphi(x,t)$ defined on a space-time cylinder of circumference $L$ (along the spatial direction). The boson field is compactified on a circle with radius $R$ in the sense that $\varphi$ and $\varphi + 2 \pi R$ are identified. Its mode expansion is~\cite{francesco1997}
\begin{equation}
    \varphi(x,t)=\varphi_{0}+\frac{4\pi}{L}(\pi_{0}vt+\tilde{\pi}_{0}x)+\mathrm{i}~\underset{k\neq0}{\sum}~\frac{1}{k}(a_{k}\mathrm{e}^{2\pi\mathrm{i}k\frac{x-vt}{L}}+\bar{a}_{k}\mathrm{e}^{-2\pi\mathrm{i}k\frac{x+vt}{L}}),
\end{equation}
where $k \in \mathbb{Z}$, and $v$ is the ``propagating velocity'' of the modes. The operators with $k \neq 0$ satisfy the commutation relations,
\begin{equation}
\label{eq:heisenberg-algebra}
[a_{k},a_{k^{\prime}}]=[\bar{a}_{k},\bar{a}_{k^{\prime}}]=k\delta_{k+k^{\prime}}, \quad
[a_{k},\bar{a}_{k^{\prime}}]=0,
\end{equation}
and the zero modes satisfy the canonical commutation relation,
\begin{equation}
[\varphi_{0},\pi_{0}] = [\tilde{\varphi}_{0},\tilde{\pi}_{0}] = \mathrm{i},
\end{equation}
in which the operator $\tilde{\varphi}_{0}$ is formally introduced to be the canonical conjugate of $\tilde{\pi}_{0}$. Due to the compactness of $\varphi$, field configurations with a winding number $m \in \mathbb{Z}$, i.e.,
\begin{equation}
    \varphi(x + L,t) \equiv \varphi(x,t) - 2 \pi mR,
\end{equation}
are allowed. Hence the operator $\tilde{\pi}_{0}$ has eigenvalues $-mR/2$. Moreover, as the operator $\pi_{0}$ is conjugate to $\varphi_{0}$, its eigenvalues are quantized to be $n/R$ with $n \in \mathbb{Z}$. When recombining the zero-mode operators as
\begin{align}
    Q&=\frac{1}{2}(\varphi_{0}-\tilde{\varphi}_{0}), \quad
    P=\pi_{0}-\tilde{\pi}_{0}; \\
    \bar{Q}&=\frac{1}{2}(\varphi_{0}+\tilde{\varphi}_{0}), \quad \bar{P}=\pi_{0}+\tilde{\pi}_{0},
\end{align}
the canonical commutation relations hold for the new operators:
\begin{equation}
\label{eq:canonical-commutation-relation}
    [Q,P] = 
    [\bar{Q},\bar{P}] = \mathrm{i}.
\end{equation}
Finally, the complex coordinates
\begin{equation}
    z=-\frac{x-vt}{L}, \quad
    \bar{z}=-\frac{x+vt}{L}
\end{equation}
are introduced to rewrite the mode expansion as
\begin{align}
\label{eq:mode-expansion-compactified-boson}
     \varphi(x,t) \equiv \varphi(z,\bar{z})
     &= Q+2\pi Pz+\mathrm{i}~\underset{k\neq0}{\sum}~\frac{1}{k}a_{k}\mathrm{e}^{-2\pi\mathrm{i}kz} \nonumber \\
     &+\bar{Q}-2\pi\bar{P}\bar{z}+\mathrm{i}~\underset{k\neq0}{\sum}~\frac{1}{k}\bar{a}_{k}\mathrm{e}^{2\pi\mathrm{i}k\bar{z}} \nonumber \\
     &\equiv\phi(z)+\bar{\phi}(\bar{z}),
\end{align}
i.e., the free boson field $\varphi(z,\bar{z})$ decomposes into holomorphic component $\phi(z)$ and anti-holomorphic component $\bar{\phi}(\bar{z})$.

We focus on the holomorphic sector of the CFT. It is straightforward to see from~\eqref{eq:mode-expansion-compactified-boson} that $a_{k}$ (with $a_{0} \equiv P$) are the Laurent modes of the current $j(z) = \mathrm{i} \partial \varphi(z,\bar{z}) = \mathrm{i} \partial \phi(z)$. The algebra defined by~\eqref{eq:heisenberg-algebra} is the affine extension of $\mathfrak{u}(1)$ algebra, also called the Heisenberg algebra. There is a series of highest weight states $\vert n,m;\{0\} \rangle$ that are labelled by their $a_{0}$ (``momentum'') eigenvalues:
\begin{equation}
    a_{0}\vert n,m;\{0\}\rangle=\left(\frac{n}{R}+\frac{1}{2}mR\right)\vert n,m;\{0\}\rangle.
\end{equation}
By definition, these highest weight states are annihilated by any $a_{k}$ with $k>0$. The representation labelled by $(n,m)$ is built upon these highest weight states by acting the $a_{k}$'s with $k<0$:
\begin{equation}
\label{eq:states-normalized}
    \vert n,m;n_{1},n_{2},\ldots\rangle=\left(\prod_{k=1}^{\infty}\frac{1}{\sqrt{n_{k}!~k^{n_{k}}}}\right)a_{-1}^{n_{1}}a_{-2}^{n_{2}}\cdots\vert n,m;\{0\} \rangle,
\end{equation}
where the factor in the bracket ensures normalization of the state. The holomorphic energy-momentum tensor for the free massless boson is
\begin{equation}
    T(z) = -\frac{1}{2}:\partial \phi(z) \partial \phi(z):,
\end{equation}
and the zeroth Virasoro generator, based on the Laurent expansion~\eqref{eq:virasoro-generators-definition} and the mode expansion~\eqref{eq:mode-expansion-compactified-boson}, reads
\begin{equation}
    L_{0} = \frac{1}{2}a_{0}^{2}+\underset{k\geq1}{\sum}~a_{-k}a_{k}.
\end{equation}
It satisfies
\begin{equation}
    L_{0} \vert n,m;\{0\} \rangle = \frac{1}{2}\left(\frac{n}{R}+\frac{1}{2}mR\right)^2 \vert n,m;\{0\} \rangle \equiv h_{n,m} \vert n,m;\{0\} \rangle
\end{equation}
with $h_{n,m}=(n/R+mR/2)^2/2$ being the conformal weight of the highest weight state $\vert n,m;\{0\} \rangle$. The commutation relations~\eqref{eq:heisenberg-algebra} can be used to show that
\begin{equation}
\label{eq:energy-eigenstates}
    L_{0} \vert n,m;n_{1},n_{2},\ldots\rangle = \left( h_{n,m} + \underset{k\geq1}{\sum}~kn_{k}\right) \vert n,m;n_{1},n_{2},\ldots\rangle.
\end{equation}
Then, according to the definition~\eqref{eq:character-definition}, the character of the representation built on $\vert n,m;\{0\} \rangle$ is found to be
\begin{align}
\label{eq:u(1)-character}
    \mathrm{ch}_{(n,m)}(q) &= q^{-\frac{1}{24}}\underset{n_{1},n_{2},\ldots\geq0}{\sum}\langle n,m;n_{1},n_{2},\ldots\vert q^{L_{0}}\vert n,m;n_{1},n_{2},\ldots\rangle \nonumber \\
    &=\frac{1}{\eta(\tau)}q^{\frac{1}{2}\left(n/R+mR/2\right)^{2}},
\end{align}
where $\eta(\tau)$ is Dedekind's eta-function (see appendix~\ref{appdx:derivation-correlator}). Obviously, there are infinitely many characters $\mathrm{ch}_{(n,m)}(q)$ corresponding to representations of the Heisenberg algebra of the compactified boson CFT. Nevertheless, if some generators other than the $\mathfrak{u}(1)$ current are introduced to extend the algebra for $R = \sqrt{p}$, these characters can be reorganized into a finite number of generalized ones~\cite{francesco1997}. The choice of these generators should be justified by physical requirements. To this end, we distinguish between the cases with even $p$ and odd $p$, which correspond to bosonic and fermionic Laughlin states, respectively.

For even $p$ (bosonic case), the algebra is extended by the ``particle operator'' and its conjugate $:\mathrm{e}^{\pm \mathrm{i}\sqrt{p}\phi(z)}:$. In this theory, the allowed chiral primary fields must have trivial monodromy with these generators. They are the vertex operators
\begin{equation}
\label{eq:primary-fields-vertex}
    \Phi_{l}(z) = :\mathrm{e}^{\mathrm{i}\frac{l}{\sqrt{p}}\phi(z)}:
\end{equation}
with $l = 0,1,\ldots,p-1$, as shifting $l$ to $l + p$ amounts to inserting a ``particle operator''. Physically, $\Phi_{1}$ is the quasihole operator since its charge is $1/p$ of that carried by the ``particle operator''. The conformal weight of $\Phi_{l}$ is
\begin{equation}
    h_{l} = \frac{l^2}{2p},
\end{equation}
and the associated primary state (or ``highest weight'' state) is
\begin{equation}
    \vert l;\{0\} \rangle \equiv \Phi_{l}(0) \vert 0 \rangle.
\end{equation}
The representation labelled by $l$ is built upon $\vert l;\{0\} \rangle$ and the corresponding chiral character reads
\begin{align}
    \mathrm{ch}^{(p)}_{l} (q) &= q^{-\frac{1}{24}}~\underset{u \in \mathbb{Z}}{\sum}~ \underset{n_{1},n_{2},\ldots\geq0}{\sum}\langle l+up;n_{1},n_{2},\ldots\vert q^{L_{0}}\vert l+up;n_{1},n_{2},\ldots\rangle \nonumber \\
    &= \frac{1}{\eta(\tau)}~\underset{u \in \mathbb{Z}}{\sum}~q^{\frac{(l+up)^2}{2p}}.
\label{eq:generalized-character-even}
\end{align}
In fact, these generalized characters can be obtained from $\mathrm{ch}_{(n,m)}(q)$ [see eq.~\eqref{eq:u(1)-character}] by setting $n = up + r$ with $u \in \mathbb{Z}$, $r = 0,1,\ldots,p-1$, introducing $l = r + mp/2$ and summing over $u$.

For odd $p$ (fermionic case), the generators introduced to extend the algebra are $:\mathrm{e}^{\pm 2 \mathrm{i}\sqrt{p}\phi(z)}:$. This choice is dictated by the fact that for odd $p$ the underlying particles are fermions and the monodromy of a fermion is trivial only up to a sign. In mathematical terms, it is because fermion fields are spinors that live on the double cover of the punctured plane (or, in the spin bundle over some higher-genus Riemann surface)~\cite{ginsparg1988}. Consequently, the allowed chiral primary fields are still $\Phi_{l}$ in~\eqref{eq:primary-fields-vertex}, but with $l=0,1/2,\ldots,2p - 1/2$. The chiral character of the corresponding representation of the extended algebra is
\begin{equation}
\label{eq:generalized-character-odd}
    \mathrm{ch}^{(p)}_{l} (q) = \frac{1}{\eta(\tau)}~\underset{u \in 2\mathbb{Z}}{\sum}~q^{\frac{(l+up)^2}{2p}}.
\end{equation}
These generalized characters can be obtained from $\mathrm{ch}_{(n,m)}(q)$ by setting $n = up + r$ with $u \in 2\mathbb{Z}$, $r = 0,1,\ldots,2p-1$, introducing $l = r + mp/2$ and summing over $u$.

Having established the ``reorganization'' of the chiral characters, we are now ready to write down the ansatz for the anyon eigenbasis on the torus, which are labelled by $l$, since the same ``reorganization'' also applies to the (unnormalized) chiral correlators. The general form of the ansatz~\eqref{eq:ansatz} now reads
\begin{equation}
    \vert \psi_{l} \rangle = \sum_{n_{1},\ldots,n_{N}} \psi_{l}(n_{1},\ldots,n_{N}) \vert n_{1},\ldots,n_{N} \rangle,
\end{equation}
where the wave-function coefficients are given by the (unnormalized) chiral correlators
\begin{equation}
\label{eq:N-point-chiral-correlator-p}
    \psi_{l}(\{n\}) \equiv \psi_{l}(n_{1},\ldots,n_{N}) = \langle V_{n_1}(z_1) \cdots V_{n_N}(z_N) \rangle^{\prime}_{l},
\end{equation}
in which $n_{i} = 0, 1$ is the occupation number at site $i$, and
\begin{align}
    V_{n_i}(z_i) &=~:\mathrm{e}^{\mathrm{i}\frac{p n_{i}-1}{\sqrt{p}}\phi(z_i)}: \nonumber \\
    &\equiv~:\mathrm{e}^{\mathrm{i}\alpha_{i}\phi(z_i)}:
\end{align}
are certain vertex operators. These correlators are non-vanishing only if the ``charge-neutrality condition'' $\sum_{i=1}^{N}\alpha_{i} = 0$ is satisfied (see appendix~\ref{appdx:derivation-correlator}). This condition translates into the requirement that the filling factor is $1/p$: $\sum_{i=1}^{N}n_{i} = N/p$.

To compute the chiral correlators, we again distinguish between the cases with even and odd $p$. For even $p$, the wave function~\eqref{eq:N-point-chiral-correlator-p} can be calculated as follows:
\begin{align}
\label{eq:N-point-chiral-correlator-even-p}
    \psi_{l}(\{n\}) &= q^{-\frac{1}{24}}~\underset{u \in \mathbb{Z}}{\sum}~ \underset{n_{1},n_{2},\ldots\geq0}{\sum}\langle l+up; n_{1},n_{2},\ldots\vert V_{n_1}(z_1) \cdots V_{n_N}(z_N) q^{L_{0}} \vert l+up;n_{1},n_{2},\ldots\rangle \nonumber \\
    &= q^{-\frac{1}{24}}~\underset{u \in \mathbb{Z}}{\sum}~q^{\frac{(l+up)^2}{2p}}~ \underset{n_{1},n_{2},\ldots\geq0}{\sum} q^{\sum_{{k\geq1}}kn_{k}}  \nonumber \\
    &\phantom{=} \times \langle l+up;n_{1},n_{2},\ldots\vert V_{n_1}(z_1) \cdots V_{n_N}(z_N) \vert l+up;n_{1},n_{2},\ldots\rangle.
\end{align}
To evaluate the right hand side, we first calculate the expectation value in a particular momentum eigenstate $\vert n,m;n_{1},n_{2},\ldots\rangle$,
\begin{align}
\label{eq:correlator-momentum-eigenstate}
    &~\underset{n_{1},n_{2},\ldots\geq0}{\sum} q^{\sum_{{k\geq1}}kn_{k}} \langle n,m;n_{1},n_{2},\ldots\vert V_{n_1}(z_1) \cdots V_{n_N}(z_N) \vert n,m;n_{1},n_{2},\ldots\rangle \nonumber \\
    =&~\delta_{\alpha}~\frac{q^{\frac{1}{24}}}{\eta(\tau)}~\mathrm{e}^{2\pi\mathrm{i} \frac{1}{\sqrt{p}}(n + mp/2) \sum_{i=1}^{N} \alpha_{i} z_{i}}~\underset{i<j}{\prod}\left( E(z_{1}-z_{2}\vert\tau) \right)^{\alpha_{i}\alpha_{j}},
\end{align}
where the function $E(z_{1}-z_{2}\vert\tau)$ is defined in terms of Jacobi's theta function $\vartheta_{1}(z_{1}-z_{2}\vert\tau)$ as
\begin{equation}
    E(z_{1}-z_{2}\vert\tau)=\frac{\vartheta_{1}(z_{1}-z_{2}\vert\tau)}{\partial_{z}\vartheta_{1}(z\vert\tau)\vert_{z=0}},
\end{equation}
and $\delta_{\alpha}=1$ if $\sum_{i=1}^{N}\alpha_{i} = 0$ (or equivalently, $\sum_{i=1}^{N}n_{i} = N/p$) and $0$ otherwise. The derivation of eq.~\eqref{eq:correlator-momentum-eigenstate} is presented in appendix~\ref{appdx:derivation-correlator}. It should be pointed out that this derivation is actually valid for both parity of $p$. The difference between even and odd $p$ lies in the next step, where we sum over certain combinations of $n$ and $m$ following the prescription explained previously. After redefining $n = up + r$ and $l = r + mp/2$ ($u \in \mathbb{Z}$, $l=0,1,\ldots,p-1$), we obtain
\begin{align}
    &~\underset{n_{1},n_{2},\ldots\geq0}{\sum} q^{\sum_{{k\geq1}}kn_{k}}
    \langle l+up;n_{1},n_{2},\ldots\vert V_{n_1}(z_1) \cdots V_{n_N}(z_N) \vert l+up;n_{1},n_{2},\ldots\rangle \nonumber \\
    =&~\delta_{\alpha}~\frac{q^{\frac{1}{24}}}{\eta(\tau)}~\mathrm{e}^{2\pi\mathrm{i} \frac{1}{\sqrt{p}}(up + l) \sum_{i=1}^{N} \alpha_{i} z_{i}}~\underset{i<j}{\prod}\left( E(z_{1}-z_{2}\vert\tau) \right)^{\alpha_{i}\alpha_{j}}.
\end{align}
This can be substituted in~\eqref{eq:N-point-chiral-correlator-even-p} to yield
\begin{align}
\label{eq:N-point-chiral-correlator-even-p-result}
    \psi_{l}(\{n\})  &= \frac{\delta_{\alpha}}{\eta(\tau)} ~\underset{u\in\mathbb{Z}}{\sum}~q^{\frac{(l+up)^2}{2p}}~\mathrm{e}^{2\pi\mathrm{i} \frac{1}{\sqrt{p}}(up + l) \sum_{i=1}^{N} \alpha_{i} z_{i}}~\underset{i<j}{\prod}\left( E(z_{1}-z_{2}\vert\tau) \right)^{\alpha_{i}\alpha_{j}} \nonumber \\
    &= \frac{\delta_{n}}{\eta(\tau)} ~\vartheta\left[\begin{array}{c}
    l/p\\
    0
    \end{array}\right] \left( \sum_{i=1}^{N} (pn_{i} - 1)z_{i},p\tau \right)~\underset{i<j}{\prod}\left( E(z_{1}-z_{2}\vert\tau) \right)^{pn_{i}n_{j} - n_{i} - n_{j} + 1/p},
\end{align}
where $\delta_{n} \equiv \delta_{\alpha}$ and we have used the~\emph{theta function with characteristics}:
\begin{equation}
    \vartheta\left[\begin{array}{c}
    \gamma\\
    \beta
    \end{array}\right](z,\tau) \equiv \sum_{n \in \mathbb{Z}} \mathrm{e}^{\pi\mathrm{i}\tau(n+\gamma)^{2}+2\pi\mathrm{i}(n+\gamma)(z+\beta)}.
\end{equation}
We note in passing that Jacobi's theta functions~\eqref{eq:theta-sum-begin} to~\eqref{eq:theta-sum-end} correspond to special values of $\beta$ and $\gamma$:
\begin{align}
    \vartheta_{1}(z\vert\tau)&=\vartheta\left[\begin{array}{c}
    1/2\\
    1/2
    \end{array}\right](z,\tau), \\
    \vartheta_{2}(z\vert\tau)&=\vartheta\left[\begin{array}{c}
    1/2\\
    0
    \end{array}\right](z,\tau), \\
    \vartheta_{3}(z\vert\tau)&=\vartheta\left[\begin{array}{c}
    0\\
    0
    \end{array}\right](z,\tau), \\
    \vartheta_{4}(z\vert\tau)&=\vartheta\left[\begin{array}{c}
    0\\
    1/2
    \end{array}\right](z,\tau).
\end{align}
For odd $p$, the only difference is that we have $u \in 2\mathbb{Z}$, $l=0,1/2,\ldots,2p-1/2$ in the parameters $n = up + r$ and $l = r + mp/2$. Hence the wave function reduces to
\begin{align}
    \psi_{l}(\{n\}) &= \frac{\delta_{\alpha}}{\eta(\tau)} ~\underset{u \in 2\mathbb{Z}}{\sum}~q^{\frac{(l+up)^2}{2p}}~\mathrm{e}^{2\pi\mathrm{i} \frac{1}{\sqrt{p}}(up + l) \sum_{i=1}^{N} \alpha_{i} z_{i}}~\underset{i<j}{\prod}\left( E(z_{1}-z_{2}\vert\tau) \right)^{\alpha_{i}\alpha_{j}} \nonumber \\
    &= \frac{\delta_{\alpha}}{\eta(\tau)} ~\underset{u \in \mathbb{Z}}{\sum}~q^{\frac{(l+up)^2}{2p}}~\mathrm{e}^{2\pi\mathrm{i} \frac{1}{\sqrt{p}}(up + l) \sum_{i=1}^{N} \alpha_{i} z_{i}}~\frac{1 + \mathrm{e}^{\pi\mathrm{i}u}}{2}~\underset{i<j}{\prod}\left( E(z_{1}-z_{2}\vert\tau) \right)^{\alpha_{i}\alpha_{j}}.
\end{align}
As one can readily verify, this is equivalent to
\begin{align}
   \psi_{l}^{\beta\gamma}(\{n\}) 
    = \frac{\delta_{n}}{\eta(\tau)} ~\vartheta\left[\begin{array}{c}
    l/p + \gamma \\
    \beta
    \end{array}\right] \left( \sum_{i=1}^{N} (pn_{i} - 1)z_{i},p\tau \right) \underset{i<j}{\prod}\left( E(z_{1}-z_{2}\vert\tau) \right)^{pn_{i}n_{j} - n_{i} - n_{j} + 1/p},
\end{align}
where $l$ can only take the values $0,1,\ldots,p-1$. Two extra variables $\beta$ and $\gamma$, which could be either $0$ or $1/2$, need to be introduced to facilitate this transformation.

Finally, we note that, for the Kalmeyer-Laughlin case with $p=2$, an extra ``Marshall sign factor''
\begin{equation}
    \chi_{n} = (-1)^{\sum_{i=1}^{N}(i-1)n_{i}}
\end{equation}
can be added to the wave functions~\eqref{eq:N-point-chiral-correlator-p} so that the states are $\mathrm{SU}(2)$ singlets~\cite{nielsen2014b}.

\subsection{States from the~\texorpdfstring{$\mathrm{SU}(n)_1$}{SU(n)1} WZW model}
\label{subsec:sun}

After validating our general formalism using the CFT of a single compactified boson, we proceed to the $\mathrm{SU}(n)_{1}$ WZW model, which has central charge $c=n-1$ and can be represented in terms of $n-1$ free bosons. Each primary field of the $\mathrm{SU}(n)_{1}$ WZW model is in correspondence with an integrable affine weight of $\widehat{\mathfrak{su}}(n)$~\footnote{The notation $\widehat{\mathfrak{g}}$ is used to denote the affine extension of the Lie algebra $\mathfrak{g}$.} at level $1$, and the subspace labelled by this primary field is nothing but the integrable module of the corresponding highest weight~\cite{francesco1997}. Thus, in order to describe these primary fields, we briefly review some relevant aspects of the integrable highest-weight representations of $\widehat{\mathfrak{su}}(n)$ at level $1$.

\begin{figure}
\centering
\includegraphics[width=0.60\textwidth]{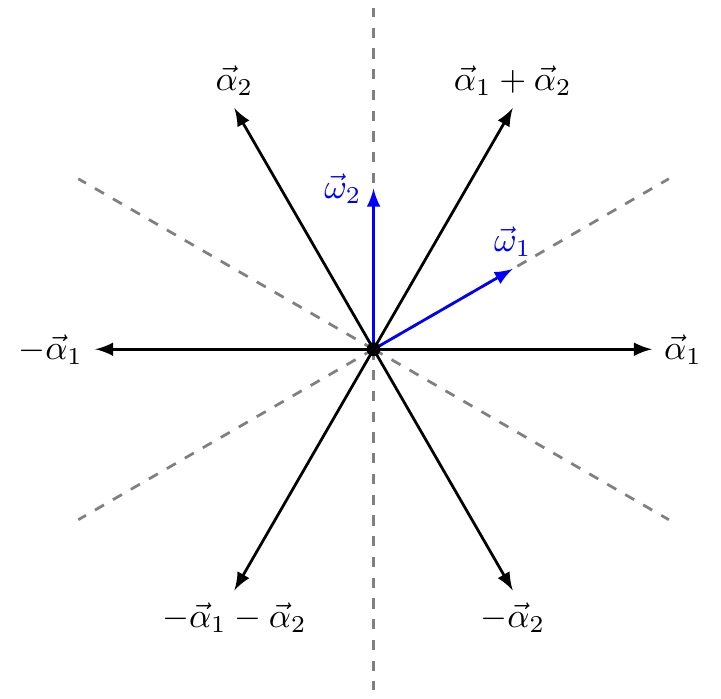}
\caption{Root system of $\mathfrak{su}(3)$. The simple roots, $\Vec{\alpha}_{1}$ and $\Vec{\alpha}_{2}$, are related to the fundamental weights by $\Vec{\alpha}_{1} = 2\Vec{\omega}_{1} - \Vec{\omega}_{2}$ and $\Vec{\alpha}_{2} = -\Vec{\omega}_{1} + 2\Vec{\omega}_{2}$. In general, for a simple Lie algebra $\mathfrak{g}$, the relation between the simple roots and the fundamental weights reads $\Vec{\alpha}_{i} = \sum_{j}\mathfrak{A}_{ij}\Vec{\omega}_{j}$, where $\mathfrak{A}$ is the Cartan matrix of $\mathfrak{g}$ (see appendix~\ref{appdx:derivation-modular}).}
\label{Figure3}
\end{figure}

For notational simplicity, we will use a vector symbol $\Vec{\phi}(z) \equiv (\phi^{(1)}(z), \ldots, \phi^{(n-1)}(z))$. The mode expansion of $\Vec{\phi}(z)$ is a multi-component generalization of~\eqref{eq:mode-expansion-compactified-boson}:
\begin{equation}
\label{eq:mode-expansion-multi-boson}
    \Vec{\phi}(z) = \Vec{Q} + 2\pi \Vec{P} z+\mathrm{i}~\underset{k\neq0}{\sum}~\frac{1}{k}~\Vec{a}_{k}~\mathrm{e}^{-2\pi\mathrm{i}kz},
\end{equation}
where $\Vec{Q}$, $\Vec{P}$, and $\Vec{a}_{k}$ are also vectors with $n-1$ components labelled by a superscript $(r)$ with $r=1,2,\ldots,n-1$. The primary states $\vert \Vec{\omega}_{l}; \{0\} \rangle$ of this theory are denoted as $\Vec{\omega}_{l}$ ($l = 0, 1, \ldots, n-1$), where $\Vec{\omega}_{0} \equiv \Vec{0}$ and $\Vec{\omega}_{1}, \ldots, \Vec{\omega}_{n-1}$ are the fundamental weights of $\mathfrak{su}(n)$. The associated module $\mathcal{H}_{\Vec{\omega}_{l}}$ is generated by the states
\begin{align}
\vert \Vec{\omega}_{l} + \sum_{l^{\prime}=1}^{n-1} m_{l^{\prime}}\Vec{\alpha}_{l^{\prime}}; \{n^{(1)}_{k}\}, \ldots, \{n^{(n-1)}_{k}\}\rangle
= & \prod_{r=1}^{n-1} \left[ \left(\prod_{k=1}^{\infty} \frac{1}{\sqrt{n^{(r)}_{k}!~k^{n^{(r)}_{k}}}} \right)
    \left(a^{(r)}_{-1}\right)^{n^{(r)}_{1}} \left(a^{(r)}_{-2}\right)^{n^{(r)}_{2}}\cdots \right]  \nonumber \\
& \times \vert \Vec{\omega}_{l} + \sum_{l^{\prime}=1}^{n-1} m_{l^{\prime}}\Vec{\alpha}_{l^{\prime}}; \{0\}\rangle
\end{align}
with $m_{l} \in \mathbb{Z}$ and $\Vec{n}_{k} \in (\mathbb{N}^{0})^{\otimes (n-1)}$ ($\mathbb{N}^{0}$: non-negative integers), where $\Vec{\alpha}_{1}, \ldots, \Vec{\alpha}_{n-1}$ are the simple roots of $\mathfrak{su}(n)$. Figure~\ref{Figure3} shows the root system of $\mathfrak{su}(3)$ as an example. These states are eigenstates of the momentum part of the free-boson zero modes in~\eqref{eq:mode-expansion-multi-boson},
\begin{align}
    \Vec{P}~\vert \Vec{\omega}_{l} + \sum_{l^{\prime}=1}^{n-1} m_{l^{\prime}}\Vec{\alpha}_{l^{\prime}}; \{n^{(1)}_{k}\}, \ldots, \{n^{(n-1)}_{k}\}\rangle =& \left( \Vec{\omega}_{l} + \sum_{l^{\prime}=1}^{n-1} m_{l^{\prime}}\Vec{\alpha}_{l^{\prime}} \right) \nonumber \\
    & \times \vert \Vec{\omega}_{l} + \sum_{l^{\prime}=1}^{n-1} m_{l^{\prime}}\Vec{\alpha}_{l^{\prime}}; \{n^{(1)}_{k}\}, \ldots, \{n^{(n-1)}_{k}\}\rangle,
\end{align}
and their eigenvalues with respect to $L_{0}$ are
\begin{align}
\label{eq:L0-eigenvalue}
    L_{0}~\vert \Vec{\omega}_{l} + \sum_{l^{\prime}=1}^{n-1} m_{l^{\prime}}\Vec{\alpha}_{l^{\prime}}; \{n^{(1)}_{k}\}, \ldots, \{n^{(n-1)}_{k}\}\rangle =& \left[ \frac{1}{2}\left( \Vec{\omega}_{l} + \sum_{l^{\prime}=1}^{n-1} m_{l^{\prime}}\Vec{\alpha}_{l^{\prime}} \right)^{2} + \underset{k\geq1}{\sum} \sum_{r=1}^{n-1} k n^{(r)}_{k} \right] \nonumber \\
    & \times \vert \Vec{\omega}_{l} + \sum_{l^{\prime}=1}^{n-1} m_{l^{\prime}}\Vec{\alpha}_{l^{\prime}}; \{n^{(1)}_{k}\}, \ldots, \{n^{(n-1)}_{k}\}\rangle.
\end{align}
From the definition~\eqref{eq:character-definition} and~\eqref{eq:L0-eigenvalue}, the character of the highest-weight representation labelled by $\Vec{\omega}_{l}$ is found to be
\begin{align}
    \chi_{\Vec{\omega}_{l}} (\tau) &= \mathrm{Tr}_{\mathcal{H}_{\Vec{\omega}_{l}}} \left( q^{L_{0} - c/24} \right) \nonumber \\
    &= \frac{1}{\left( \eta(\tau) \right)^{n-1}} \underset{m_{1},\ldots,m_{n-1}}{\sum} q^{\frac{1}{2}\left( \Vec{\omega}_{l} + \sum_{l^{\prime}=1}^{n-1} m_{l^{\prime}}\Vec{\alpha}_{l^{\prime}} \right)^{2}},
\end{align}
where we have used the fact that the $\mathrm{SU}(n)_1$ WZW model has central charge $c = n-1$ and the definition of Dedekind's eta function (see appendix~\ref{appdx:derivation-correlator}). For $n=2$, we have $\omega_{1} = 1/\sqrt{2}$ and $\alpha_{1} = \sqrt{2}$, so the characters reduce to those of a compactified boson with radius $R = \sqrt{2}$. Hence the Kalmeyer-Laughlin states discussed in last subsection can also be obtained in this framework.

\begin{figure}
\centering
\includegraphics[width=0.60\textwidth]{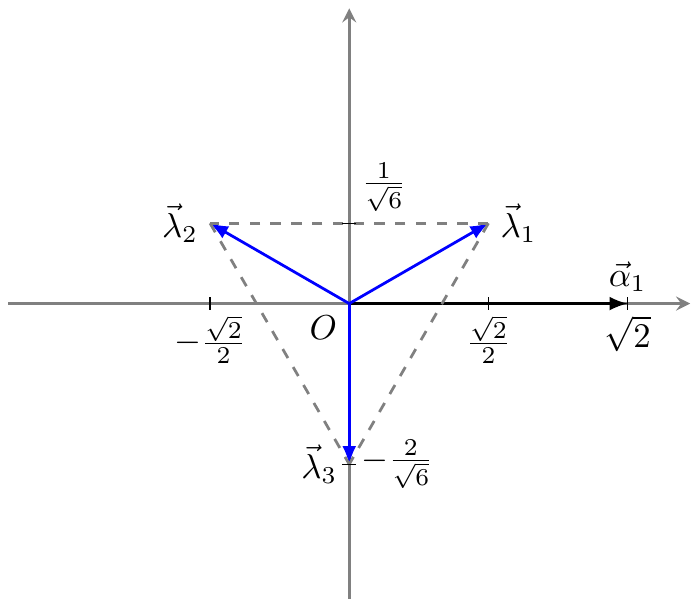}
\caption{The weight vectors $\Vec{\lambda}_{1}$, $\Vec{\lambda}_{2}$ and $\Vec{\lambda}_{3}$ in the fundamental representation of $\mathfrak{su}(3)$, in which the highest weight $\Vec{\lambda}_{1} = \Vec{\omega}_{1}$. Note that the coordinate system is chosen such that $\Vec{\alpha}_{1} = (\sqrt{2},0)$.}
\label{Figure4}
\end{figure}

The primary fields of the $\mathrm{SU}(n)_1$ WZW model can be represented as chiral vertex operators constructed from the $n-1$ free bosons in~\eqref{eq:mode-expansion-multi-boson}. For the highest-weight representation labelled by $\Vec{\omega}_{l}$, the associated primary field has $\mathrm{dim} \mathcal{H}_{\Vec{\omega}_{l}} = \binom{n}{l}$ components ($\Vec{\omega}_{0}$ is the trivial one-dimensional representation):
\begin{equation}
    V_{\Vec{\omega}_{l},s}(z)~\propto~:\mathrm{e}^{\mathrm{i} \Vec{\lambda}_{s} \cdot \Vec{\phi}(z)}:,
\end{equation}
where $\Vec{\lambda}_{s}$ ($s = 1, \ldots, \mathrm{dim} \mathcal{H}_{\Vec{\omega}_{l}}$) are the weight vectors in this representation. In the chiral correlators to be computed, the primary fields correspond to the fundamental representation with dimension $n$, i.e., the representation labelled by $\Vec{\omega}_{1}$. Hence we shall suppress the index $\Vec{\omega}_{1}$ of the vertex operators and simply write
\begin{equation}
\label{eq:sun-vertex}
    V_{s}(z) = \kappa_{s}:\mathrm{e}^{\mathrm{i} \Vec{\lambda}_{s} \cdot \Vec{\phi}(z)}:
\end{equation}
with $s = 1, \ldots, n$. $\kappa_{s}$ is a Klein factor which commutes with the vertex operators and satisfies Majorana-like anticommutation relations
\begin{equation}
\label{eq:Majorana-like-anticommutation-relations}
    \{ \kappa_{s},\kappa_{s^{\prime}} \} = 2 \delta_{s s^{\prime}}.
\end{equation}
The Klein factor is introduced to ensure that the wave function is an SU($n$) singlet, as we shall see below. The components of the weight vectors can be explicitly written down if a specific basis is given. For the orthonormal basis in which the simple root $\Vec{\alpha}_{1} = (\sqrt{2},0,\ldots,0)$, the result is~\cite{tu2014b}
\begin{equation}
\begin{array}{rcccccccccccl}
\Vec{\lambda}_{1} & = & \sqrt{2}~\Big( & \frac{1}{2} & , & \frac{1}{2\sqrt{3}} & , & \ldots
& , & \frac{1}{\sqrt{2n(n-1)}} & \Big) &  & = ~\Vec{\omega}_{1},  \\
\Vec{\lambda}_{2} & = & \sqrt{2}~\Big( & -\frac{1}{2} & , & \frac{1}{2\sqrt{3}} & , & \ldots
& , & \frac{1}{\sqrt{2n(n-1)}} & \Big), &  &  \\
\Vec{\lambda}_{3} & = & \sqrt{2}~\Big( & 0 & , & -\frac{1}{\sqrt{3}} & , & \ldots & , &
\frac{1}{\sqrt{2n(n-1)}} & \Big), &  &  \\
& \vdots &  &  &  &  &  &  &  &  &  &  &  \\
\Vec{\lambda}_{n} & = & \sqrt{2}~\Big( & 0 & , & 0 & , & \ldots & , & -\frac{n-1}{\sqrt{
2n(n-1)}} & \Big). &  &
\end{array}
\label{eq:sunstates}
\end{equation}
The weight vectors for the $\mathfrak{su}(3)$ case are shown in figure~\ref{Figure4} for illustration. One can easily verify that $(\Vec{\lambda}_{s})^{2} = (n-1)/n $ $\forall s$, so the conformal weight of the vertex operators~\eqref{eq:sun-vertex} is $h = (\Vec{\lambda}_{s})^{2}/2 = (n-1)/2n$.

Assume that each lattice site $i$ ($i = 1, 2, \ldots, N$) hosts an $\mathfrak{su}(n)$-generalized spin $\vert s_{i} \rangle$, $s_{i} = 1, \ldots, n$, which transforms under the fundamental representation of $\mathfrak{su}(n)$. Our ansatz for the anyon eigenbasis on the torus is
\begin{equation}
    \label{eq:sun-MES}
    \vert \psi_{l} \rangle = \underset{s_{1}, \ldots, s_{N}}{\sum} \langle V_{s_{1}}(z_{1}) \cdots V_{s_{N}}(z_{N}) \rangle^{\prime}_{l}~\vert s_{1}, \ldots, s_{N} \rangle,
\end{equation}
where the coefficients
\begin{equation}
    \langle V_{s_{1}}(z_{1}) \cdots V_{s_{N}}(z_{N}) \rangle^{\prime}_{l} = \mathrm{Tr}_{\mathcal{H}_{\Vec{\omega}_{l}}} \left( V_{s_{1}}(z_{1}) \cdots V_{s_{N}}(z_{N}) q^{L_{0} - c/24} \right)
\end{equation}
are the (unnormalized) chiral correlators of the vertex operators~\eqref{eq:sun-vertex} evaluated in the subspace $\mathcal{H}_{\Vec{\omega}_{l}}$. This correlator is essentially $n-1$ copies of that in~\eqref{eq:N-point-chiral-correlator-even-p-result}, so we can directly write down the result
\begin{align}
\label{eq:sun-correlator}
    \langle V_{s_{1}}(z_{1}) \cdots V_{s_{N}}(z_{N}) \rangle^{\prime}_{l} =&~ \kappa_{s_{1}} \cdots \kappa_{s_{N}} \delta_{\sum_{i}\Vec{\lambda}_{s_{i}} = \Vec{0}}~\frac{1}{\left( \eta(\tau) \right)^{n-1}} \underset{i<i^{\prime}}{\prod} \left( E(z_{i}-z_{i^{\prime}} \vert \tau) \right)^{\Vec{\lambda}_{s_{i}} \cdot \Vec{\lambda}_{s_{i^{\prime}}}} \nonumber \\
    & \times \underset{m_{1},\ldots,m_{n-1}}{\sum} q^{\frac{1}{2}\left( \Vec{\omega}_{l} + \sum_{l^{\prime}=1}^{n-1} m_{l^{\prime}}\Vec{\alpha}_{l^{\prime}} \right)^{2}} \mathrm{e}^{2\pi\mathrm{i} \sum_{j=1}^{N} \Vec{\lambda}_{s_{j}} \cdot \left( \Vec{\omega}_{l} + \sum_{l^{\prime}=1}^{n-1} m_{l^{\prime}}\Vec{\alpha}_{l^{\prime}} \right) z_{j}},
\end{align}
where $\delta_{\sum_{i}\Vec{\lambda}_{s_{i}} = \Vec{0}}$ arises from charge neutrality. According to~\eqref{eq:sunstates}, charge neutrality requires that the number of the state $\vert s \rangle$, $N_{s}$, to be equal for all $s = 1, \ldots, n$: $N_{1} = \ldots = N_{n} = N/n$. This can be fulfilled only if $N/n$ is an integer, which we shall assume in the following. The labels of the lattice sites occupied by $\vert s \rangle$ are denoted as $i^{(s)}_{1} < \ldots < i^{(s)}_{N/n}$. Choosing $\kappa_{1}\kappa_{2} \cdots \kappa_{n} = 1$ and using~\eqref{eq:Majorana-like-anticommutation-relations}, one obtains $\kappa_{s_{1}}\kappa_{s_{2}} \cdots \kappa_{s_{N}} = \mathrm{sgn}(i^{(1)}_{1},\ldots,i^{(1)}_{N/n},\ldots,i^{(n)}_{1},\ldots,i^{(n)}_{N/n})$. Similar to the construction on the plane~\cite{tu2014b}, this sign factor ensures that the wave functions are $\mathrm{SU}(n)$ singlets.

As for the $\mathrm{SO}(n)_1$ WZW model discussed previously, the modular matrices of the topologically degenerate ground states constructed from the $\mathrm{SU}(n)_1$ WZW model can also be derived analytically. To simplify the calculations, we consider an $L \times L$ square lattice with $L \in 2n\mathbb{Z}$ embedded on a square with a unit side length. The final results are presented here and the details are explained in appendix~\ref{appdx:derivation-modular}.

The modular $S$ transformation changes the wave functions as
\begin{align}
    &\vert \psi_{0} \rangle \rightarrow \frac{1}{\sqrt{n}}  ~\underset{l^{\prime}=0}{\overset{n-1}{\sum}}~\vert \psi_{l^{\prime}} \rangle, \\
    &\vert \psi_{l} \rangle \rightarrow \frac{1}{\sqrt{n}} \left( \vert \psi_{0} \rangle + \underset{l^{\prime}=1}{\overset{n-1}{\sum}} \mathrm{e}^{-2\pi\mathrm{i}\mathfrak{F}_{ll^{\prime}}} \vert \psi_{l^{\prime}} \rangle \right), \quad l = 1, \ldots, n-1.
\end{align}
The modular $T$ transformation changes the wave functions as
\begin{align}
     &\vert \psi_{0} \rangle \rightarrow \mathrm{e}^{-\frac{\pi\mathrm{i}}{12}(n-1)} \vert \psi_{0} \rangle, \\
     &\vert \psi_{l} \rangle \rightarrow \mathrm{e}^{-\pi\mathrm{i} \left( \frac{n-1}{12} - \mathfrak{F}_{ll} \right)} \vert \psi_{l} \rangle, \quad l = 1, \ldots, n-1.
\end{align}
Here $\mathfrak{F}$ is an $(n-1) \times (n-1)$ matrix
\begin{equation}
\label{eq:F-explicit-1}
    \mathfrak{F} = \frac{1}{n}~\left( \begin{array}{cccccc}
        n-1 & ~n-2 & ~n-3 & ~\cdots & ~2 & ~1 \\
        n-2 & ~2(n-2) & ~2(n-3) & ~\cdots & ~4 & ~2 \\
        n-3 & ~2(n-3) & ~3(n-3) & ~\cdots & ~6 & ~3 \\
        \vdots & ~\vdots & ~\vdots & ~\ddots & ~\vdots & ~\vdots \\
        2 & ~4 & ~6 & ~\cdots & ~2(n-2) & ~n-2 \\
        1 & ~2 & ~3 & ~\cdots & ~n-2 & ~n-1
    \end{array} \right),
\end{equation}
which is nothing but the quadratic form matrix of the $\mathfrak{su}(n)$ Lie algebra (see appendix~\ref{appdx:derivation-modular} for more details). This means that the modular matrices can be expressed as
\begin{eqnarray}
\label{eq:modular-S-sun}
\mathcal{S}_{ll^{\prime}}=\begin{cases}
\begin{array}{c}
\frac{1}{\sqrt{n}} \mathrm{e}^{-2\pi\mathrm{i}\mathfrak{F}_{ll^{\prime}}} \\
\frac{1}{\sqrt{n}}
\end{array} & \begin{array}{c}
l \neq 0, l^{\prime} \neq 0 \\
\textrm{otherwise}
\end{array} \end{cases},
\end{eqnarray}
and
\begin{equation}
\label{eq:modular-T-sun}
    \mathcal{T}_{ll^{\prime}} = \mathrm{e}^{-2\pi\mathrm{i} \left( \frac{n-1}{24} - \frac{l(n-l)}{2n} \right)} \delta_{ll^{\prime}},
\end{equation}
which are precisely those describing the $\mathrm{SU}(n)_{1}$ TQFT. In particular, the modular $T$ matrix is diagonal, so the topologically degenerate ground states indeed correspond to anyon eigenbasis on the torus. As a concrete example, the quadratic form matrix for the $n=3$ case is
\begin{equation}
\mathfrak{F} = \frac{1}{3}\left(\begin{array}{cc}
2~ & ~1\\
1~ & ~2
\end{array}\right),
\end{equation}
so one obtains
\begin{equation}
\mathcal{S} = \frac{1}{\sqrt{3}}
\begin{pmatrix}
1~ & 1 & 1\\
1~ & \mathrm{e}^{\frac{2\pi\mathrm{i}}{3}} & \mathrm{e}^{\frac{4\pi\mathrm{i}}{3}}\\
1~ & \mathrm{e}^{\frac{4\pi\mathrm{i}}{3}} & \mathrm{e}^{\frac{2\pi\mathrm{i}}{3}}
\end{pmatrix}, \qquad
\mathcal{T} = \mathrm{e}^{-\frac{\pi\mathrm{i}}{6}}
\begin{pmatrix}
1~ & 0 & 0\\
0~ & ~\mathrm{e}^{\frac{2\pi\mathrm{i}}{3}} & 0\\
0~ & 0 & \mathrm{e}^{\frac{2\pi\mathrm{i}}{3}}
\end{pmatrix}
\end{equation}
using~\eqref{eq:modular-S-sun} and~\eqref{eq:modular-T-sun}.

\section{Summary and outlook}
\label{sec:summary}

In summary, we have proposed a new approach which generates model wave functions for chiral topological states on the torus. The construction utilizes the correspondence between chiral primary fields of the CFT at the edge and anyon eigenbasis of the topological order in the bulk, which endows it with a clear physical meaning and makes it conceptually simpler as compared to the previous approach. The anyon eigenbasis of the topologically degenerate ground states on the torus can be obtained \emph{automatically} in this approach. This is confirmed by explicit calculations of the model wave functions and the associated modular matrices from the compactified boson CFT, the $\mathrm{SU}(n)_{1}$ WZW models and the $\mathrm{SO}(n)_{1}$ WZW models. In particular, the model wave functions constructed from the $\mathrm{SO}(n)_{1}$ WZW models provide a realization of Kitaev's sixteenfold way on the torus. Our work has further elevated the power of CFT-based methods in understanding the physics of strongly correlated quantum many-body systems.

An exceptional open problem that has not been solved in this paper is how to find the parent Hamiltonian for a given set of degenerate ground states on the torus. While parent Hamiltonians for many model wave functions constructed from CFT correlators on the plane can be derived using the null-field technique, a suitable generalization that would work on the torus has not been found. The approach developed in this paper might shed some light in this direction, and we hope to report interesting results in future works.
\\
\\
\textit{Note added.}-- After submission of our manuscript, it was pointed out to us that the chiral CFT approach has already been proposed in Ref.~\cite{crepel2019}. This fact calls for a brief comparison of Ref.~\cite{crepel2019} and the present work. Both papers utilized chiral correlators of CFT to construct topologically ordered states on the torus, but the specific systems are different: Ref.~\cite{crepel2019} studied fractional quantum Hall states in the continuum Landau levels and the present work studied chiral spin liquids in lattice models. The two important progresses of the present work are: i) a one-to-one correspondence between elements in the anyon eigenbasis and sectors of the chiral CFT is established; ii) the modular $S$ and $T$ matrices for the topologically degenerate ground states are derived analytically. We thank the authors of Ref.~\cite{crepel2019} for illuminating discussions regarding their paper.

\acknowledgments

We are grateful to Germ\'an Sierra for helpful discussions. The authors are supported by the National Key Research and Development Project of China (Grant No.~2017YFA0302901), the National Natural Science Foundation of China (Grants No.~11888101, No.~11874095, No.~11804107), the Strategic Priority Research Program of Chinese Academy of Sciences (Grant No.~XDB33000000), startup grant of HUST, and the Deutsche Forschungsgemeinschaft (DFG) through project A06 of SFB 1143 (project-id 247310070). H.C.Z. acknowledges funding from the CAS-DAAD Doctoral Joint Scholarship Program.

\appendix
\section{Chiral correlators on the torus: derivation in the operator formalism}
\label{appdx:derivation-correlator}

In this appendix, we derive several multipoint chiral correlators for free fermion and compactified boson CFTs on the torus. The direct computation of such chiral correlators seems to be scarce in the literature. Here, we utilize the operator formalism to achieve this goal.

\subsection{The free fermion CFT}

In this subsection, we derive the holomorphic partition functions~\eqref{eqs:chiral-partition-function-begin} to~\eqref{eqs:chiral-partition-function-end} and chiral correlators~\eqref{eq:multi-point-1} with various spin structures of the free fermion CFT on the torus.

Let us take the case $\nu =$ AA as an example to illustrate the derivation. The holomorphic partition function is defined in eq.~\eqref{eqs:chiral-partition-function-begin},
\begin{equation}
    Z_{\textrm{AA}}(\tau) = q^{-\frac{1}{48}}~\mathrm{Tr}_{\textrm{NS}} \left(q^{L_0}\right),
\end{equation}
where $L_{0}$ is the zeroth Virasoro generator for the NS sector
\begin{equation}
\label{eq:virasoro-generator-NS-repeated}
    L_{0}=\underset{r\in\mathbb{Z}+1/2,r>0}{\sum}r\chi_{-r}\chi_{r}.
\end{equation}
The trace over all modes is equivalent to the product of the traces over each mode:
\begin{align}
    Z_{\textrm{AA}}(\tau) &= q^{-\frac{1}{48}}~\mathrm{Tr}\left(\underset{r\in\mathbb{Z}+1/2,r>0}{\bigotimes}q^{r\chi_{-r}\chi_{r}}\right) \nonumber \\
    &= q^{-\frac{1}{48}}\underset{r\in\mathbb{Z}+1/2,r>0}{\prod}\mathrm{Tr}\left(q^{r\chi_{-r}\chi_{r}}\right).
\end{align}
For the fermionic mode labelled by $r$, the corresponding subspace is spanned by the basis $\{\vert0\rangle,\chi_{-r}\vert0\rangle\}$. In this basis, the operator to be traced over has the matrix representation
\begin{equation}
\label{eq:matrix-representation}
    q^{r\chi_{-r}\chi_{r}}=\left(\begin{array}{cc}
1 & 0\\
0 & q^{r}
\end{array}\right),
\end{equation}
and the trace can immediately be calculated, yielding the result
\begin{equation}
    Z_{\textrm{AA}}(\tau) = q^{-\frac{1}{48}}\underset{r\in\mathbb{Z}+1/2,r>0}{\prod}\left(1+q^{r}\right).
\end{equation}
The calculations for other spin structures are almost identical to that for $\nu =$ AA, except that the insertion of the fermion parity operator $(-1)^{F}=\prod_{r}(-1)^{F_{r}}$ changes the trace over each fermionic mode to
\begin{equation}
    \mathrm{Tr}\left((-1)^{F_{r}}q^{r\chi_{-r}\chi_{r}}\right) = 1-q^{r}.
\end{equation}
The results are
\begin{align}
    Z_{\textrm{AP}}(\tau) &= q^{-\frac{1}{48}}\underset{r\in\mathbb{Z}+1/2,r>0}{\prod}\left(1-q^{r}\right), \\
    Z_{\textrm{PA}}(\tau) &= \frac{1}{\sqrt{2}}~q^{\frac{1}{24}}\underset{r\in\mathbb{Z},r\geq0}{\prod}\left(1+q^{r}\right), \\
    Z_{\textrm{PP}}(\tau) &= \frac{1}{\sqrt{2}}~q^{\frac{1}{24}}\underset{r\in\mathbb{Z},r\geq0}{\prod}\left(1-q^{r}\right) = 0.
\end{align}
Obviously, $Z_{\textrm{PP}}$ vanishes due to the existence of the fermionic zero mode with $r = 0$ in the R sector. To obtain non-vanishing results for the PP boundary condition, consider the holomorphic partition function with the insertion of a fermion field $\chi(z_{i_{0}})$. This quantity is actually an (unnormalized) single-point correlator; however, it is clear that only the zero mode of $\chi(z_{i_{0}})$ will have non-vanishing contribution as $\mathrm{Tr}\left((-1)^{F_{0}}\chi_{0}\right)=1$, thus the result does not depend on the position $z_{i_{0}}$ of the insertion. In other words, the effect of the insertion of a fermion field is just to compensate the trace over the zero mode, giving the result
\begin{align}
    Z^{\prime}_{\textrm{PP}}(\tau) &= q^{-\frac{1}{48}}~\mathrm{Tr}^{\prime}_{\textrm{R}} \left((-1)^{F}q^{L_0}\right) \nonumber \\
    &= q^{\frac{1}{24}}\underset{r\in\mathbb{Z},r>0}{\prod}\left(1-q^{r}\right),
\end{align}
where $\mathrm{Tr}^{\prime}$ indicates that the zero mode is excluded in the trace.

Let us now turn to the chiral correlators. The two-point correlators between fermion fields $\chi(z)$ and $\chi(w)$ are defined by
\begin{align}
\label{eq:2-point-correlators-begin}
    g_{\textrm{AA}}(z-w \vert \tau)~&=~ \frac{1}{Z_{\textrm{AA}}(\tau)}~q^{-\frac{1}{48}}~ \mathrm{Tr}_{\textrm{NS}} \left( \chi(z) \chi(w) q^{L_{0}} \right), \\
    g_{\textrm{AP}}(z-w \vert \tau)~&=~ \frac{1}{Z_{\textrm{AP}}(\tau)}~q^{-\frac{1}{48}}~ \mathrm{Tr}_{\textrm{NS}} \left( (-1)^{F} \chi(z) \chi(w) q^{L_{0}} \right), \\
    g_{\textrm{PA}}(z-w \vert \tau)~&=~ \frac{1}{Z_{\textrm{PA}}(\tau)}~q^{-\frac{1}{48}}~ \frac{1}{\sqrt{2}} \mathrm{Tr}_{\textrm{R}} \left( \chi(z) \chi(w) q^{L_{0}} \right).
\label{eq:2-point-correlators-end}
\end{align}
For $\nu =$ AA, substituting the mode expansion~\eqref{eq:fermion-field-cylinder} into the correlator yields
\begin{align}
\label{eq:2-point-correlator-trace}
    q^{-\frac{1}{48}}\mathrm{Tr}_{\textrm{NS}}\left(\chi(z)\chi(w)q^{L_{0}}\right) &= -2\pi\mathrm{i}\cdot q^{-\frac{1}{48}} \underset{m,n\in\mathbb{Z}+1/2}{\sum}\mathrm{Tr}_{\textrm{NS}}\left(\chi_{n}\chi_{m}q^{L_{0}}\right)\mathrm{e}^{2\pi\mathrm{i}(nz+mw)} \nonumber \\
    &= -2\pi\mathrm{i}\cdot q^{-\frac{1}{48}} \underset{n\in\mathbb{Z}+1/2}{\sum}\mathrm{Tr}_{\textrm{NS}}\left(\chi_{n}\chi_{-n}q^{L_{0}}\right)\mathrm{e}^{2\pi\mathrm{i}n(z-w)} \nonumber \\
    &= -2\pi\mathrm{i}\cdot q^{-\frac{1}{48}} \underset{n\in\mathbb{Z}+1/2,n>0}{\sum} \big( \mathrm{Tr}_{\textrm{NS}}\left(\chi_{n}\chi_{-n}q^{L_{0}}\right)\mathrm{e}^{2\pi\mathrm{i}n(z-w)} \nonumber \\
    &~~~~~~~~~~~~~~~~~~~~~~~~~~~~~~~+ \mathrm{Tr}_{\textrm{NS}}\left(\chi_{-n}\chi_{n}q^{L_{0}}\right)\mathrm{e}^{-2\pi\mathrm{i}n(z-w)} \big),
\end{align}
where we have made use of the fact that the trace is non-vanishing only if $m+n=0$. Now let us consider the two terms on the right hand side of~\eqref{eq:2-point-correlator-trace} separately. Substituting~\eqref{eq:virasoro-generator-NS-repeated} into the first term, we find
\begin{align}
\label{eq:trace-first-term}
    &~q^{-\frac{1}{48}} \underset{n\in\mathbb{Z}+1/2,n>0}{\sum} \mathrm{Tr}_{\textrm{NS}}\left(\chi_{n}\chi_{-n}q^{L_{0}}\right)\mathrm{e}^{2\pi\mathrm{i}n(z-w)} \nonumber \\
    =&~q^{-\frac{1}{48}}\underset{n\in\mathbb{Z}+1/2,n>0}{\sum}\mathrm{Tr}\left(\chi_{n}\chi_{-n}\underset{r\in\mathbb{Z}+1/2,r>0}{\bigotimes}q^{r\chi_{-r}\chi_{r}}\right)\mathrm{e}^{2\pi\mathrm{i}n(z-w)} \nonumber \\
    =&~q^{-\frac{1}{48}}\underset{r\in\mathbb{Z}+1/2,r>0}{\prod}\left(1+q^{r}\right)\underset{n\in\mathbb{Z}+1/2,n>0}{\sum}\frac{1}{1+q^{n}}\mathrm{e}^{2\pi\mathrm{i}n(z-w)} \nonumber \\
    =&~Z_{\textrm{AA}}(\tau) \underset{n\in\mathbb{Z}+1/2,n>0}{\sum}\frac{1}{1+q^{n}}\mathrm{e}^{2\pi\mathrm{i}n(z-w)},
\end{align}
where we have made use of the matrix representation~\eqref{eq:matrix-representation} and
\begin{equation}
    \chi_{n}\chi_{-n}~q^{n\chi_{-n}\chi_{n}}=\left(\begin{array}{cc}
1 & 0\\
0 & 0
\end{array}\right).
\end{equation}
Similar calculations for the second term on the right hand side of~\eqref{eq:2-point-correlator-trace} yield
\begin{equation}
\label{eq:trace-second-term}
    q^{-\frac{1}{48}} \underset{n\in\mathbb{Z}+1/2,n>0}{\sum} \mathrm{Tr}_{\textrm{NS}}\left(\chi_{-n}\chi_{n}q^{L_{0}}\right)\mathrm{e}^{-2\pi\mathrm{i}n(z-w)}
    = Z_{\textrm{AA}}(\tau) \underset{n\in\mathbb{Z}+1/2,n<0}{\sum}\frac{1}{1+q^{n}}~\mathrm{e}^{2\pi\mathrm{i}n(z-w)}.
\end{equation}
Substituting~\eqref{eq:trace-first-term} and~\eqref{eq:trace-second-term} into~\eqref{eq:2-point-correlator-trace} and then~\eqref{eq:2-point-correlator-trace} into~\eqref{eq:2-point-correlators-begin}, we conclude that
\begin{equation}
\label{eq:2-point-correlator-series-AA}
    g_{\textrm{AA}}(z-w \vert \tau) = -2\pi\mathrm{i}\cdot \underset{n\in\mathbb{Z}+1/2}{\sum}\frac{1}{1+q^{n}}~\mathrm{e}^{2\pi\mathrm{i}n(z-w)}.
\end{equation}
The calculations for $\nu =$ AP, PA are quite similar to that for $\nu =$ AA, with results
\begin{align}
\label{eq:2-point-correlator-series-AP}
    g_{\textrm{AP}}(z-w \vert \tau) &= -2\pi\mathrm{i}\cdot \underset{n\in\mathbb{Z}+1/2}{\sum}\frac{1}{1-q^{n}}~\mathrm{e}^{2\pi\mathrm{i}n(z-w)}, \\
    g_{\textrm{PA}}(z-w \vert \tau) &= -2\pi\mathrm{i}\cdot \underset{n\in\mathbb{Z}}{\sum}~\frac{1}{1+q^{n}}~\mathrm{e}^{2\pi\mathrm{i}n(z-w)}.
\label{eq:2-point-correlator-series-PA}
\end{align}
For the PP sector, the expectation value
\begin{equation}
    q^{-\frac{1}{48}} \frac{1}{\sqrt{2}} \mathrm{Tr}_{\textrm{R}} \left( (-1)^{F} \chi(z) \chi(w) q^{L_{0}} \right)
\end{equation}
vanishes due to the zero mode. Similar to the character~\eqref{eq:chiral-partition-function-pp}, we define the  correlator in the PP sector
\begin{equation}
    g^{\prime}_{\textrm{PP}}(z-w \vert \tau)= \frac{1}{Z^{\prime}_{\textrm{PP}}(\tau)}~ q^{-\frac{1}{48}}~\mathrm{Tr}^{\prime}_{\textrm{R}} \left( (-1)^{F} \chi(z) \chi(w) q^{L_{0}} \right)
\end{equation}
by excluding the zero mode. Its explicit expression can be evaluated as follows:
\begin{align}
    & \phantom{=} \quad q^{-\frac{1}{48}} \mathrm{Tr}^{\prime}_{\textrm{R}} \left( (-1)^{F} \chi(z) \chi(w) q^{L_{0}} \right) \nonumber \\
    &= -2\pi\mathrm{i}\cdot q^{\frac{1}{24}}\underset{r\in\mathbb{Z},r>0}{\prod}\left(1-q^{r}\right)\cdot\left(\underset{n\in\mathbb{Z},n\neq0}{\sum}\frac{1}{1-q^{n}}\mathrm{e}^{2\pi\mathrm{i}n(z-w)}+\frac{1}{2}\right) \nonumber \\
    &= -2\pi\mathrm{i}\cdot Z^{\prime}_{\textrm{PP}}(\tau) \left(\underset{n\in\mathbb{Z},n\neq0}{\sum}\frac{1}{1-q^{n}}\mathrm{e}^{2\pi\mathrm{i}n(z-w)}+\frac{1}{2}\right),
\end{align}
hence
\begin{equation}
\label{eq:2-point-correlator-series-PP}
    g^{\prime}_{\textrm{PP}}(z-w \vert \tau)= -2\pi\mathrm{i}\cdot \left(\underset{n\in\mathbb{Z},n\neq0}{\sum}\frac{1}{1-q^{n}}\mathrm{e}^{2\pi\mathrm{i}n(z-w)}+\frac{1}{2}\right).
\end{equation}

Let us proceed to calculate multipoint chiral correlators. Recall that in the derivation of the two-point chiral correlator with AA boundary condition,~\eqref{eq:2-point-correlator-trace}, the indices of the two fermionic modes must satisfy $m+n=0$, otherwise the trace will vanish. Repeating this argument for a multipoint chiral correlator with $\nu = \mathrm{AA}$, it is clear that only even-point correlators are non-vanishing. To be concrete, consider the $2N$-point (unnormalized) chiral correlator defined by
\begin{equation}
    \langle \chi(z_1) \cdots \chi(z_{2N}) \rangle^{\prime}_{\textrm{AA}} = q^{-\frac{1}{48}}~ \mathrm{Tr}_{\textrm{NS}} \left( \chi(z_1) \cdots \chi(z_{2N}) q^{L_0} \right).
\end{equation}
When using the mode expansion~\eqref{eq:fermion-field-cylinder} for all fermion fields, the modes $\chi_{n_{1}},~\chi_{n_{2}},\ldots,~\chi_{n_{2N}}$ should be put into $N$ pairs under some permutation $\mathcal{P} \in S_{2N}$ (the permutation group of $2N$ elements) such that $n_{\mathcal{P}(2i-1)} + n_{\mathcal{P}(2i)} = 0$ for $i=1,\ldots,N$. For a specific way of pairing, we get a product of $N$ two-point correlators, and the final result is the summation over all possible ways of pairing:
\begin{align}
\label{eq:pairing-pattern}
    \langle \chi(z_1) \cdots \chi(z_{2N}) \rangle^{\prime}_{\textrm{AA}} & = Z_{\textrm{AA}}(\tau) \cdot \frac{1}{N! 2^{N}} \underset{\mathcal{P} \in S_{2N}}{\sum}~\mathrm{sgn}(\mathcal{P}) ~\underset{i=1}{\overset{N}{\prod}}~ g_{\textrm{AA}}(z_{\mathcal{P}(2i-1)}-z_{\mathcal{P}(2i)} \vert \tau) \nonumber \\
    & = Z_{\textrm{AA}}(\tau)~\mathrm{Pf} \left( g_{\textrm{AA}}(z_{i}-z_{j} \vert \tau) \right),
\end{align}
where $\mathrm{sgn}(\mathcal{P})$ is the signature of the permutation arising from exchanging fermionic operators and the factor $1/(N! 2^{N})$ compensates the overcounting. To arrive at the Pfaffian form, we have just used the definition of the Pfaffian of a $(2N) \times (2N)$ antisymmetric matrix whose $(i,j)$ entry is $g_{\textrm{AA}}(z_{i}-z_{j} \vert \tau)$ for $i \neq j$ and $0$ for $i = j$.

We note that the pattern of pairing that leads to~\eqref{eq:pairing-pattern} is exactly the same as that in Wick's theorem~\cite{francesco1997}; in fact, the above steps can be viewed as a rederivation of the Wick's theorem within the operator formalism. Similar calculations for $\nu =$ AP and PA give
\begin{equation}
    \langle \chi(z_1) \cdots \chi(z_{2N}) \rangle^{\prime}_{\textrm{AP}} = Z_{\textrm{AP}}(\tau)~\mathrm{Pf} \left( g_{\textrm{AP}}(z_{i}-z_{j} \vert \tau) \right),
\end{equation}
and
\begin{equation}
    \langle \chi(z_1) \cdots \chi(z_{2N}) \rangle^{\prime}_{\textrm{PA}} = Z_{\textrm{PA}}(\tau)~\mathrm{Pf} \left( g_{\textrm{PA}}(z_{i}-z_{j} \vert \tau) \right).
\end{equation}
For $\nu =$ PP, on the contrary, only odd-point correlators are non-vanishing. In the $(2N + 1)$-point (unnormalized) chiral correlator
\begin{equation}
    \langle \chi(z_1) \cdots \chi(z_{2N + 1}) \rangle^{\prime}_{\textrm{PP}} = q^{-\frac{1}{48}}~ \mathrm{Tr}_{\textrm{R}} \left( (-1)^{F} \chi(z_1) \cdots \chi(z_{2N + 1}) q^{L_0} \right),
\end{equation}
one of the fermionic fields, $\chi(z_{i_{0}})$, $i_{0} = 1,2,\ldots,(2N + 1)$, plays the role of providing the zero mode to compensate the trace over it, and the remaining $2N$ fermion fields follow the same calculations that leading to~\eqref{eq:pairing-pattern}. The result is given by
\begin{equation}
\label{eq:odd-point-correlator}
    \langle \chi(z_1) \cdots \chi(z_{2N + 1}) \rangle^{\prime}_{\textrm{PP}} = Z^{\prime}_{\textrm{PP}}(\tau) \sum_{i_{0}=1}^{2N+1}(-1)^{i_{0}-1}\mathrm{Pf}\left(g^{\prime}_{\textrm{PP}}(z_{i}-z_{j}\vert\tau)\right)
\end{equation}
with $i,j=1,\cdots,\hat{i}_{0},\cdots,(2N+1)$ in the Pfaffian. Here the hat `` $\hat{}$ '' on an object indicates that this object is to be omitted.

Finally, we note that the results we have derived for the holomorphic partition functions and chiral correlators on the torus can be expressed in terms of certain special functions, namely Jacobi's theta functions, Dedekind's eta function and (generalized) Weierstrass functions. For completeness, we list here the definition of these special functions. Further details about these functions can be found in refs.~\cite{francesco1997,blumenhagen2009}.

Jacobi's theta functions are defined by
\begin{align}
\label{eq:theta-sum-begin}
\vartheta_{1}(z\vert\tau)&=-\mathrm{i}~\underset{n\in\mathbb{Z}}{\sum}~(-1)^{n}y^{n+1/2}q^{(n+1/2)^{2}/2}, \\
\vartheta_{2}(z\vert\tau)&=\underset{n\in\mathbb{Z}}{\sum}~y^{n+1/2}q^{(n+1/2)^{2}/2}, \\
\vartheta_{3}(z\vert\tau)&=\underset{n\in\mathbb{Z}}{\sum}~y^{n}q^{n^{2}/2}, \\
\vartheta_{4}(z\vert\tau)&=\underset{n\in\mathbb{Z}}{\sum}~(-1)^{n}y^{n}q^{n^{2}/2},
\label{eq:theta-sum-end}
\end{align}
where $q=\mathrm{e}^{2\pi\mathrm{i}\tau}$ and $y=\mathrm{e}^{2\pi\mathrm{i}z}$. The argument of these functions is the complex variable $z$, whilst $\tau$ is a complex parameter with $\mathrm{Im}\tau>0$. By using Jacobi's triple product identity~\cite{francesco1997,ginsparg1988,blumenhagen2009}
\begin{equation}
\prod_{n=1}^{\infty}(1-q^{n})(1+yq^{n-1/2})(1+y^{-1}q^{n-1/2})=\underset{n\in\mathbb{Z}}{\sum}~y^{n}q^{n^{2}/2},
\end{equation}
Jacobi's theta functions can also be expressed in the form of infinite products:
\begin{align}
\label{eq:theta1-product}
\vartheta_{1}(z\vert\tau)&=-\mathrm{i}y^{\frac{1}{2}}q^{\frac{1}{8}}\prod_{n=1}^{\infty}(1-q^{n})(1-yq^{n})(1-y^{-1}q^{n-1}), \\
\vartheta_{2}(z\vert\tau)&=y^{\frac{1}{2}}q^{\frac{1}{8}}\prod_{n=1}^{\infty}(1-q^{n})(1+yq^{n})(1+y^{-1}q^{n-1}), \\
\vartheta_{3}(z\vert\tau)&=~\prod_{n=1}^{\infty}(1-q^{n})(1+yq^{n-1/2})(1+y^{-1}q^{n-1/2}), \\
\vartheta_{4}(z\vert\tau)&=~\prod_{n=1}^{\infty}(1-q^{n})(1-yq^{n-1/2})(1-y^{-1}q^{n-1/2}).
\end{align}
Jacobi's theta functions at $z=0$ are termed standard Jacobi's theta functions
\begin{equation}
\vartheta_{\nu}(\tau)\equiv\vartheta_{\nu}(0\vert\tau)
\end{equation}
for $\nu=1,2,3,4$. Hence we have
\begin{align}
\vartheta_{2}(\tau)&=\underset{n\in\mathbb{Z}}{\sum}~q^{(n+1/2)^{2}/2}, \\
\vartheta_{3}(\tau)&=\underset{n\in\mathbb{Z}}{\sum}~q^{n^{2}/2}, \\
\vartheta_{4}(\tau)&=\underset{n\in\mathbb{Z}}{\sum}~(-1)^{n}q^{n^{2}/2},
\end{align}
or, equivalently,
\begin{align}
\label{eq:standard-Jacobi-theta-begin}
\vartheta_{2}(\tau)&=2q^{1/8}\prod_{n=1}^{\infty}(1-q^{n})(1+q^{n})^{2}, \\
\vartheta_{3}(\tau)&=~\prod_{n=1}^{\infty}(1-q^{n})(1+q^{n-1/2})^{2}, \\
\vartheta_{4}(\tau)&=~\prod_{n=1}^{\infty}(1-q^{n})(1-q^{n-1/2})^{2}.
\label{eq:standard-Jacobi-theta-end}
\end{align}
Note that $\vartheta_{1}(\tau)=0$, as can easily be seen from its infinite product form.
Dedekind's eta function is defined by
\begin{equation}
    \eta(\tau)=q^{\frac{1}{24}}~\prod_{n=1}^{\infty}~(1-q^{n}).
\label{eq:Dedekind}
\end{equation}
From~\eqref{eq:standard-Jacobi-theta-begin} to~\eqref{eq:standard-Jacobi-theta-end} and~\eqref{eq:Dedekind}, it is easy to see that the holomorphic partition functions can be expressed as
\begin{align}
     Z_{\textrm{AA}}(\tau) &= \sqrt{\frac{\vartheta_{3}(\tau)}{\eta(\tau)}}, \\
    Z_{\textrm{AP}}(\tau) &= \sqrt{\frac{\vartheta_{4}(\tau)}{\eta(\tau)}}, \\
    Z_{\textrm{PA}}(\tau) &= \sqrt{\frac{\vartheta_{2}(\tau)}{\eta(\tau)}}.
\end{align}
The (generalized) Weierstrass functions are defined in terms of Jacobi's theta functions (for $\nu=2,3,4$):
\begin{equation}
\wp_{\nu}(z_{i}-z_{j}\vert\tau)=\frac{\vartheta_{\nu}(z_{i}-z_{j}\vert\tau)\partial_{z}\vartheta_{1}(z\vert\tau)\vert_{z=0}}{\vartheta_{\nu}(\tau)\vartheta_{1}(z_{i}-z_{j}\vert\tau)}.
\label{eq:Weierstrass}
\end{equation}
One should note that the series expressions of the two-point chiral correlators, \eqref{eq:2-point-correlator-series-AA}, \eqref{eq:2-point-correlator-series-AP}, \eqref{eq:2-point-correlator-series-PA} and \eqref{eq:2-point-correlator-series-PP}, converge in the so-called principal interval $0<\mathrm{Im}z<1$ (which we assumed throughout the derivation);
to define the correlation functions on the entire complex
plane, one has to continue these expressions according to the corresponding periodic/anti-periodic boundary condition along the imaginary direction. It has been verified numerically that these (properly continued)
expressions can be expressed as
\begin{align}
\label{eq:two-point-correlator-result-AA}
    g_{\textrm{AA}}(z-w \vert \tau) &= \wp_{3}(z-w\vert\tau), \\
\label{eq:two-point-correlator-result-AP}
    g_{\textrm{AP}}(z-w \vert \tau) &= \wp_{4}(z-w\vert\tau), \\
\label{eq:two-point-correlator-result-PA}
    g_{\textrm{PA}}(z-w \vert \tau) &= \wp_{2}(z-w\vert\tau), \\
\label{eq:two-point-correlator-result-PP}
    g^{\prime}_{\textrm{PP}}(z-w \vert \tau) &= \frac{\partial_{z}\vartheta_{1}(z-w\vert\tau)}{\vartheta_{1}(z-w\vert\tau)}.
\end{align}
As a byproduct,~\eqref{eq:2-point-correlator-series-AA},~\eqref{eq:2-point-correlator-series-AP},~\eqref{eq:2-point-correlator-series-PA} and~\eqref{eq:2-point-correlator-series-PP} may serve as convenient series expressions for these special functions.

\subsection{The compactified boson CFT}

This subsection is devoted to the detailed derivation of the multipoint correlator of chiral vertex operators of a compactified boson evaluated in a certain momentum eigenstate, i.e., eq.~\eqref{eq:correlator-momentum-eigenstate}.

For simplicity, let us begin with the two-point case and calculate
\begin{equation}
\label{eq:momentum-eigenstate-expectation}
    \underset{n_{1},n_{2},\ldots\geq0}{\sum} q^{\sum_{{k\geq1}}kn_{k}} \langle n,m;n_{1},n_{2},\ldots\vert :\mathrm{e}^{\mathrm{i}\alpha_{1}\phi(z_1)}: :\mathrm{e}^{\mathrm{i}\alpha_{2}\phi(z_2)}: \vert n,m;n_{1},n_{2},\ldots\rangle,
\end{equation}
where the chiral boson has the mode expansion (see~\eqref{eq:mode-expansion-compactified-boson})
\begin{equation}
\label{eq:mode-expansion-chiral}
    \phi(z) = Q + 2\pi Pz + \mathrm{i}~\underset{k\neq0}{\sum}~\frac{1}{k}a_{k}\mathrm{e}^{-2\pi\mathrm{i}kz} \equiv Q+2\pi Pz + \Hat{\phi}(z),
\end{equation}
where $\Hat{\phi}(z)$ has no zero-mode. The zero-mode parts can be separated out:
\begin{equation}
\label{eq:product-of-vertex-operators}
    :\mathrm{e}^{\mathrm{i}\alpha_{1}\phi(z_1)}: :\mathrm{e}^{\mathrm{i}\alpha_{2}\phi(z_2)}: = \mathrm{e}^{\mathrm{i}\alpha_{1}(Q + 2\pi Pz_{1})} \mathrm{e}^{\mathrm{i}\alpha_{2}(Q + 2\pi Pz_{2})} :\mathrm{e}^{\mathrm{i}\alpha_{1}\Hat{\phi}(z_1)}: :\mathrm{e}^{\mathrm{i}\alpha_{2}\Hat{\phi}(z_2)}:.
\end{equation}
Now we invoke the Baker-Campbell-Hausdorff (BCH) formula, which states that if the commutator $[O_{1},O_{2}]$ of two operators commutes with both $O_{1}$ and $O_{2}$, then
\begin{equation}
\label{eq:BCH}
    \mathrm{e}^{O_{1}} \mathrm{e}^{O_{2}} = \mathrm{e}^{O_{1}+O_{2}} \mathrm{e}^{[O_{1},O_{2}]/2} = \mathrm{e}^{O_{2}} \mathrm{e}^{O_{1}} \mathrm{e}^{[O_{1},O_{2}]}.
\end{equation}
Choosing $O_{1} \rightarrow \mathrm{i}\alpha_{1}(Q + 2\pi Pz_{1})$, $O_{2} \rightarrow \mathrm{i}\alpha_{2}(Q + 2\pi Pz_{2})$ and making use of the commutation relation~\eqref{eq:canonical-commutation-relation},~\eqref{eq:product-of-vertex-operators} becomes
\begin{equation}
\label{eq:BCH-result}
    :\mathrm{e}^{\mathrm{i}\alpha_{1}\phi(z_1)}: :\mathrm{e}^{\mathrm{i}\alpha_{2}\phi(z_2)}: = \mathrm{e}^{\mathrm{i}(\alpha_{1} + \alpha_{2})Q + 2\pi\mathrm{i}P (\alpha_{1}z_{1} + \alpha_{2}z_{2})} ~\mathrm{e}^{\pi\mathrm{i} \alpha_{1}\alpha_{2} (z_{1} - z_{2})} :\mathrm{e}^{\mathrm{i}\alpha_{1}\Hat{\phi}(z_1)}: :\mathrm{e}^{\mathrm{i}\alpha_{2}\Hat{\phi}(z_2)}:.
\end{equation}
From the canonical commutation relation~\eqref{eq:canonical-commutation-relation}, it is easy to see that $\mathrm{e}^{\mathrm{i}\alpha Q}$ is the ``translation operator'' for the ``momentum'': $[P,\mathrm{e}^{\mathrm{i}\alpha Q}] = \alpha\cdot\mathrm{e}^{\mathrm{i}\alpha Q}$; if $\vert \alpha_{0} \rangle$ is an arbitrary ``momentum'' eigenstate, $P \vert \alpha_{0} \rangle = \alpha_{0} \vert \alpha_{0} \rangle$, then we have $P~\mathrm{e}^{\mathrm{i}\alpha Q} \vert \alpha_{0} \rangle = (\alpha_{0} + \alpha) \mathrm{e}^{\mathrm{i}\alpha Q} \vert \alpha_{0} \rangle$. In other words, $\mathrm{e}^{\mathrm{i}\alpha Q} \vert \alpha_{0} \rangle = \vert \alpha_{0} + \alpha \rangle$. Thus, it is clear that the two-point correlator~\eqref{eq:momentum-eigenstate-expectation} vanishes unless we have $\alpha_{1} + \alpha_{2} = 0$ in~\eqref{eq:BCH-result}, which is just the ``charge-neutrality condition''~\cite{francesco1997}. Without loss of generality, in the following we choose $\alpha_{1} = -\alpha_{2} = \alpha$.
Then, substituting~\eqref{eq:BCH-result} into~\eqref{eq:momentum-eigenstate-expectation} and noticing that, with $R = \sqrt{p}$,
\begin{equation}
    P \vert n,m \rangle = \left( \frac{n}{R} + \frac{1}{2}mR \right) \vert n,m \rangle = \frac{1}{\sqrt{p}} \left( n + \frac{1}{2}mp \right) \vert n,m \rangle,
\end{equation}
we find
\begin{align}
\label{eq:momentum-eigenstate-expectation-continued}
    &~\underset{n_{1},n_{2},\ldots\geq0}{\sum} q^{\sum_{{k\geq1}}kn_{k}} \langle n,m;n_{1},n_{2},\ldots\vert :\mathrm{e}^{\mathrm{i}\alpha\phi(z_1)}: :\mathrm{e}^{-\mathrm{i}\alpha\phi(z_2)}: \vert n,m;n_{1},n_{2},\ldots\rangle \nonumber \\
    =&~\mathrm{e}^{2\pi\mathrm{i} \frac{1}{\sqrt{p}}(n + mp/2) \alpha (z_{1} - z_{2})} \mathrm{e}^{-\pi\mathrm{i} \alpha^{2} (z_{1} - z_{2})} \nonumber \\
    & \times \underset{n_{1},n_{2},\ldots\geq0}{\sum} q^{\sum_{{k\geq1}}kn_{k}} \langle n_{1},n_{2},\ldots\vert :\mathrm{e}^{\mathrm{i}\alpha\Hat{\phi}(z_1)}: :\mathrm{e}^{-\mathrm{i}\alpha\Hat{\phi}(z_2)}: \vert n_{1},n_{2},\ldots\rangle.
\end{align}

Firstly, let us focus on the expectation value in the occupation-number eigenstates. To explicitly evaluate its expression, the basic idea is to commute the modes of $\Hat{\phi}(z)$ with $k>0$ to the right and those with $k<0$ to the left using the BCH formula~\eqref{eq:BCH}. Below we carry out this calculation:
\begin{align}
\label{eq:expectation-value-occupation-number-eigenstates}
    & \phantom{=} \; \langle n_{1},n_{2},\ldots\vert :\mathrm{e}^{\mathrm{i}\alpha\Hat{\phi}(z_1)}: :\mathrm{e}^{-\mathrm{i}\alpha\Hat{\phi}(z_2)}: \vert n_{1},n_{2},\ldots\rangle \nonumber \\
    &= \exp\left(\alpha^{2}\sum_{k=1}^{\infty}\frac{1}{k}\mathrm{e}^{-2\pi\mathrm{i}k(z_{1}-z_{2})}\right) \nonumber \\
    &~~~~\times \prod_{k=1}^{\infty}\langle n_{k}\vert\exp\left(\frac{\alpha}{k}a_{-k}\left(\mathrm{e}^{2\pi\mathrm{i}kz_{1}}-\mathrm{e}^{2\pi\mathrm{i}kz_{2}}\right)\right)\exp\left(-\frac{\alpha}{k}a_{k}\left(\mathrm{e}^{-2\pi\mathrm{i}kz_{1}}-\mathrm{e}^{-2\pi\mathrm{i}kz_{2}}\right)\right)\vert n_{k}\rangle \nonumber \\
    &= \exp\left(\alpha^{2}\sum_{k=1}^{\infty}\frac{1}{k}\mathrm{e}^{-2\pi\mathrm{i}k(z_{1}-z_{2})}\right) \nonumber \\
    &~~~~\times \prod_{k=1}^{\infty}\sum_{n=0}^{n_{k}}\frac{1}{\left(n!\right)^{2}}\left(\frac{\alpha}{k}\left(\mathrm{e}^{2\pi\mathrm{i}kz_{1}}-\mathrm{e}^{2\pi\mathrm{i}kz_{2}}\right)\right)^{n}\left(-\frac{\alpha}{k}\left(\mathrm{e}^{-2\pi\mathrm{i}kz_{1}}-\mathrm{e}^{-2\pi\mathrm{i}kz_{2}}\right)\right)^{n}\langle n_{k}\vert a_{-k}^{n}a_{k}^{n}\vert n_{k}\rangle \nonumber \\
    &= \exp\left(\alpha^{2}\sum_{k=1}^{\infty}\frac{1}{k}\mathrm{e}^{-2\pi\mathrm{i}k(z_{1}-z_{2})}\right) \nonumber \\
    &~~~~\times \prod_{k=1}^{\infty}\sum_{n=0}^{n_{k}}\frac{1}{n!}\left(\frac{\alpha}{k}\left(\mathrm{e}^{2\pi\mathrm{i}kz_{1}}-\mathrm{e}^{2\pi\mathrm{i}kz_{2}}\right)\right)^{n}\left(-\frac{\alpha}{k}\left(\mathrm{e}^{-2\pi\mathrm{i}kz_{1}}-\mathrm{e}^{-2\pi\mathrm{i}kz_{2}}\right)\right)^{n}k^{n}\binom{n_{k}}{n}.
\end{align}
To get the second equality, we Taylor-expanded the exponentials inside the expectation value; the third equality uses
\begin{equation}
    a_{k}^{n}\vert n_{k}\rangle=\sqrt{k^{n}\frac{n_{k}!}{(n_{k}-n)!}}\vert n_{k}-n\rangle
\end{equation}
for $n\leq n_{k}$ (which follows from~\eqref{eq:states-normalized}) and
\begin{equation}
    a_{k}^{n}\vert n_{k}\rangle=0
\end{equation}
for $n>n_{k}$, and the definition of the number of combinations
\begin{equation}
    \binom{n_{k}}{n} \equiv \frac{n_{k}!}{n!(n_{k}-n)!}.
\end{equation}
Next, using the result obtained in~\eqref{eq:expectation-value-occupation-number-eigenstates}, we obtain
\begin{align}
    &~\mathrm{e}^{-\pi\mathrm{i} \alpha^{2} (z_{1} - z_{2})}~\underset{n_{1},n_{2},\ldots\geq0}{\sum} q^{\sum_{{k\geq1}}kn_{k}}~\langle n_{1},n_{2},\ldots\vert :\mathrm{e}^{\mathrm{i}\alpha\Hat{\phi}(z_1)}: :\mathrm{e}^{-\mathrm{i}\alpha\Hat{\phi}(z_2)}: \vert n_{1},n_{2},\ldots\rangle \nonumber \\
    =&~\mathrm{e}^{-\pi\mathrm{i} \alpha^{2} (z_{1} - z_{2})}~\exp\left(\alpha^{2}\sum_{k=1}^{\infty}\frac{1}{k}\mathrm{e}^{-2\pi\mathrm{i}k(z_{1}-z_{2})}\right) \nonumber \\
    & \times \prod_{k=1}^{\infty}\sum_{n_{k}=0}^{\infty}\sum_{n=0}^{n_{k}}\frac{1}{n!}\left[\frac{\alpha}{k}\left(\mathrm{e}^{2\pi\mathrm{i}kz_{1}}-\mathrm{e}^{2\pi\mathrm{i}kz_{2}}\right)\right]^{n}\left[-\frac{\alpha}{k}\left(\mathrm{e}^{-2\pi\mathrm{i}kz_{1}}-\mathrm{e}^{-2\pi\mathrm{i}kz_{2}}\right)\right]^{n}k^{n}\binom{n_{k}}{n}q^{kn_{k}} \nonumber \\
    =&~\mathrm{e}^{-\pi\mathrm{i} \alpha^{2} (z_{1} - z_{2})}~\exp\left(\alpha^{2}\sum_{k=1}^{\infty}\frac{1}{k}\mathrm{e}^{-2\pi\mathrm{i}k(z_{1}-z_{2})}\right) \nonumber \\
    & \times \prod_{k=1}^{\infty}\sum_{n=0}^{\infty}\frac{1}{n!}\left[\frac{\alpha}{k}\left(\mathrm{e}^{2\pi\mathrm{i}kz_{1}}-\mathrm{e}^{2\pi\mathrm{i}kz_{2}}\right)\right]^{n}\left[-\frac{\alpha}{k}\left(\mathrm{e}^{-2\pi\mathrm{i}kz_{1}}-\mathrm{e}^{-2\pi\mathrm{i}kz_{2}}\right)\right]^{n}k^{n}
    \sum_{n_{k}=n}^{\infty}\binom{n_{k}}{n}q^{kn_{k}} \nonumber \\
    =&~\mathrm{e}^{-\pi\mathrm{i} \alpha^{2} (z_{1} - z_{2})}~\exp\left(\alpha^{2}\sum_{k=1}^{\infty}\frac{1}{k}\mathrm{e}^{-2\pi\mathrm{i}k(z_{1}-z_{2})}\right) \nonumber \\
    & \times \prod_{k=1}^{\infty}\sum_{n=0}^{\infty}\frac{1}{n!}\left[\frac{\alpha}{k}\left(\mathrm{e}^{2\pi\mathrm{i}kz_{1}}-\mathrm{e}^{2\pi\mathrm{i}kz_{2}}\right)\right]^{n}\left[-\frac{\alpha}{k}\left(\mathrm{e}^{-2\pi\mathrm{i}kz_{1}}-\mathrm{e}^{-2\pi\mathrm{i}kz_{2}}\right)\right]^{n}k^{n}q^{kn}\left(1-q^{k}\right)^{-n-1} \nonumber \\
    =&~\frac{q^{\frac{1}{24}}}{\eta(\tau)}~ \mathrm{e}^{-\pi\mathrm{i} \alpha^{2} (z_{1} - z_{2})}~\exp\left(\alpha^{2}\sum_{k=1}^{\infty}\frac{1}{k}\mathrm{e}^{-2\pi\mathrm{i}k(z_{1}-z_{2})}\right) \nonumber \\
    & \times \exp\left[-\alpha^{2}\sum_{k=1}^{\infty}\frac{1}{k}\left(2-\mathrm{e}^{2\pi\mathrm{i}k(z_{1}-z_{2})}-\mathrm{e}^{-2\pi\mathrm{i}k(z_{1}-z_{2})}\right)\frac{q^{k}}{1-q^{k}}\right] \nonumber \\
    =&~\frac{q^{\frac{1}{24}}}{\eta(\tau)}~ \mathrm{e}^{-\pi\mathrm{i} \alpha^{2} (z_{1} - z_{2})}~\exp\left(\alpha^{2}\sum_{k=1}^{\infty}\frac{1}{k}\mathrm{e}^{-2\pi\mathrm{i}k(z_{1}-z_{2})}\right) \nonumber \\
    & \times \prod_{n=1}^{\infty}\exp\left[-\alpha^{2}\sum_{k=1}^{\infty}\frac{1}{k}\left(2-\mathrm{e}^{2\pi\mathrm{i}k(z_{1}-z_{2})}-\mathrm{e}^{-2\pi\mathrm{i}k(z_{1}-z_{2})}\right)q^{kn}\right] \nonumber \\
    =&~\frac{q^{\frac{1}{24}}}{\eta(\tau)} \left[ \frac{\mathrm{e}^{-\pi\mathrm{i} (z_{1} - z_{2})}}{1 - \mathrm{e}^{-2\pi\mathrm{i} (z_{1} - z_{2})}} \prod_{n=1}^{\infty} \frac{(1-q^{n})^{2}}{\left( 1 - \mathrm{e}^{2\pi\mathrm{i}(z_{1}-z_{2})}q^{n} \right) \left( 1 - \mathrm{e}^{-2\pi\mathrm{i}(z_{1}-z_{2})}q^{n} \right)} \right]^{\alpha^{2}}.
\label{eq:expansion-no-zero-mode}
\end{align}
To get the second equality we changed the ordering of the summation over $n$ and that over $n_{k}$; to get the third one we introduced
a new summation variable $n_{k}^{\prime}=n_{k}-n$ and rewrote the summation over $n_{k}$ as
\begin{equation}
    \sum_{n_{k}=n}^{\infty}\binom{n_{k}}{n}q^{kn_{k}}=q^{kn}\sum_{n_{k}^{\prime}=0}^{\infty}\binom{n_{k}^{\prime}+n}{n}q^{kn_{k}^{\prime}},
\end{equation}
then used the property of the number of combinations
\begin{equation}
    \binom{n_{k}^{\prime}+n}{n} = (-1)^{n_{k}^{\prime}}\binom{-n-1}{n_{k}^{\prime}},
\end{equation}
and the binomial theorem; to get the fourth one we re-exponentiated the expression and made use of the definition of Dedekind eta-function~\eqref{eq:Dedekind}; the fifth one uses the expansion
\begin{equation}
    \frac{q^{k}}{1-q^{k}}=~\sum_{n=1}^{\infty}q^{kn};
\end{equation}
and to get the final equality, we used the Taylor expansion of the function $\ln(1-x)$. The right hand side of~\eqref{eq:expansion-no-zero-mode} can be rewritten in a more compact form by introducing the function
\begin{equation}
    E(z_{1}-z_{2}\vert\tau)=\frac{\vartheta_{1}(z_{1}-z_{2}\vert\tau)}{\partial_{z}\vartheta_{1}(z\vert\tau)\vert_{z=0}}.
\end{equation}
Using the infinite-product expression of $\vartheta_{1}(z\vert\tau)$,~\eqref{eq:theta1-product}, one can readily obtain
\begin{equation}
    \partial_{z}\vartheta_{1}(z\vert\tau)\vert_{z=0}\equiv\underset{z\rightarrow0}{\lim}\partial_{z}\vartheta_{1}(z\vert\tau)=2\pi\mathrm{i}\left(-\mathrm{i}q^{\frac{1}{8}}\prod_{n=1}^{\infty}(1-q^{n})\prod_{n=1}^{\infty}(1-q^{n})^{2}\right).
\end{equation}
Thus, we have
\begin{equation}
    \left( E(z_{1}-z_{2}\vert\tau) \right)^{-1} = 2\pi\mathrm{i} \cdot \frac{\mathrm{e}^{-\pi\mathrm{i}(z_{1}-z_{2})}}{1-\mathrm{e}^{-2\pi\mathrm{i}(z_{1}-z_{2})}}\prod_{n=1}^{\infty}\frac{(1-q^{n})^{2}}{(1-\mathrm{e}^{2\pi\mathrm{i}(z_{1}-z_{2})}q^{n})(1-\mathrm{e}^{-2\pi\mathrm{i}(z_{1}-z_{2})}q^{n})},
\end{equation}
(from which it is clear that $E(z_{1}-z_{2}\vert\tau) \rightarrow z_{1}-z_{2}$ when $z_{1}-z_{2} \rightarrow 0$) and if we drop the unimportant factor of $2\pi\mathrm{i}$,~\eqref{eq:expansion-no-zero-mode} can be rewritten as
\begin{align}
    &\mathrm{e}^{-\pi\mathrm{i} \alpha^{2} (z_{1} - z_{2})}~\underset{n_{1},n_{2},\ldots\geq0}{\sum} q^{\sum_{{k\geq1}}kn_{k}}~\langle n_{1},n_{2},\ldots\vert :\mathrm{e}^{\mathrm{i}\alpha\Hat{\phi}(z_1)}: :\mathrm{e}^{-\mathrm{i}\alpha\Hat{\phi}(z_2)}: \vert n_{1},n_{2},\ldots\rangle \nonumber \\
    &=~\frac{q^{\frac{1}{24}}}{\eta(\tau)}~\left( E(z_{1}-z_{2}\vert\tau) \right)^{-\alpha^{2}}.
\end{align}
Finally, by substituting this into~\eqref{eq:momentum-eigenstate-expectation-continued}, we find
\begin{align}
    &~\underset{n_{1},n_{2},\ldots\geq0}{\sum} q^{\sum_{{k\geq1}}kn_{k}} \langle n,m;n_{1},n_{2},\ldots\vert :\mathrm{e}^{\mathrm{i}\alpha\phi(z_1)}: :\mathrm{e}^{-\mathrm{i}\alpha\phi(z_2)}: \vert n,m;n_{1},n_{2},\ldots\rangle \nonumber \\
    =&~\frac{q^{\frac{1}{24}}}{\eta(\tau)}~\mathrm{e}^{2\pi\mathrm{i} \frac{1}{\sqrt{p}}(n + mp/2) \alpha (z_{1} - z_{2})}~\left( E(z_{1}-z_{2}\vert\tau) \right)^{-\alpha^{2}}.
\end{align}

The derivation for the general $N$-point case can be carried out in the same manner. In particular, one has to repeatedly use the BCH formula to commute the $k<0$ modes of $\Hat{\phi}(z_{j})$ through the $k>0$ modes of $\Hat{\phi}(z_{i})$ with $i<j$ to the left. Therefore, eq.~\eqref{eq:correlator-momentum-eigenstate}
\begin{align}
    &~\underset{n_{1},n_{2},\ldots\geq0}{\sum} q^{\sum_{{k\geq1}}kn_{k}} \langle n,m;n_{1},n_{2},\ldots\vert V_{n_1}(z_1) \cdots V_{n_N}(z_N) \vert n,m;n_{1},n_{2},\ldots\rangle \nonumber \\
    =&~\delta_{\alpha}~\frac{q^{\frac{1}{24}}}{\eta(\tau)}~\mathrm{e}^{2\pi\mathrm{i} \frac{1}{\sqrt{p}}(n + mp/2) \sum_{i=1}^{N} \alpha_{i} z_{i}}~\underset{i<j}{\prod}\left( E(z_{1}-z_{2}\vert\tau) \right)^{\alpha_{i}\alpha_{j}}
\end{align}
follows by induction. Again, $\delta_{\alpha}$ ensures that the charge-neutrality condition $\sum_{i=1}^{N}\alpha_{i} = 0$ is satisfied.

\section{Resonating valence bond approach to $\mathrm{SO}(n)_1$ states on the torus}
\label{appdx:RVB}

In this appendix, we show that the model wave functions~\eqref{eq:MESs-son} constructed using chiral correlators of the $\mathrm{SO}(n)_1$ WZW models can also be represented as a series of RVB states. Here, the term ``valence bond'' (VB) refers to a singlet formed by the $\mathfrak{so}(n)$-spins on two lattice sites, and an RVB state is a coherent superposition of different VB configurations. More concretely, the RVB formulation rewrites the $\mathrm{SO}(n)_1$ states~\eqref{eq:MESs-son} as Gutzwiller projected Bardeen-Cooper-Schrieffer (BCS) states, where the BCS state describes $p+\mathrm{i}p$ pairing of some auxiliary fermions (``fermionic partons'') and the Gutzwiller projection imposes a single-occupancy constraint for the fermions at each site. The RVB approach has the advantage that the singlet property of these wave functions becomes manifest and sheds light on how to construct anyon eigenbasis in the parton approach.

The RVB approach starts with a fermionic parton representation of the $\mathfrak{so}(n)$ Lie algebra in terms of fermionic partons $c_{A}^{\dagger}$, where $A=1,\ldots,n$ is a ``color'' index. The $\mathfrak{so}(n)$ generators are represented as
\begin{equation}
    L_{AB}=\mathrm{i}\left(c_{A}^{\dagger}c_{B}-c_{B}^{\dagger}c_{A}\right),
\end{equation}
where $1 \leq A < B \leq n$. With the aid of anticommutation relations of the fermionic operators, i.e., $\{c_{A},c_{B}\} = \{c^\dag_{A},c^\dag_{B}\} =0$ and $\{c_{A},c^\dag_{B}\}=\delta_{AB}$, one can easily confirm that these generators satisfy the $\mathfrak{so}(n)$ algebra,
\begin{equation}
\label{eq:son-commutation-relation}
    [L_{AB},L_{CD}]=\mathrm{i}\left(\delta_{AD}L_{BC}+\delta_{BC}L_{AD}-\delta_{AC}L_{BD}-\delta_{BD}L_{AC}\right).
\end{equation}
Assume that the partons reside at each site of the $2$D lattice (with arbitrary lattice geometry; as in section~\ref{sec:son}, the total number of lattice sites is $2N$):~$c^{\dagger}_{i,A_{i}}$,~$i = 1,2,\ldots,2N$. The $\mathfrak{so}(n)$-spin basis~\eqref{eq:local-spin-basis} is written as
\begin{equation}
\label{eq:local-spin-basis-repeated}
    \vert A_{1},A_{2},\ldots,A_{2N}\rangle = c_{1,A_{1}}^{\dagger} c_{2,A_{2}}^{\dagger} \cdots c_{2N,A_{2N}}^{\dagger}\vert0\rangle,
\end{equation}
where $\vert0\rangle$ is the parton vacuum. This basis spans our physical Hilbert space. Obviously, each lattice site should be occupied by precisely one parton, and other states are unphysical. The unphysical states can be removed by applying a Gutzwiller projector $P_{\textrm{G}}$, which imposes the single-occupancy constraint $\sum_{A=1}^{n}c_{i,A}^{\dagger}c_{i,A}=1$ at each site.

For $n=2r+1$, the projected BCS wave functions of our interest are defined by
\begin{equation}
\label{eq:RVB-proposal-odd}
    \vert \Psi_{\nu} \rangle = P_{\textrm{G}} \exp\left(\underset{i<j}{\sum}g_{\nu}(z_{i}-z_{j}\vert\tau)\sum_{A=1}^{n}c_{i,A}^{\dagger}c_{j,A}^{\dagger}\right)\vert0\rangle,
\end{equation}
where $\sum_{A=1}^{n}c_{i,A}^{\dagger}c_{j,A}^{\dagger}$ is a VB operator creating an $\mathrm{SO}(n)$ singlet between sites $i$ and $j$, and $g_{\nu}(z_{i}-z_{j}\vert\tau)$ is the corresponding pairing function. In the construction on the plane~\cite{tu2013a}, the pairing function is chosen to be $1/(z_{i}-z_{j})$. With this choice, the wave function before Gutzwiller projection is a BCS wave function describing a $p+\mathrm{i}p$ superconductor in its weak-pairing topological phase~\cite{read2000}. This pairing function coincides with the two-point correlator of a chiral Majorana field in the Ising CFT. For our construction on the torus, it is therefore natural to choose the pairing functions to be the two-point correlators on the torus, which are $g_{\nu}(z - w \vert \tau)$ (see appendix~\ref{appdx:derivation-correlator}). The index $\nu = 2, 3, 4$ corresponds to the fermionic spin structures PA, AA, AP, respectively. As expected, the pairing functions corresponding to these three spin structures reproduce the three-fold topologically degenerate ground states on the torus. The result of these correlators is given in eqs.~\eqref{eq:two-point-correlator-result-AA} to~\eqref{eq:two-point-correlator-result-PA} as (generalized) Weierstrass functions. One could verify that the short-distance ($z - w \rightarrow 0$) behavior of these functions recovers the two-point correlator on the plane, $1/(z - w)$.
To establish the relation with the chiral correlator result~\eqref{eq:MESs-odd-n}, we expand the exponential in~\eqref{eq:RVB-proposal-odd}:
\begin{align}
    \vert \Psi_{\nu} \rangle &= P_{\textrm{G}} \prod_{A=1}^{n}\prod_{i<j}\left( 1+g_{\nu}(z_{i}-z_{j}\vert\tau)c_{i,A}^{\dagger}c_{j,A}^{\dagger} \right) \vert0\rangle \nonumber \\
    &= P_{\mathrm{G}}\prod_{A=1}^{n}\left[ \sum_{N_{A}=0}^{N}~\sum_{i_{1}^{(A)}<i_{2}^{(A)}\cdots <i_{2N_{A}}^{(A)}}\mathrm{Pf}_{A}(g_{\nu}(z_{i}-z_{j}\vert\tau))c_{i_{1}^{(A)},A}^{\dagger}c_{i_{2}^{(A)},A}^{\dagger} \cdots c_{i_{2N_{A}}^{(A)},A}^{\dagger}\right] \vert0\rangle,
\end{align}
where $\mathrm{Pf}_{A}(g_{\nu}(z_{i}-z_{j}\vert\tau))$ is the Pfaffian of a $(2N_{A}) \times (2N_{A})$ antisymmetric matrix, whose diagonal entries are zero and off-diagonal entries are $g_{\nu}(z_{i}-z_{j} \vert \tau)$, in which the complex coordinates are restricted to the positions of the spin state $\vert A \rangle$. Next, we implement the Gutzwiller projection. Since $P_{\textrm{G}}$ keeps only those configurations with exactly one fermion at each site, after the Gutzwiller projection the positions of the fermions $i_{1}^{(1)},\ldots,i_{2N_{1}}^{(1)},i_{1}^{(2)},\ldots,i_{2N_{2}}^{(2)},\ldots,i_{1}^{(n)},\ldots,i_{2N_{n}}^{(n)}$ must be all different from each other, i.e., they correspond to a permutation of the site index $\{1,2,\ldots,2N\}$. Finally, we rearrange these fermion operators so that the resulting basis state can be identified as a ``spin'' state~\eqref{eq:local-spin-basis-repeated}. This permutation yields a sign factor $\xi = \mathrm{sgn}(i_{1}^{(1)},\ldots,i_{2N_{1}}^{(1)},i_{1}^{(2)},\ldots,i_{2N_{2}}^{(2)},\ldots,i_{1}^{(n)},\ldots,i_{2N_{n}}^{(n)})$. We therefore obtain the result
\begin{equation}
    \vert \Psi_{\nu} \rangle = \sum_{A_{1},A_{2},\ldots,A_{2N} = 1}^{n}\xi \left[ \prod_{A=1}^{n} \mathrm{Pf}_{A}(g_{\nu}(z_{i}-z_{j}\vert\tau)) \right]~\vert A_{1},A_{2},\ldots,A_{2N}\rangle.
\end{equation}
In comparing with~\eqref{eq:MESs-son} to~\eqref{eq:MESs-odd-n}, we see that up to overall factors $(Z_{\nu}(\tau))^{n}$, the linear combinations of the RVB states we constructed above are identical to the anyon eigenbasis constructed from the $\mathrm{SO}(n)_{1}$ WZW model with odd $n$:
\begin{align}
    & \vert \psi_{\boldsymbol{1}} \rangle = \frac{1}{2} \left( Z_{3}^{n} \vert \Psi_{3} \rangle + Z_{4}^{n} \vert \Psi_{4} \rangle \right), \\
    & \vert \psi_{\boldsymbol{v}} \rangle = \frac{1}{2} \left( Z_{3}^{n} \vert \Psi_{3} \rangle - Z_{4}^{n} \vert \Psi_{4} \rangle \right), \\
    & \vert \psi_{\boldsymbol{s}} \rangle = \frac{1}{\sqrt{2}} Z_{2}^{n} \vert \Psi_{2} \rangle.
\end{align}
If the model wave functions constructed in section~\ref{sec:son} were defined with~\emph{normalized} chiral correlators, these factors would have disappeared. However, as we shall see in appendix~\ref{appdx:derivation-modular}, keeping these factors is essential for understanding the modular-transformation properties of these wave functions.

For $n = 2r$, we expect four distinct wave functions on the torus. Obviously, the three RVB states with $\nu = 2, 3, 4$ for the odd $n$ case directly carry over to the even $n$ case. The subtlety is the construction of the remaining one: since the BCS pairing functions of the RVB states labelled by $\nu = 2, 3, 4$ correspond to the two-point correlators with spin structures PA, AA, AP, respectively, it is natural to guess that the ``two-point correlator''~\eqref{eq:two-point-correlator-result-PP} in the PP sector (with the zero-mode removed; see appendix~\ref{appdx:derivation-correlator}) should be used as the pairing function for the remaining RVB state. However, the ansatz for this state, labelled by $\nu = 1$, needs a small adjustment:
\begin{equation}
    \vert \Psi_{1} \rangle = P_{\textrm{G}} \prod_{A=1}^{n}\left[ \left(\sum_{j_{0}=1}^{2N} c_{j_{0},A}^{\dagger} \right)\exp\left(\underset{i<j}{\sum}g_{1}^{\prime}(z_{i}-z_{j}\vert\tau)c_{i,A}^{\dagger}c_{j,A}^{\dagger}\right)\right]\vert0\rangle,
\end{equation}
namely, apart from the fermionic partons forming Cooper pairs, there is one \emph{unpaired} fermionic mode for each parton species. The operator $c_{j_{0},A}^{\dagger}$ creates such an unpaired fermion with color $A$
at site $j_{0}$. After expanding the exponential, we obtain
\begin{equation}
    \vert \Psi_{1} \rangle = P_{\textrm{G}} \prod_{A=1}^{n}\left[ \sum_{j^{(A)}_{0}=1}^{2N} c_{j^{(A)}_{0},A}^{\dagger}
    \sum_{N_{A}=0}^{N-1}~\sum_{i_{1}^{(A)}<\cdots <i_{2N_{A}}^{(A)}}\mathrm{Pf}_{A}(g_{1}^{\prime}(z_{i}-z_{j}\vert\tau))c_{i_{1}^{(A)},A}^{\dagger} \cdots c_{i_{2N_{A}}^{(A)},A}^{\dagger}
    \right]\vert0\rangle.
\end{equation}
The position $j^{(A)}_{0}$ of the unpaired fermion must be different from any element in the set $\{ i_{1}^{(A)}, i_{2}^{(A)}, \ldots, i_{2N_{A}}^{(A)} \}$ due to the fermionic anticommutation relation. Now we rename the indices of these ($2N_{A}+1$) sites as $i_{1}^{(A)}<i_{2}^{(A)}<\cdots<i_{2N_{A}+1}^{(A)}$, one of which is the site index for the unpaired fermion, $j_{0}^{(A)}=i_{M_{A}}^{(A)}$, where $M_A = 1,2,\ldots, 2N_A+1$. With the Gutzwiller projection, the positions of the fermions $i_{1}^{(1)},\ldots,i_{2N_{1}+1}^{(1)},\ldots,i_{1}^{(n)},\ldots,i_{2N_{n}+1}^{(n)}$ are all different and correspond to a permutation of $1,2,\ldots,2N$. After implementing the Gutzwiller projection and rearranging the fermionic operators, we arrive at
\begin{equation}
    \vert \Psi_{1} \rangle = \sum_{A_{1},A_{2},\ldots,A_{2N} = 1}^{n} \xi^{\prime} \prod_{A=1}^{n}\left[\sum_{M_{A}=1}^{2N_{A}+1} (-1)^{M_{A}-1}\mathrm{Pf}_{A}\left(g_{1}^{\prime}(z_{i}-z_{j}\vert\tau)\right)\right] \vert A_{1},A_{2},\ldots,A_{2N} \rangle,
\end{equation}
where the permutation sign $\xi^{\prime} = \mathrm{sgn}(i_{1}^{(1)},\ldots,i_{2N_{1}+1}^{(1)},\ldots,i_{1}^{(n)},\ldots,i_{2N_{n}+1}^{(n)})$ comes from the rearrangement of the fermionic operators, and the Pfaffian factor $\mathrm{Pf}_{A}\left(g^{\prime}_{\textrm{PP}}(z_{i}-z_{j}\vert\tau)\right)$ is restricted to $i,j=i_{1}^{(A)},\cdots,\hat{i}_{M_{A}}^{(A)},\cdots,i_{2N_{A}+1}^{(A)}$. Note that the hat `` $\hat{}$ '' on an object indicates that the object is to be omitted. Thus, for even $n$, the linear combinations of the RVB states we constructed above are again identical to the anyon eigenbasis that we have constructed from the $\mathrm{SO}(n)_{1}$ WZW model in section~\ref{sec:son}:
\begin{align}
    & \vert \psi_{\boldsymbol{1}} \rangle = \frac{1}{2} \left( Z_{3}^{n} \vert \Psi_{3} \rangle + Z_{4}^{n} \vert \Psi_{4} \rangle \right), \\
    & \vert \psi_{\boldsymbol{v}} \rangle = \frac{1}{2} \left( Z_{3}^{n} \vert \Psi_{3} \rangle - Z_{4}^{n} \vert \Psi_{4} \rangle \right), \\
    & \vert \psi_{\boldsymbol{s}_{+}} \rangle = \frac{1}{2} \left( Z_{2}^{n} \vert \Psi_{2} \rangle + (Z_{1}^{\prime})^{n} \vert \Psi_{1} \rangle \right), \\
    & \vert \psi_{\boldsymbol{s}_{-}} \rangle = \frac{1}{2} \left( Z_{2}^{n} \vert \Psi_{2} \rangle - (Z_{1}^{\prime})^{n} \vert \Psi_{1} \rangle \right).
\end{align}

To summarize, we have proven that the $\mathrm{SO}(n)_{1}$ states on the torus can be represented as RVB states. An advantage of the RVB form is that the translation invariance of the wave functions on a regular lattice can be easily proven. The proof of translation invariance was given for the wave functions with $n=3$ on the square lattice~\cite{zhang2021}. The generalization to cases with $n\geq 4$ is straightforward and will not be repeated here.

\section{Derivation of the modular matrices}
\label{appdx:derivation-modular}

In this appendix, we provide the derivation of the modular $S$ and $T$ matrices for the wave functions constructed from the $\mathrm{SO}(n)_{1}$ WZW model (see~\eqref{eq:MESs-son}) and the $\mathrm{SU}(n)_{1}$ WZW model (see~\eqref{eq:sun-MES}). For $\mathrm{SO}(n)_{1}$ states, we shall focus on the cases with even $n$, since the $n=3$ results in ref.~\cite{zhang2021} directly generalize to all cases with odd $n$. In the derivation, we consider an $L \times L$ square lattice embedded on a square with a unit side length, so the modular parameter is $\tau=\mathrm{i}$. Each site $j$ is assigned a complex coordinate $z_{j} = \tfrac{1}{L}(x_j + \mathrm{i} y_j)$ with $x_{j}, y_{j} \in \{0,1,\ldots,L-1\}$. For $\mathrm{SO}(n)_{1}$ and $\mathrm{SU}(n)_{1}$ states, we choose even $L$ and $L \in 2n\mathbb{Z}$, respectively.

\begin{figure}
\centering
\includegraphics[width=0.80\textwidth]{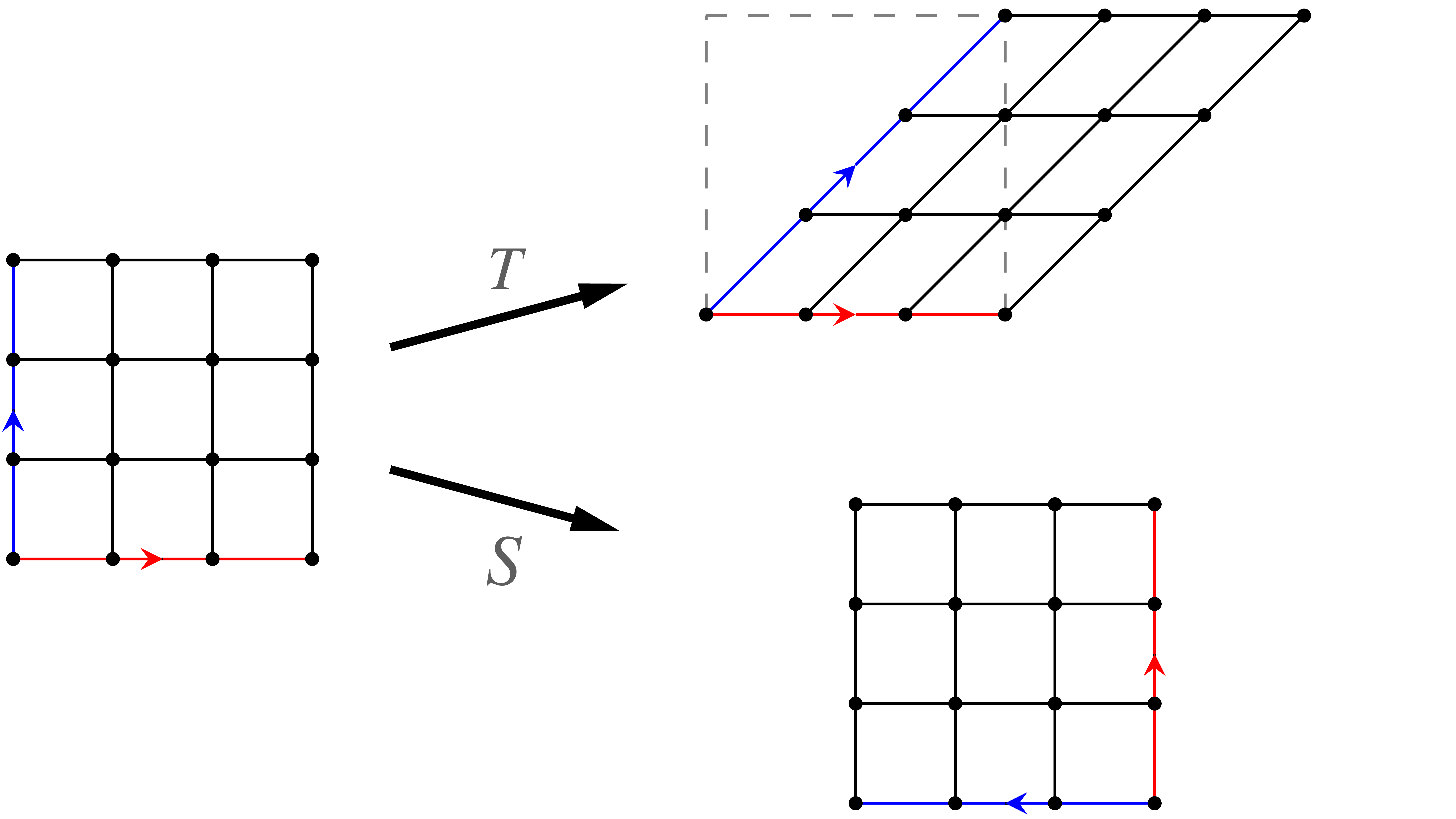}
\caption{Illustration of the modular transformations on a $4 \times 4$ square lattice, which is periodic along both directions. The modular $T$ transformation corresponds to a Dehn twist, while the modular $S$ transformation corresponds to a counterclockwise rotation by 90 degrees. The arrows indicate the orientation of the lattice.}
\label{Figure5}
\end{figure}

The modular $T$ transformation is the Dehn twist (see figure~\ref{Figure5}), which changes $\tau$ to $\tau + 1$ and the action on the coordinates is $(x_{j},y_{j})\rightarrow(x_{j}+y_{j},y_{j})$. Due to the periodicity of the lattice, the addition is modulo $L$ and the lattice maps to itself after Dehn twist. The modular $S$ transformation is a counterclockwise rotation of the lattice by 90 degrees and its action on the coordinates is $(x_{j},y_{j})\rightarrow(y_{j},L-1-x_{j})$ or $z_{j}\rightarrow-\mathrm{i}z_{j}$. This is realized by changing the modular parameter $\tau$ to $-1/\tau$. Within the “holonomy equals monodromy” framework~\cite{read2009}, the calculation of the modular matrices reduces to the study of the properties of the wave functions under analytical continuations $\mathrm{i} \rightarrow \mathrm{i} + 1$ (for the modular $T$ transformation) and $z_{j}\rightarrow-\mathrm{i}z_{j}$ (for the modular $S$ transformation).

\subsection{Modular-transformation properties of the special functions}

In this subsection, we list the modular-transformation properties of Jacobi's theta functions, Dedekind's eta function and (generalized) Weierstrass functions, which are used in the derivation of the modular matrices of the wave functions constructed from the $\mathrm{SO}(n)_{1}$ WZW model.

Under the modular $T$ transformation ($\tau\rightarrow\tau+1$), Jacobi's theta functions (\eqref{eq:theta-sum-begin} to~\eqref{eq:theta-sum-end}) and Dedekind's eta function~\eqref{eq:Dedekind} transform as
\begin{align}
\label{eq:theta-T-begin}
\vartheta_{1}(z\vert\tau+1)&=\mathrm{e}^{\frac{\mathrm{i}\pi}{4}}\vartheta_{1}(z\vert\tau), \\
\vartheta_{2}(z\vert\tau+1)&=\mathrm{e}^{\frac{\mathrm{i}\pi}{4}}\vartheta_{2}(z\vert\tau), \\
\vartheta_{3}(z\vert\tau+1)&=\vartheta_{4}(z\vert\tau), \\
\vartheta_{4}(z\vert\tau+1)&=\vartheta_{3}(z\vert\tau), \\
\eta(\tau+1)&=\mathrm{e}^{\frac{\mathrm{i}\pi}{12}}\eta(\tau).
\label{eq:eta_T}
\end{align}
Under the modular $S$ transformation ($\tau\rightarrow-1/\tau, z\rightarrow z/\tau$), they transform as
\begin{align}
\label{eq:theta-S-begin}
\vartheta_{1}(\tfrac{z}{\tau}\vert\tfrac{-1}{\tau}) &= -\mathrm{i}\mathrm{e}^{\mathrm{i}\pi\frac{z^{2}}{\tau}}(-\mathrm{i}\tau)^{\frac{1}{2}}\vartheta_{1}(z\vert\tau), \\
\vartheta_{2}(\tfrac{z}{\tau}\vert\tfrac{-1}{\tau}) &= \mathrm{e}^{\mathrm{i}\pi\frac{z^{2}}{\tau}}(-\mathrm{i}\tau)^{\frac{1}{2}}\vartheta_{4}(z\vert\tau), \\
\vartheta_{3}(\tfrac{z}{\tau}\vert\tfrac{-1}{\tau}) &= \mathrm{e}^{\mathrm{i}\pi\frac{z^{2}}{\tau}}(-\mathrm{i}\tau)^{\frac{1}{2}}\vartheta_{3}(z\vert\tau), \\
\vartheta_{4}(\tfrac{z}{\tau}\vert\tfrac{-1}{\tau}) &= \mathrm{e}^{\mathrm{i}\pi\frac{z^{2}}{\tau}}(-\mathrm{i}\tau)^{\frac{1}{2}}\vartheta_{2}(z\vert\tau), \\
\eta(-1/\tau) &= (-\mathrm{i}\tau)^{\frac{1}{2}}\eta(\tau).
\label{eq:eta_S}
\end{align}
By using the above results, the (generalized) Weierstrass functions $\wp_{\nu}(z_{i}-z_{j}\vert\tau)$ defined in (\ref{eq:Weierstrass}) can be shown to satisfy
\begin{align}
\wp_{2}(z_{i}-z_{j}\vert\mathrm{i}+1)&=\wp_{2}(z_{i}-z_{j}\vert\mathrm{i}), \\
\wp_{3}(z_{i}-z_{j}\vert\mathrm{i}+1)&=\wp_{4}(z_{i}-z_{j}\vert\mathrm{i}), \\
\wp_{4}(z_{i}-z_{j}\vert\mathrm{i}+1)&=\wp_{3}(z_{i}-z_{j}\vert\mathrm{i}),
\end{align}
and
\begin{align}
\wp_{2}(-\mathrm{i}(z_{i}-z_{j})\vert\mathrm{i})&=\mathrm{i}\wp_{4}(z_{i}-z_{j}\vert\mathrm{i}), \\
\wp_{3}(-\mathrm{i}(z_{i}-z_{j})\vert\mathrm{i})&=\mathrm{i}\wp_{3}(z_{i}-z_{j}\vert\mathrm{i}), \\
\wp_{4}(-\mathrm{i}(z_{i}-z_{j})\vert\mathrm{i})&=\mathrm{i}\wp_{2}(z_{i}-z_{j}\vert\mathrm{i}).
\end{align}

According to~\eqref{eq:wave-function-PP-sector}, the $\mathrm{SO}(n)_{1}$ wave function with $n=2r$ in the PP sector is written as
\begin{equation}
    \psi_{1}(\{A_{j}\};\{z_{j}\};\tau) = \xi^{\prime} \left( Z^{\prime}_{\textrm{PP}} \right)^{2r} \prod_{A=1}^{2r} \left[ \underset{j_{0}^{(A)}=i_{1}^{(A)},\ldots,i_{2N_{A}+1}^{(A)}}{\sum}(-1)^{M(j_{0}^{(A)})-1} \mathrm{Pf}_{A}\left(g^{\prime}_{\textrm{PP}}(z_{i}-z_{j}\vert\tau)\right) \right]
\end{equation}
with $Z^{\prime}_{\textrm{PP}}(\tau) = \eta(\tau)$ and
\begin{equation}
    g^{\prime}_{\textrm{PP}}(z_{i} - z_{j} \vert \tau) = \frac{\partial_{z_{i}}\vartheta_{1}(z_{i} - z_{j}\vert\tau)}{\vartheta_{1}(z_{i} - z_{j}\vert\tau)}.
\end{equation}
By using the modular-transformation properties of $\vartheta_{1}(z\vert\tau)$, one finds
\begin{equation}
    g^{\prime}_{\textrm{PP}}(z_{i} - z_{j} \vert \mathrm{i}+1) = g^{\prime}_{\textrm{PP}}(z_{i} - z_{j} \vert \mathrm{i}),
\end{equation}
and
\begin{equation}
    g^{\prime}_{\textrm{PP}}(-\mathrm{i}(z_{i} - z_{j}) \vert \mathrm{i}) = \mathrm{i} g^{\prime}_{\textrm{PP}}(z_{i} - z_{j} \vert \mathrm{i}) + 2\pi\mathrm{i}(z_{i} - z_{j}).
\end{equation}
Actually, the extra term $2\pi\mathrm{i}(z_{i} - z_{j})$ has no contributions to the Pfaffian (see ref.~\cite{chung2007} for the proof). The modular $S$ transformation is a relabelling of the lattice sites which consists of $N(2N-2L+1)/2$ exchanges of two labels. Since $L$ is even and $L^{2} = 2N$, the resulting extra sign factor $(-1)^{N(2N-2L+1)/2}$ exactly cancels the factor $\mathrm{i}^{N}$ coming from the term $\mathrm{i} g^{\prime}_{\textrm{PP}}(z_{i} - z_{j} \vert \mathrm{i})$. Thus, $\psi_{1}$ transforms under modular $S$ and $T$ transformations as follows:
\begin{align}
    \psi_{1}(\{A_{j}\};\{-\mathrm{i}z_{j}\};\mathrm{i}) &= \psi_{1}(\{A_{j}\};\{z_{j}\};\mathrm{i}), \\
    \psi_{1}(\{A_{j}\};\{z_{j}\};\mathrm{i}+1) &= \mathrm{e}^{\frac{r\pi\mathrm{i}}{6}}\psi_{1}(\{A_{j}\};\{z_{j}\};\mathrm{i}).
\end{align}

\subsection{Modular transformations of the anyon eigenbasis constructed from the $\mathrm{SU}(n)_{1}$ WZW model}

Before studying the modular transformation properties of the $\mathrm{SU}(n)_{1}$ wave functions, it is necessary to review a few results in Lie algebra theory.

The full information of a finite-dimensional simple Lie algebra of rank $r$ is encoded in its Cartan matrix, which is an $r \times r$ matrix defined in terms of the simple roots:
\begin{equation}
\label{eq:A-definition}
    \mathfrak{A}_{ij} = \frac{2 \Vec{\alpha}_{i} \cdot \Vec{\alpha}_{j}}{(\Vec{\alpha}_{j})^2}.
\end{equation}
Obviously, all diagonal elements of the Cartan matrix are equal to 2. The algebra $\mathfrak{su}(n)$ belongs to the set of the so-called simply-laced algebras, i.e., the algebras of which all roots have the same length (which is normalized to be $\sqrt{2}$). For a simply-laced algebra, the definition of its Cartan matrix simplifies to $\mathfrak{A}_{ij} = \Vec{\alpha}_{i} \cdot \Vec{\alpha}_{j}$. The Cartan matrix of $\mathfrak{su}(n)$ (which is of rank $n-1$) reads~\cite{francesco1997}
\begin{equation}
\label{eq:A-explicit}
    \mathfrak{A} = \left( \begin{array}{cccccc}
        2 & -1 & 0 & \cdots & 0 & 0 \\
        -1 & 2 & -1 & \cdots & 0 & 0 \\
        0 & -1 & 2 & \cdots & 0 & 0 \\
        \vdots & \vdots & \vdots & \ddots & \vdots & \vdots \\
        0 & 0 & 0 & \cdots & 2 & -1 \\
        0 & 0 & 0 & \cdots & -1 & 2
    \end{array} \right).
\end{equation}
By definition, the fundamental weights satisfy
\begin{equation}
    \frac{2 \Vec{\omega}_{i} \cdot \Vec{\alpha}_{j}}{(\Vec{\alpha}_{j})^2} = \delta_{ij}.
\end{equation}
The scalar products between the fundamental weights define a symmetric quadratic form matrix
\begin{equation}
\label{eq:F-definition}
    \mathfrak{F}_{ij} = \Vec{\omega}_{i} \cdot \Vec{\omega}_{j}.
\end{equation}
For simply-laced algebras, it is straightforward to see that the quadratic form matrix is nothing but the inverse of the Cartan matrix. In particular, for $\mathfrak{su}(n)$, it reads
\begin{equation}
\label{eq:F-explicit}
    \mathfrak{F} = \frac{1}{n}~\left( \begin{array}{cccccc}
        n-1 & ~n-2 & ~n-3 & ~\cdots & ~2 & ~1 \\
        n-2 & ~2(n-2) & ~2(n-3) & ~\cdots & ~4 & ~2 \\
        n-3 & ~2(n-3) & ~3(n-3) & ~\cdots & ~6 & ~3 \\
        \vdots & ~\vdots & ~\vdots & ~\ddots & ~\vdots & ~\vdots \\
        2 & ~4 & ~6 & ~\cdots & ~2(n-2) & ~n-2 \\
        1 & ~2 & ~3 & ~\cdots & ~n-2 & ~n-1
    \end{array} \right).
\end{equation}
Using~\eqref{eq:A-definition} and~\eqref{eq:F-definition}, one can rewrite~\eqref{eq:sun-correlator} as
\begin{align}
    \langle V_{s_{1}}(z_{1}) \cdots V_{s_{N}}(z_{N}) \rangle_{0} =&~ \kappa_{s_{1}} \cdots \kappa_{s_{N}} \delta_{\sum_{j}\Vec{\lambda}_{s_{j}} = \Vec{0}}~\frac{1}{\left( \eta(\tau) \right)^{n-1}} \underset{j<j^{\prime}}{\prod} \left( E(z_{j}-z_{j^{\prime}} \vert \tau) \right)^{\Vec{\lambda}_{s_{j}} \cdot \Vec{\lambda}_{s_{j^{\prime}}}} \nonumber \\
    & \times \underset{m_{1},\ldots,m_{n-1}}{\sum} \mathrm{e}^{\pi\mathrm{i}\tau\sum_{i,i^{\prime}}m_{i}\mathfrak{A}_{ii^{\prime}}m_{i^{\prime}} + 2\pi\mathrm{i} \sum_{j} \left( \sum_{i^{\prime}} m_{i^{\prime}} (\Vec{\lambda}_{s_{j}} \cdot \Vec{\alpha}_{i^{\prime}}) \right) z_{j}},
\end{align}
and
\begin{align}
    \langle V_{s_{1}}(z_{1}) \cdots V_{s_{N}}(z_{N}) \rangle_{l} =&~ \kappa_{s_{1}} \cdots \kappa_{s_{N}} \delta_{\sum_{j}\Vec{\lambda}_{s_{j}} = \Vec{0}}~\frac{1}{\left( \eta(\tau) \right)^{n-1}} \underset{j<j^{\prime}}{\prod} \left( E(z_{j}-z_{j^{\prime}} \vert \tau) \right)^{\Vec{\lambda}_{s_{j}} \cdot \Vec{\lambda}_{s_{j^{\prime}}}} \nonumber \\
    & \times \underset{m_{1},\ldots,m_{n-1}}{\sum} \mathrm{e}^{\pi\mathrm{i}\tau(\mathfrak{F}_{ll}+2m_{l}+\sum_{i,i^{\prime}}m_{i}\mathfrak{A}_{ii^{\prime}}m_{i^{\prime}}) + 2\pi\mathrm{i} \sum_{j} \Vec{\lambda}_{s_{j}} \cdot \left( \Vec{\omega}_{l} + \sum_{i^{\prime}} m_{i^{\prime}}\Vec{\alpha}_{i^{\prime}} \right) z_{j}}
\end{align}
for $l = 1, \ldots, n-1$.

Let us consider first the modular $T$ transformation: $\tau \rightarrow \tau + 1$.
Using~\eqref{eq:A-explicit}, one finds that
\begin{equation}
    \sum_{i,i^{\prime}=1}^{n-1}m_{i}\mathfrak{A}_{ii^{\prime}}m_{i^{\prime}} = 2 \left( \sum_{i=1}^{n-1}m_{i}^{2} - \sum_{i=1}^{n-2}m_{i}m_{i+1}\right)
\end{equation}
is an even number. Therefore, under the modular $T$ transformation,
\begin{align}
    &\vert \psi_{0} \rangle \rightarrow \mathrm{e}^{-\frac{\pi\mathrm{i}}{12}(n-1)} \vert \psi_{0} \rangle, \\
    &\vert \psi_{l} \rangle \rightarrow \mathrm{e}^{-\pi\mathrm{i} \left( \frac{n-1}{12} - \mathfrak{F}_{ll} \right)} \vert \psi_{l} \rangle,  \qquad (l = 1, \ldots, n-1) \; .
\end{align}
According to~\eqref{eq:F-explicit}, one has $\mathfrak{F}_{ll} = l(n-l)/n$, hence the modular $T$ matrix can be expressed as
\begin{equation}
    \mathcal{T}_{ll^{\prime}} = \mathrm{e}^{-2\pi\mathrm{i} \left( \frac{c}{24} - \frac{l(n-l)}{2n} \right)} \delta_{ll^{\prime}}
\end{equation}
for $l, l^{\prime} = 0, 1, \ldots, n-1$, where the chiral central charge is determined to be $c=n-1$ (modulo $8$) and the term proportional to the central charge $c$ comes from the transformation of Dedekind's eta functions. Thus, the wave functions~\eqref{eq:sun-MES} are indeed the anyon eigenbasis, as the modular $T$ matrix is diagonal.

The derivation of transforming properties of the wave functions under the modular $S$ transformation, $\tau\rightarrow-1/\tau, z\rightarrow z/\tau$, is more involved. We observe that the wave function consists of three parts: the exponent, the Dedekind's eta functions and the ``Jastrow'' factor. Firstly, let us focus on the function on the exponent. For $l = 1, \ldots, n-1$, it is
\begin{equation}
\label{eq:function-exponents}
    \mathrm{e}^{-\frac{\pi\mathrm{i}}{\tau} F_{ll} + \frac{2\pi\mathrm{i}}{\tau} \sum_{j}(\Vec{\lambda}_{s_{j}} \cdot \Vec{\omega}_{l})z_{j}}
    \underset{m_{1},\ldots,m_{n-1}}{\sum} \mathrm{e}^{-\frac{2\pi\mathrm{i}}{\tau}\left(m_{l}+\frac{1}{2}\sum_{i,i^{\prime}}m_{i}\mathfrak{A}_{ii^{\prime}}m_{i^{\prime}} - \sum_{j} \Vec{\lambda}_{s_{j}} \cdot \left( \sum_{i^{\prime}} m_{i^{\prime}}\Vec{\alpha}_{i^{\prime}} \right) z_{j}\right)}.
\end{equation}
To proceed, we invoke the multi-variable Poisson resummation formula
\begin{equation}
    \underset{m_{1},\ldots,m_{n-1}}{\sum} \mathrm{e}^{-\pi \boldsymbol{m}^{\mathrm{T}}\mathbf{A}\boldsymbol{m} + \boldsymbol{b}^{\mathrm{T}}\boldsymbol{m}} = (\mathrm{det}\mathbf{A})^{-\frac{1}{2}} \underset{k_{1},\ldots,k_{n-1}}{\sum} \mathrm{e}^{-\pi \left( \boldsymbol{k} + \frac{\boldsymbol{b}}{2\pi\mathrm{i}} \right)^{\mathrm{T}}\mathbf{A}^{-1}\left( \boldsymbol{k} + \frac{\boldsymbol{b}}{2\pi\mathrm{i}} \right)},
\end{equation}
where $\mathbf{A}$ is an $(n-1)\times(n-1)$ symmetric invertible matrix, $\boldsymbol{b} \equiv (b_{1},\ldots,b_{n-1})^{\mathrm{T}}$, $\boldsymbol{m} \equiv (m_{1},\ldots,m_{n-1})^{\mathrm{T}} \in \mathbb{Z}^{n-1}$ and $\boldsymbol{k} \equiv (k_{1},\ldots,k_{n-1})^{\mathrm{T}} \in \mathbb{Z}^{n-1}$. Choosing
\begin{equation}
    \mathbf{A}_{ii^{\prime}} = \frac{\mathrm{i}}{\tau}\mathfrak{A}_{ii^{\prime}}~~~~\textrm{and}~~~~(\boldsymbol{b}_{i})_{i^{\prime}} = -\frac{2\pi\mathrm{i}}{\tau} \left( \delta_{ii^{\prime}} - \underset{j=1}{\overset{N}{\sum}}(\Vec{\lambda}_{s_{j}} \cdot \Vec{\alpha}_{i^{\prime}})z_{j} \right),
\end{equation}
introducing a new set of summation variables $k^{\prime}_{1},\ldots,k^{\prime}_{n-1} \in \mathbb{Z}$ by
\begin{equation}
    k_{i} = \underset{i^{\prime}=1}{\overset{n-1}{\sum}}~\mathfrak{A}_{ii^{\prime}}k^{\prime}_{i^{\prime}} + \delta_{il^{\prime}},
\end{equation}
substituting into~\eqref{eq:function-exponents} and summing over $l^{\prime} = 0, 1, \ldots, n-1$, one finds
\begin{align}
    \eqref{eq:function-exponents} =~& (-\mathrm{i}\tau)^{\frac{n-1}{2}}\frac{1}{\sqrt{n}}~\mathrm{e}^{\frac{\pi\mathrm{i}}{\tau}\sum_{jj^{\prime}}(\Vec{\lambda}_{s_{j}}\cdot\Vec{\lambda}_{s_{j^{\prime}}})z_{j}z_{j^{\prime}}} \Big[ \underset{k^{\prime}_{1},\ldots,k^{\prime}_{n-1}}{\sum} \mathrm{e}^{\pi\mathrm{i}\tau\sum_{i,i^{\prime}}k^{\prime}_{i}\mathfrak{A}_{ii^{\prime}}k^{\prime}_{i^{\prime}} + 2\pi\mathrm{i} \sum_{j} \left( \sum_{i} k^{\prime}_{i} ( \Vec{\lambda}_{s_{j}} \cdot  \Vec{\alpha}_{i} ) \right) z_{j}} \nonumber \\
    &+ \underset{l^{\prime}=1}{\overset{n-1}{\sum}} \Big( \mathrm{e}^{-2\pi\mathrm{i}\mathfrak{F}_{ll^{\prime}}} \underset{k^{\prime}_{1},\ldots,k^{\prime}_{n-1}}{\sum} \mathrm{e}^{\pi\mathrm{i}\tau(\mathfrak{F}_{l^{\prime}l^{\prime}}+2k^{\prime}_{l^{\prime}}+\sum_{i,i^{\prime}}k^{\prime}_{i}\mathfrak{A}_{ii^{\prime}}k^{\prime}_{i^{\prime}}) + 2\pi\mathrm{i} \sum_{j} \Vec{\lambda}_{s_{j}} \cdot \left( \Vec{\omega}_{l^{\prime}} + \sum_{i} k^{\prime}_{i}\Vec{\alpha}_{i} \right) z_{j}}
    \Big) \Big],
\end{align}
where $\mathrm{det}\mathfrak{A} = n$ has been used. Next, the transformation of Dedekind's eta functions gives
\begin{equation}
    \frac{1}{\left( \eta(-1/\tau) \right)^{n-1}} = (-\mathrm{i}\tau)^{-\frac{n-1}{2}} \frac{1}{\left( \eta(\tau) \right)^{n-1}}.
\end{equation}
Finally, the transformation of the ``Jastrow'' factor is given by
\begin{equation}
    E((z_{j}-z_{j^{\prime}})/\tau \vert -1/\tau) = \frac{1}{\tau} \mathrm{e}^{\frac{\pi\mathrm{i}}{\tau}(z_{j}-z_{j^{\prime}})^2} E(z_{j}-z_{j^{\prime}} \vert \tau).
\end{equation}
Using the charge-neutrality condition, one finds
\begin{equation}
    \sum_{j<j^{\prime}}(\Vec{\lambda}_{s_{j}}\cdot\Vec{\lambda}_{s_{j^{\prime}}}) = -\frac{1}{2} ~\underset{j=1}{\overset{N}{\sum}}~(\Vec{\lambda}_{s_{j}})^2 = -\frac{1}{2} N~\frac{n-1}{n},
\end{equation}
and
\begin{equation}
    \sum_{j<j^{\prime}}(\Vec{\lambda}_{s_{j}}\cdot\Vec{\lambda}_{s_{j^{\prime}}}) (z_{j}-z_{j^{\prime}})^2 =  - \sum_{jj^{\prime}}(\Vec{\lambda}_{s_{j}}\cdot\Vec{\lambda}_{s_{j^{\prime}}}) z_{j}z_{j^{\prime}}.
\end{equation}
Thus,
\begin{align}
    \underset{j<j^{\prime}}{\prod} \left( E((z_{j}-z_{j^{\prime}})/\tau \vert -1/\tau) \right)^{\Vec{\lambda}_{s_{j}} \cdot \Vec{\lambda}_{s_{j^{\prime}}}} =~& \tau^{\frac{1}{2} N\frac{n-1}{n}}~ \mathrm{e}^{-\frac{\pi\mathrm{i}}{\tau}\sum_{jj^{\prime}}(\Vec{\lambda}_{s_{j}}\cdot\Vec{\lambda}_{s_{j^{\prime}}})z_{j}z_{j^{\prime}}} \nonumber \\
    & \times \underset{j<j^{\prime}}{\prod} \left( E(z_{j}-z_{j^{\prime}} \vert \tau) \right)^{\Vec{\lambda}_{s_{j}} \cdot \Vec{\lambda}_{s_{j^{\prime}}}}.
\end{align}
The modular parameter of our torus is $\tau = \mathrm{i}$. With $L \in 2n\mathbb{Z}$, we have $\tau^{\frac{1}{2} N\frac{n-1}{n}} = 1$, and the permutation sign (coming from the Klein factors) is unchanged under the modular $S$ transformation.
Putting these three pieces together, one obtains
\begin{equation}
    \vert \psi_{l} \rangle \rightarrow \frac{1}{\sqrt{n}} \left( \vert \psi_{0} \rangle + \underset{l^{\prime}=1}{\overset{n-1}{\sum}} \mathrm{e}^{-2\pi\mathrm{i}\mathfrak{F}_{ll^{\prime}}} \vert \psi_{l^{\prime}} \rangle \right)
\end{equation}
under the modular $S$ transformation. For $l=0$, a similar but simpler calculation gives
\begin{equation}
    \vert \psi_{0} \rangle \rightarrow \frac{1}{\sqrt{n}}  ~\underset{l^{\prime}=0}{\overset{n-1}{\sum}}~\vert \psi_{l^{\prime}} \rangle.
\end{equation}
Therefore, the modular $S$ matrix can be expressed as
\begin{eqnarray}
\mathcal{S}_{ll^{\prime}}=\begin{cases}
\begin{array}{c}
\frac{1}{\sqrt{n}} \mathrm{e}^{-2\pi\mathrm{i}\mathfrak{F}_{ll^{\prime}}} \\
\frac{1}{\sqrt{n}}
\end{array} & \begin{array}{c}
l \neq 0, \, l^{\prime} \neq 0 \\
\textrm{otherwise}
\end{array} \end{cases} .
\end{eqnarray}

\bibliographystyle{JHEP}
\bibliography{chiral}

\end{document}